\documentclass[journal,comsoc]{IEEEtran}

\usepackage[usestackEOL]{stackengine}
\usepackage{footnote}

\usepackage{hyperref}
\hypersetup{
    bookmarksopen=true,
    bookmarksnumbered=true,
    colorlinks=true,
    linkcolor=blue,
    filecolor=magenta,
    citecolor=red,
}
\usepackage{tikz,times}
\usetikzlibrary{mindmap,backgrounds}

\usepackage[backend=bibtex, sorting=none, style=numeric-comp, defernumbers,maxbibnames=99]{biblatex}
\usepackage{pgfplots}
\usepackage{booktabs}

\usepackage{multirow}

\usepackage{color, colortbl}
\definecolor{Gray}{gray}{0.9}

\usepackage{scalerel}
\usetikzlibrary{svg.path}

\definecolor{orcidlogocol}{HTML}{A6CE39}
\tikzset{
    orcidlogo/.pic={
        \fill[orcidlogocol] svg{M256,128c0,70.7-57.3,128-128,128C57.3,256,0,198.7,0,128C0,57.3,57.3,0,128,0C198.7,0,256,57.3,256,128z};
        \fill[white] svg{M86.3,186.2H70.9V79.1h15.4v48.4V186.2z}
        svg{M108.9,79.1h41.6c39.6,0,57,28.3,57,53.6c0,27.5-21.5,53.6-56.8,53.6h-41.8V79.1z M124.3,172.4h24.5c34.9,0,42.9-26.5,42.9-39.7c0-21.5-13.7-39.7-43.7-39.7h-23.7V172.4z}
        svg{M88.7,56.8c0,5.5-4.5,10.1-10.1,10.1c-5.6,0-10.1-4.6-10.1-10.1c0-5.6,4.5-10.1,10.1-10.1C84.2,46.7,88.7,51.3,88.7,56.8z};
    }
}

\newcommand\orcidicon[1]{\href{https://orcid.org/#1}{\mbox{\scalerel*{
                \begin{tikzpicture}[yscale=-1,transform shape]
                \pic{orcidlogo};
                \end{tikzpicture}
            }{|}}}}

\usepackage[edges]{forest}
\forestset{
  direction switch/.style={
    for tree={edge+=thick},
    where level>=1{folder, grow'=0}{for children=forked edge},
    where level=2{}{draw},
  },
}
\usepackage{newtxmath}
\usepackage{caption}

\title{Deep Learning for Radio-based Human Sensing: Recent Advances and Future Directions}

\author{\IEEEauthorblockN{Isura Nirmal$^{\textsuperscript{\orcidicon{0000-0002-2526-4542}}}$\,, \IEEEmembership{Student Member,~IEEE},
Abdelwahed Khamis$^{\textsuperscript{\orcidicon{0000-0002-3475-3479}}}$\,,~\IEEEmembership{Member,~IEEE},
Mahbub Hassan$^{\textsuperscript{\orcidicon{0000-0002-3417-8590}}}$\,,~\IEEEmembership{Senior Member,~IEEE}, 
Wen Hu$^{\textsuperscript{\orcidicon{0000-0002-4076-1811}}}$\,,~\IEEEmembership{Senior Member,~IEEE} and
Xiaoqing Zhu$^{\textsuperscript{\orcidicon{0000-0001-9413-7240}}}$\,,~\IEEEmembership{Senior Member,~IEEE}}
}
\begin{document}
\thispagestyle{empty}
\onecolumn
©2021 IEEE. Personal use of this material is permitted. Permission from IEEE must be obtained for all other uses, in any current or future media, including reprinting/republishing this material for advertising or promotional purposes, creating new collective works, for resale or redistribution to servers or lists, or reuse of any copyrighted component of this work in other works.
\newline
Journal: IEEE Communications Surveys and Tutorials 
\newpage
\twocolumn
\maketitle
\setcounter{page}{1}
\begin{abstract}

While decade-long research has clearly demonstrated the vast potential of radio frequency (RF) for many human sensing tasks, scaling this technology to large scenarios remained problematic with conventional approaches. Recently, researchers have successfully applied deep learning to take radio-based sensing to a new level. Many different types of deep learning models have been proposed to achieve high sensing accuracy over a large population and activity set, as well as in unseen environments. Deep learning has also enabled detection of novel human sensing phenomena that were previously not possible. In this survey, we
provide a comprehensive review and taxonomy of recent research efforts on deep learning based RF sensing. We also identify and compare several publicly released labeled RF sensing datasets that can facilitate such deep learning research.  Finally, we summarize the lessons learned and discuss the current limitations and future directions of deep learning based RF sensing. 
\end{abstract}

\section{Introduction}
\label{introduction}

\begin{table*}[h!]
    \caption{Summary on related surveys}
    \label{table:surveys}
    \centering
    \begin{tabular}[b]{|m{1.3cm}|m{3cm}|m{3cm}|m{6.5cm}|m{1.3cm}|}
        \hline
        \textbf{Reference}& \textbf{Application Scope} & \textbf{Technology Scope} & \textbf{Topic Focus and Taxonomy}& \textbf{Reviewed DL works}\\\hline
        \hline
        
        Ma et al.\newline \cite{ma2019wifisurvey} & 
        Any human as well as beyond human (object, animal, environment) sensing & CSI only & Signal processing techniques and algorithms of WiFi sensing in three categories: detection, recognition, and estimation. 
        &$<$ 5 \\\hline
        
        Liu et al.\newline \cite{Liu2019} & Any human sensing & Any RF based technique (RSS, CSI, FMCW, Doppler, etc.) & RF sensing technologies and their use in human sensing categorized by different applications & $<$ 10 \\\hline 
        
        Yousefi et al. \newline \cite{Yousefi2017} & Any human activity and behavior recognition &
        CSI only & Succinct review of CSI based human sensing techniques and demonstration of performance improvement achieved with LSTM-RNN-based deep learning compared to conventional machine learning & None \\\hline
        
        Farhana et al. \newline \cite{Farhana2020} &  Gesture recognition & CSI only & Comprehensive review of CSI-based gesture recognition based on two approaches: model-based and learning-based; both ML and DL are covered under learning-based approach. & $<15$ \\ \hline
        
        He et al. \newline \cite{iot2020} &  Wifi imaging and all types of human sensing  & CSI only & Concise review of CSI-based sensing applications including imaging and human sensing; the taxonomy is based on applications with minimal coverage of literature involving deep learning  & $<10$ \\ \hline 
        Wang et al. \newline \cite{Wang2019Survey} & Through-wall human sensing & CSI only & Principles, methods and applications of through-the-wall human sensing & $<$ 5 \\\hline
        
        Wang et al. \newline \cite{Wang2019survey2} & Any type of human sensing & CSI only & Comprehensive review of CSI-based human sensing applications based on three categories of classification techniques: model-based (no ML), pattern-based (including conventional ML), and deep learning-based. 
        & $<$ 25 \\\hline
        
        This survey & Any type of human sensing & Any RF based technique (CSI, FMCW, Doppler, Radar, RFID, etc. ) & A systematic review of
        the application of deep learning to RF-based human sensing classified based on the types of employed deep learning techniques. Publicly available datasets are also identified and reviewed &   83 \\\hline
    
    \end{tabular}
\end{table*}

As we increasingly focus on creating smart environments that are ubiquitously aware of their inhabitants, the need for sensing humans in those environments is becoming ever more pressing~\cite{special_issue2019}. Human-sensing refers to
obtaining a range of spatio-temporal information regarding the human, such as the current and past locations, or some
actions performed by the human, such as a gesture or falling to the ground. Such information then can be used by
a range of smart-home or smart-building applications such as turning on/off heating and cooling systems when humans
enter/leave certain areas of the building, detecting trespassers, or monitoring the daily activities of an independently
living elderly resident or patient undergoing rehabilitation from an injury or illness.

Two fundamental approaches to human-sensing are (a) device-based, which requires the person to wear or carry a
device/sensor, such as smartphones or inertial sensors~\cite{Gu2010,Autokey2020}, stretch sensors~\cite{Ejacket2020}, radio frequency (RF)
identification tags~\cite{Rana2014}, and so on, and (b) device-free, which uses sensing elements located in the ambient environment
to monitor human actions without requiring the human to carry any device or sensor at all. Device-based approaches,
although generally accurate, are not practical or convenient in many important real-life scenarios, e.g., requiring the
elderly or a dementia patient to carry a device at all times. Device-free human sensing provides clear advantage for such scenarios.

For device-free human sensing, there is a wide range of existing sensing technology including ultrasound motion
sensors, thermal imaging, microphones/speakers, cameras, light sensors, and so on. Some of these sensors, i.e., motion detectors, thermal imagers, and cameras are not typically available ubiquitously, so must
be pre-deployed specifically for human sensing. Some sensors, such as microphones and camera raise privacy issues. Compared to these sensors, radio signals provide unique advantages as they are often available ubiquitously, such as the WiFi signals at home, and unlike cameras and microphones, they are not privacy-intrusive. Radio signals can `see' behind the walls and in the dark.  Indeed, RF-based device-free human sensing has become an active area of research with significant advancements reported in recent years. Several start-ups~\cite{celeno,emerald,walabot,xkcorp,origin,linksys} now offer commercial RF sensing solutions for sleep monitoring, vital sign monitoring, fall detection, localization and tracking, activity monitoring, people counting, and so on.

\begin{figure}[h]
\includegraphics[width=8cm]{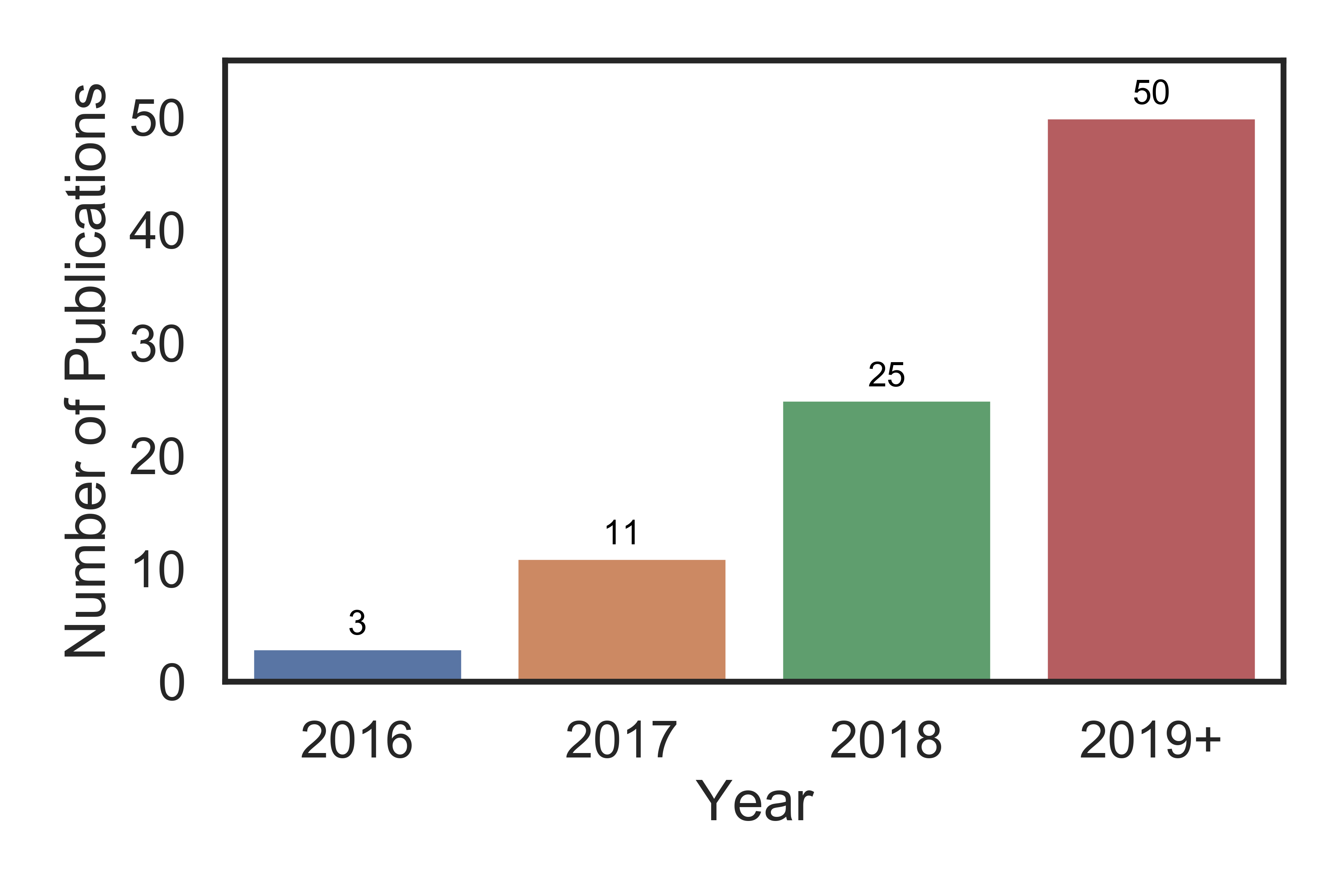}
\centering
\caption{Recent growth in the number of scientific publications reporting the application of deep learning for RF human sensing.}  
\label{fig:fig1}
\end{figure}

Early works in RF human sensing made extensive use of conventional machine learning algorithms to extract manually designed features from radio signals to classify human actions and contexts. Although conventional machine learning was capable of detecting many human contexts in small-scale experiments, they struggled to achieve good accuracy for large-scale deployments. Researchers are now increasingly making use of the latest developments in deep learning to further improve the accuracy, scale, and ubiquity of RF sensing. This trend is clearly evidenced by the growing number of publications in major conferences and journals, as shown in Figure \ref{fig:fig1}, that explore many different deep neural network architectures and algorithms for advancing RF-based human sensing. The success of deep learning for device-free RF human sensing calls for a comprehensive review of the literature for successive researchers to better understand the strengths and weaknesses, and application scenarios of these models and algorithms. 

\textbf{How this survey is different from existing ones?} Although there are several survey articles published in recent years on the topic of wireless device-free human sensing, none of them provides a systematic review of the advancements made in regards to the application of deep learning to this field of research. Since use of deep learning in wireless human sensing started only about five years ago, we compare our review with those surveys published in recent years. Table \ref{table:surveys} compares this survey against seven other recent surveys highlighting the differences in terms of their scope and focus as well as the number of reviewed publications that applied deep learning in wireless sensing. We can see that \textbf{none of the existing surveys focus their work on deep learning}. They rather restrict their surveys on specific radio measurement technology, such as Channel State Information (CSI)~\cite{ma2019wifisurvey,Yousefi2017,Wang2019survey2,Farhana2020,iot2020}, or on the sensing application, such as through-the-wall sensing~\cite{Wang2019Survey}, which prevents them from achieving a comprehensive analysis of the progress made in deep learning application to wireless sensing. The survey conducted by Wang et al.~\cite{Wang2019survey2} is the closest to our work as they have specifically reviewed deep learning publications as one of their categories. However, as the survey was restricted to CSI, they covered only about 25 deep learning papers and missed many important recent advancements. 

Given the rising popularity of the application of deep learning in wireless sensing, a more inclusive review would be of high value to the research community to gain deeper insight to these advancements. We conduct a systematic review without any restriction on the radio technology or human sensing application. To this end, more than 80 deep learning works have been surveyed and classified to  provide a comprehensive picture of the latest advancements in this research. We also review 20 public datasets of labeled radio measurements, which is not covered in existing surveys. Finally, we provide a more comprehensive discussion on the lessons learned and future directions for this growing field of research.  

\textbf{How did we select the papers?} Semantic Scholar and Google Scholar are the two main databases used to perform the initial search for the relevant papers using combinations of several keywords including: WiFi, wireless, device-free, activity recognition, localization, and deep learning. We also specifically inspected the proceedings of the following major conferences from 2018 onwards: MobiCom, MobiSys, Infocom, Ubicomp, PerCom, IPSN, SenSys, NSDI, and SIGCOMM. In addition, we inspected the following three specialised machine learning conferences: CVPR, ICCV, and ICML. The entire literature review was managed in Mendeley, which provided its own recommendations of relevant papers from time to time. Our search revealed in excess of 130 publications that considered some form of deep learning for RF human sensing, but we finally selected about 90, i.e., only those that were published in major conferences and journals with noteworthy contributions to the field. When preparing the ``dataset section" of our survey, we searched public academic dataset repositories such as IEEE Dataport, Harvard Dataverse, Figshare, Mendeley Data, and so on, in addition to the web pages of the authors who mentioned public data release in their publications. 

\textbf{Contributions of this survey.} The goal of this survey is to thoroughly review the literature to understand the landscape of recent advancements made in deep learning-based RF human sensing. It serves as a quick guide for the reader to understand which deep learning techniques were successful in solving which aspects of the RF sensing problem, what limitations they faced, and what are some of the future directions for research. It also serves as a `dataset guide' for those researchers who do not have the means to collect and label own data, but wishes to venture into deep learning-based RF human sensing research using only publicly available data. We believe that the detailed public dataset information provided in this survey will also be useful for researchers who have their own data, but would like to evaluate their proposed algorithms with additional independent datasets. Our survey therefore is expected to be a useful reference for future researchers and help accelerate deep learning research in RF sensing. The key contributions of this survey can be summarized as follows: 
\begin{enumerate}
    \item We provide a comprehensive review and taxonomy of recent advancements in deep learning-based RF sensing. We first classify all works based on the fundamental deep learning algorithms used. Different approaches within a given class are then compared and contrasted to provide a more fine-grained view of the application of deep learning to the specific problems of RF sensing.   
    \item We identify and review 20 recently released public datasets of radio signal measurements of labeled human activities that can be readily used by future researchers for exploring novel deep learning methods for RF sensing. 
    \item We discuss current limitations as well as opportunities and future directions of deep learning based RF sensing covering recent developments in cognate fields such as drone-mounted wireless networks and metamaterials-based programmable wireless environments. 
    
\end{enumerate}

The rest of this paper is organized as follows. Section \ref{preliminaries} introduces the preliminaries for RF sensing and deep neural networks. Section \ref{deeplearningworks} presents our classification framework and provides a detailed analysis of the state-of-the-art. Section \ref{sec:datasets} introduces the recently released RF sensing datasets that are freely available to conduct future research in this area. Lessons learned and future research directions are discussed in Section \ref{futuredirections} and Section \ref{conclusion} concludes the paper.

\begin{figure}[h]
    \includegraphics[width=8cm]{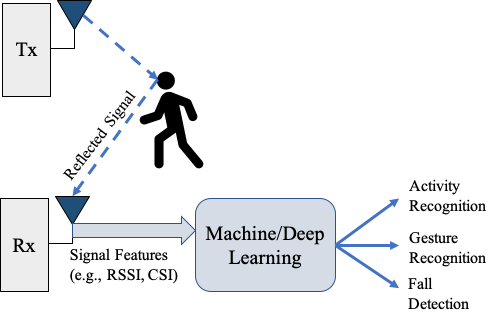}
    \centering
    \caption{Principles of RF human sensing.}  
    \label{fig:RFsensing}
\end{figure}
    
\begin{figure}[t]
    \centerline{\includegraphics[width=0.5\textwidth]{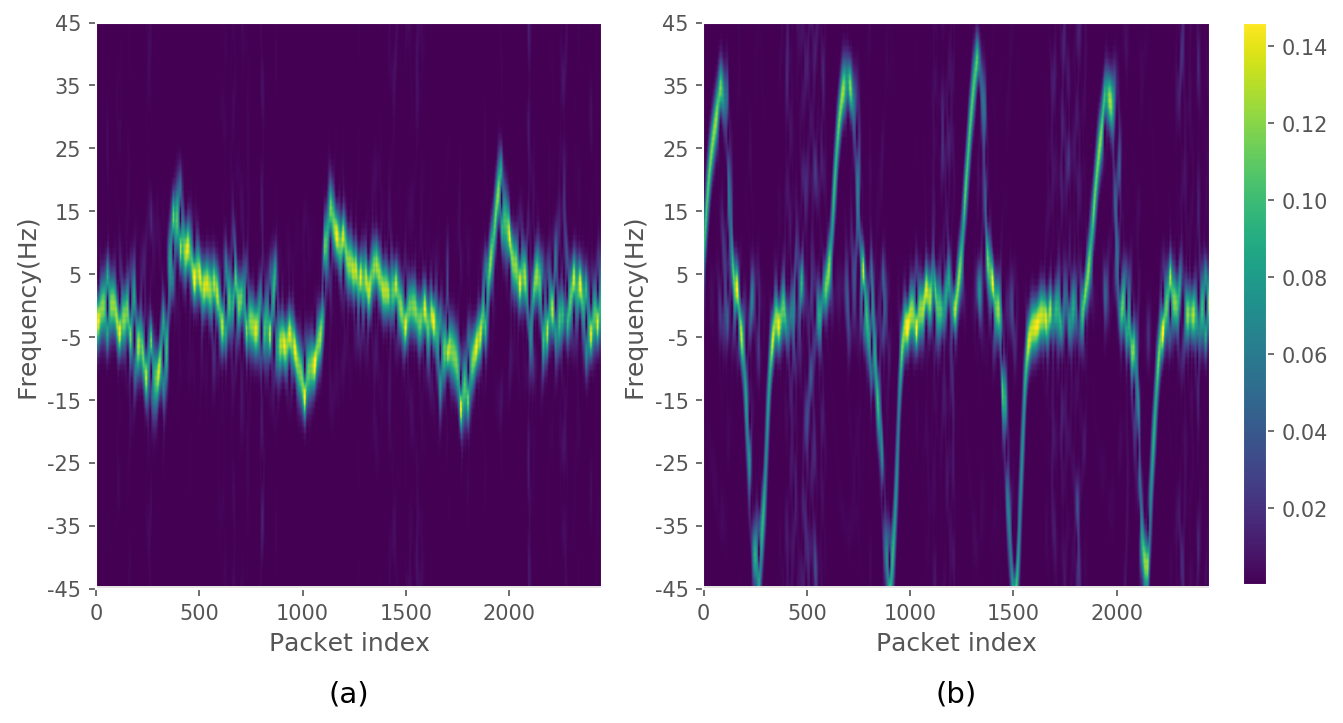}}
    \centering
    \caption{WiFi CSI spectograms obtained in our laboratory for two different gestures: (a) the right leg moving back-and-forth, and  (b) the right hand doing push-and-pull. }  
    \label{fig:spectogram}
\end{figure}
    
\begin{figure}[h]
    \includegraphics[width=8cm]{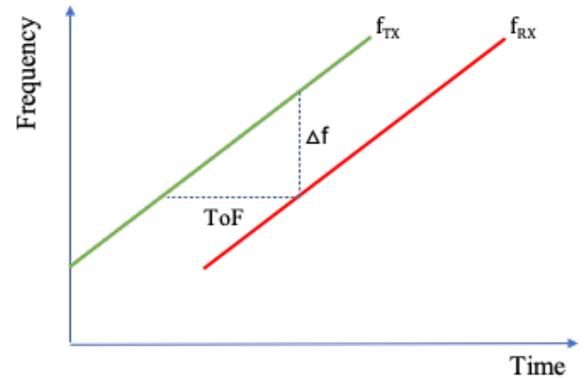}
    \centering
    \caption{Principles of FMCW.}  
    \label{fig:FMCW}
\end{figure}

\section{Overview of RF Human Sensing and Deep Learning} 
\label{preliminaries}

In this section, we first review the basic principles, instruments, and techniques for both RF human sensing and deep learning. We then briefly discuss the potential of deep learning in RF sensing. 

\subsection{RF Human Sensing}
\label{s:RFsensing}
Figure \ref{fig:RFsensing} illustrates the basic principles of RF human sensing. The presence and movements of a human in the vicinity of an ongoing wireless transmission cause changes in the wireless signal reflections, which in turn results in variation in the received signal properties, i.e., its amplitude, phase, frequency, angle of arrival (AoA), time of flight (ToF) and so on. Since different human movements and postures affect the wireless signal in unique ways, it is possible to detect a wide variety of human contexts, such as location, activity, gesture, gait, etc., by modeling the signal changes or simply learning the signal changing patterns with machine learning.   

To model changes in signal properties, they must be measured precisely.  There is a wide range of metrics of varied complexity to measure different signal properties. The RF signal metrics widely used for human and other sensing are reviewed below. 

\textbf{Received Signal Strength (RSS):} RSS is the most basic and pervasively available metric, which represents the average signal amplitude over the whole channel bandwidth. By moving in front of the wireless receiver, a human can noticeably affect the RSS, which has been successfully exploited by researchers to recognize hand gestures performed near a mobile phone fitted with WiFi~\cite{WiGest}. RSS, however, does not capture the signal phase changes caused by reflection and it varies randomly to some extent without even any changes occurring in the environment. RSS, therefore, is considered good only for detecting very basic contexts and cannot be used for fine-grained human activity recognition.  

\textbf{Channel State Information (CSI):} CSI captures the frequency response of the wireless channel, i.e., it tells us how different frequencies will attenuate and change their phases while travelling from the transmitter to the receiver. The receiver calculates the CSI by comparing the known transmitted signals in the packet preamble or pilot carriers to the received signals, and then use the CSI to accurately detect the unknown data symbols contained in the packet. In contrast to a single power value returned by RSS, CSI provides a set of values $a_{n}e^{j\theta}$, capturing signal attenuation $a_n$ and phase shift $\theta_{n}$ for each frequency (sub-carrier) that makes up the communication channel. For example, a typical 20MHz Orthogonal Frequency-Division Multiplexing (OFDM) WiFi channel has 52 data sub-carriers, which allows the receiver to compute 52 amplitude and phase values for each packet received. For human sensing, a series of packets are transmitted, which yields a time series of CSI at the receiver. The patterns in the raw CSI time series, or in their fast Fourier transforms (FFTs), which is referred to as \textbf{CSI spectogram}, reflect the corresponding human activity as illustrated in Figure \ref{fig:spectogram}. Such CSI spectograms are the popular choice for training machine learning models for the recognition of various human activities~\cite{rf-fall,Wang2019}. 

In commodity WiFi, CSI is computed and used at the physical layer of the communications protocol. Use of CSI in human sensing algorithms therefore requires additional tools and techniques for the extraction of the CSI from the physical layer to the user space. In the past, only expensive software defined radios like the Wireless Open Access Research Platform (WARP)~\cite{warp} and the Universal Software Radio Peripheral (USRP)~\cite{usrp} could provide CSI to the user application. Now publicly available software tools, such as nexmon~\cite{nexmon}, are available freely that allow WiFi CSI extraction in most commodity platforms including mobile phones, laptops, and even Raspberry Pi. A detailed comparison of all available CSI extraction tools can be found in~\cite{nexmon}. Easy access to such tools have made CSI one of the most widely used signal metric for RF human sensing~\cite{ma2019wifisurvey,Yousefi2017}.      

Although both amplitude and phase information are available in CSI, the amplitude is by far the most commonly used metric in WiFi because the returned phase values are usually very noisy in commodity WiFi platforms due to the absence of synchronization between the sender and the receiver~\cite{wang2017tensorbeat}. Simple transformations of CSI values, however, proved to be very useful. For example, phase differences between sub-carriers have been shown to mitigate the noise effect~\cite{wang2017tensorbeat} and was successfully employed in a number vital sign sensing applications~\cite{khamis2018cardiofi,wang2017phasebeat}. FullBreathe~\cite{zeng2018fullbreathe} applied conjugate multiplication of CSI from two receiver antennas to remove the phase offset, which enabled accurate detection of human respiration using CSI. 

\textbf{Time of Flight (ToF) and range estimation:} RSS and CSI cannot be used to estimate the range or distance of a person from a radio receiver. Range estimation can be very useful for human sensing because it can help localizing a person and detect the presence of multiple persons in the environment located at different distances from the receiver. If ToF of the signal is known, then the range can be estimated as the product of ToF and the speed of light. Typically, expensive and bulky radar systems are used in most military and civilian applications to detect objects and estimate their ranges by transmitting a series of ultra short pulses of duration on the order of nano or micro-seconds and then recording their reflections from the object at the receiver located in the same device and using the same clock. ToF is measured directly from the time measurements of the transmitted and received pulses. However, as short pulses consume massive bandwidth, very high sampling rate is required to process the received pulses, which in turn leads to high analog-to-digital power consumption. Due to the lack of a low-power compact radar device, use of radar technology for ubiquitous human sensing was not considered a viable option until recently. 

Frequency Modulated Continuous Wave (FMCW) is an alternative radar technology that transmits continuous waves allowing the transmitted signal to stay within a constant-power envelop (as opposed to an impulse). Use of continuous wave enables low-power and low-cost signal processing, which has recently led to the commercial development of commodity embedded  FMCW radars~\cite{TI-FMCW} that can be ubiquitously deployed in indoor spaces for human sensing. The principle of FMCW is illustrated in Figure \ref{fig:FMCW}. Basically, the transmitter sends a chirp with linearly increasing frequency and then the received signal is compared with the transmitted signal at any point of time to compute the frequency difference, $\Delta f$. Since the $slope$ of the linear chirp is known, the ToF is simply obtained as $ToF = \frac{\Delta f}{slope}$. If there are multiple persons in the environment located at different distances from the radar, FMCW can detect all of them because each person's reflection would produce a different received chirp at the radar. 

\textbf{Doppler shift:} The ability to measure the motion, i.e., the velocity of different human body parts, is critical to accurately detect human activities irrespective of the wireless environment where the activities are performed.    Doppler shift is a well-known theory~\cite{goldsmith_2005} that captures the effect of mobility on the observed frequency of the wireless signal. According to this theory, the observed frequency would appear to be higher than the transmitted frequency if the transmitter moves towards the receiver, and lower than the transmitted frequency if moving away from the receiver. The amount of frequency increase or decrease, i.e., the Doppler shift, is obtained as $$\Delta f = \pm\frac{vf}{c},$$ where $f$ is the transmitted frequency, $v$ is the velocity at which the transmitter moves towards the receiver, and $c$ is the speed of light. 

Now imagine that the person in Figure \ref{fig:RFsensing} moves his hand towards the receiver and then pulls it back as part of a complete gesture. The frequency of the reflected signal will then increase first and then decrease, which provides a unique frequency change (Doppler shift) pattern for that gesture. Indeed, Doppler shift has been exploited successfully for many human sensing applications \cite{qian2017inferring,Wisee,WiFiU}. If different users are moving at different speeds towards the receiver, then it is also possible to track \textit{multiple} people~\cite{Wisee} in the same environment, which is difficult to achieve using CSI. Unfortunately, existing commodity WiFi hardware do not explicitly report Doppler shifts. It is however possible to estimate Doppler shift from the CSI by using signals from multiple receivers located at different locations in the space~\cite{qian2017inferring,zheng2019zero}. Pu et. al.~\cite{Wisee} explains a detailed implementation of USRP-based Doppler shift extraction method from OFDM signals. Using a 2-dimensional FFT on the ToF estimates, some FMCW radar products, e.g., the mmWave industrial radar sensors from Texas Instruments~\cite{TI-FMCW}, can generate velocities as well. With access to velocity measurements, it is possible to detect and monitor multiple persons even if they are located at the same distance from the radar but moving at different speeds; such as performing different gestures.  

\textbf{Angle of Arrival (AoA)}: Human sensing could be further facilitated with the detection of the direction of arrival (DoA) or angle of arrival (AoA) of the signal reflected by the human. Fortunately, AoA can be accurately computed with an antenna array at the receiver. Although commodity WiFi hardware do not report AoA even if they are fitted with multiple antennas, the TI FMCW radar sensors provide multiple antenna options and reporting of AoA.

As different signal metrics capture different aspects of the environment, they can be combined for more detailed and complex human sensing. For example, range and Doppler effect were combined for  multi-user gait recognition~\cite{yang2020muid}, while researchers were able to significantly improve WiFi localization by combining Doppler effect, range, and AoA~\cite{xie2019mdtrack}.

\subsection{Deep Learning} 
\label{s:DL}
Deep learning refers to the branch of machine learning that employs artificial neural networks (ANNs) with many layers (hence called ``deep") of interconnected \textit{neurons} to extract relevant information from a vast amount of data. Fundamentally, each neuron employs an \textit{activation function} to produce a output signal from a set of weighted inputs coming from other neurons in adjacent layers. The key to successful learning is the iterative adjustment of all these weights as more and more data samples are fed to the network during the training phase. Historically, such deep neural networks were not considered attractive due to the massive computing resources and the enormously long time that would be required to train them. With recent advances in computing architectures, e.g., graphical processing units (GPUs), and algorithmic breakthroughs during the training procedures, e.g., works by LeCun et al.~\cite{lecun2015}, deep learning has become much more affordable. 
This has sparked intense research exploring new deep learning architectures and their use cases in many domains such as face recognition, image processing, natural language processing, and so on. 

The extensive research in recent years has produced a plethora of deep learning architectures, each with its own specific characteristics and advantages. While some of them are too specialized targeting very niche applications, others are general enough to be applied in different application areas. In this section, we provide a brief introduction to some of the widely used general architectures which also have been successfully applied to RF sensing in recent years.

Before discussing specific deep learning architectures, we would like to highlight a few fundamental concepts concerning their training and usage. A deep learning architecture is said to work \textit{unsupervised} when we do not have to label the data used for its training. On the other hand, \textit{supervised} learning refers to the situation when the input data has to be labeled. Generally speaking, data labeling is often a labour-intensive task, especially for deep learning due to the huge amount of data required for training such architectures. Unfortunately, certain use cases must employ some levels of supervised learning, although there are use cases that require only unsupervised deep learning. Finally, some deep learning architectures are called \textit{generative} as they are designed and trained to generate new data samples. Some of the impressive use cases of \textit{generative} deep learning includes generating realistic photographs of human faces, image-to-image translation, text-to-image translation, clothing translation, 3D object generation, and so on. 

In the following, we briefly examine the characteristics and use cases of the widely used deep learning architectures with a summary provided in Table \ref{table:Architectures}. For more detailed guidance on how to construct and implement these networks, readers are referred to many available tutorials on deep learning, e.g.,~\cite{Goodfellow-et-al-2016,samira2018}. Applications of these networks to RF sensing is covered in Section~\ref{deeplearningworks}.

\textbf{Multilayer Perceptron (MLP)} is the most basic and also the classical deep neural network consisting of an input layer, an output layer, and one or more hidden layers which are fully connected  as illustrated in the topology column in Table \ref{table:Architectures}. Each layer in turn consists of one or more neurons or perceptrons. The main function of the input layer is to accept the input vector from a data sample and as such the number of perceptrons in this layer is scaled to the feature vector of the problem. Each perceptron in a hidden layer uses a non-linear activation function to produce an output from the input weights and then passes the output to the next layer (forward propagation). MLPs make use of supervised learning where the labeled data is used for training. Learning occurs incrementally by updating the learned weights after each data sample is processed, based on the amount of loss in the output compared to the expected result (backward propagation).  The output layer mostly uses an activation function depending on the expected result (classification, regression, etc.)

\textbf{Restricted Boltzman Machine (RBM)} is a generative unsupervised ANN with only two layers, an input (visible) layer and one hidden layer. Neurons from one layer can communicate with neurons from another layer, but intra-layer communication is not allowed (hence the word ``restricted"), which basically makes RBM a bipartite graph. RBM has been successfully used for recommending movies for users. 

\textbf{Convolutional Neural Networks (CNN)} or ConvNets are designed to process visual images consisting of rows and columns of pixels. As such, it expects to work with 2D grid-like inputs with spatial relationships between them. CNNs employ a set of filters (or kernels) to convolve in the inputs to learn the spatial features. When multiple layers are employed, CNNs learn the hierarchical representations from the given data set. Further pooling layers are also added to reduce the learned dimentionality when designing the network.  Interestingly, although originally designed to work with images, CNNs are also found to be effective in learning spatial relationships in one-dimensional data, such as the order relationship between words in a text document or in the time steps of a time series. 

\textbf{Recurrent Neural Networks (RNNs)} were designed to work with sequence prediction problems by utilizing the feedback mechanism in each \textit{recurrent unit}. This intra-hidden-unit connections make it possible to memorize the temporal features of the inputs. However, RNNs suffer from two issues. Vanishing gradient problem occurs when gradient updates are so insignificant that the network stops learning. Exploded gradient problem occurs when the cumulative weights' gradients in back propagation result a large update to the network. Due to these shortcomings, RNNs were traditionally difficult to train and did not become popular until the variants  called Long Short-Term Memory (LSTM) and Gates Recurrent Unit (GRU) were invented. Instead of, single non-linear activation function, multiples of  functions and copying/concatenation were added to memorize long term dependencies of the inputs. The difference between LSTM and GRU comes from the number of internal activation functions and how the interconnections are handled. RNN's successors have been used successfully for many sequence detection problems, especially natural language processing.

\textbf{Autoencoder (AN)} is fundamentally a dimension reduction (or compression) technique, which contains two main components called \textit{encoder} and \textit{decoder}. Encoder transforms  input data into encoded representation with the lowest possible dimensions. The decoder then learns to reconstruct the input from this compact representation. Because the input serves as the target output, the autoencoder can self-supervise itself requiring no explicit data labeling. Variants including Denoising Autoencoders(DAE) are increasingly used to produce cleaner and sharper speech, image, and video from their noisy sources. Variational autoencoder (VAE) is a more advanced form of autoencoder designed to learn the probability distribution of the input data using principles of Bayesian statistics. The VAE thus can generate new samples with interesting use cases such as generating artificial (non-existent) fashion models, synthesizing new music or art, etc., that are drawn from the learned distribution and hence perceived as real.

\textbf{Generative Adversarial Networks (GANs)} are another type of unsupervised generative deep learning architecture designed to learn any data distribution from a training set. The main technical difference with VAE is in the method used to learn the distribution. Unlike VAE, which explicitly estimates the parameters of the distribution, GAN simultaneously trains two networks using a two-player game, hence the word ``advarsarial", to directly generate the samples without having to explicitly obtain the distribution parameters. The first network, \textit{generator}, tries to fool the second network, \textit{discriminator}, by generating new samples that look like real samples. The job of the discriminator is to detect the generated samples as fakes. The performance of the two networks improve over time and the training ends when the discriminator cannot distinguish the generated data from the real data. GANs have undoubtedly revolutionized the deep learning research with multiple variants of GAN models in state-of-the-art. It is noteworthy to mention that architectures like Domain Adversarial Neural  Networks (DANN) \cite{Ganin2016} removes the generative property but makes it possible to learn the distributions between two different domains and perform accurate classifications for both domains using a single model. Since we discuss both generative and non-generative adversarial networks in our work, we use Adversarial Networks (AN) henceforward to refer to both types of networks.

Finally, \textbf{hybrid models} contain the characteristics of more than two primary deep neural networks and hence can help overcome the hybrid nature of the problems they address. For example, CNN and LSTM are often combined to capture information latent in both spatial and temporal dimensions of the dataset. 

\begin{table*}[h!]
    \caption{Popular deep learning architectures}
    \label{table:Architectures}
    \centering
    \begin{tabular}[b]{|m{1.5cm}|m{2.5cm}|m{2.8cm}|m{1.7cm}|m{3.5cm}|m{3cm}|}
    \hline
    \textbf{Architecture} &  \textbf{Strengths}&\textbf{Weaknesses}&\textbf{Learning type}&\textbf{Use cases in RF sensing} & \textbf{Example Topology} \\ \hline
    \hline
    
        MLP & Simple structure \newline& Slow to converge \newline Modest performance & Supervised & Activity classification \newline Pattern recognition \newline Transfer learning& 
        \leavevmode\newline\includegraphics[scale=0.25]{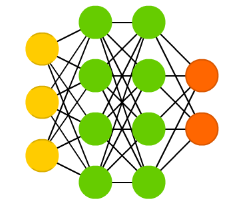}
        \\\hline
        RBM & Simple structure&Specific training requirements&Unsupervised &Feature extraction \newline Activity classification \newline Collaborative filtering&\leavevmode\newline\includegraphics[scale=0.25]{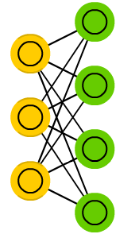} \\\hline
        CNN&Spatial feature \newline identification& High complexity in parameter tuning&Supervised &Radio image processing\newline Video  analysis \newline &\leavevmode\newline\includegraphics[scale=0.20]{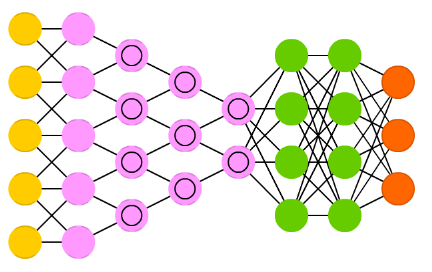} \\\hline
        RNN&Temporal feature \newline identification&High complexity in model and gradient vanishing&Supervised&Radio time series data analysis&\leavevmode\newline\includegraphics[scale=0.15]{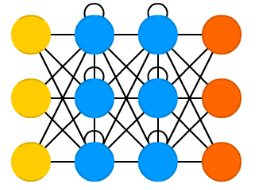} \newline\\\hline
        AE&Representational learning \newline Denosing \newline Feature compression &Costly to train &Unsupervised&Radio feature extraction\newline translation &\leavevmode\newline\includegraphics[scale=0.2]{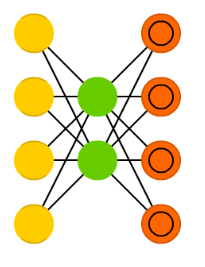} \newline\\\hline
        AN&Robust against \newline the adverserial attacks&High model complexity&Semi-supervised \newline Reinforcement &Signal  feature extraction\newline feature synthesis \newline Classification  &\leavevmode\newline\includegraphics[scale=0.25]{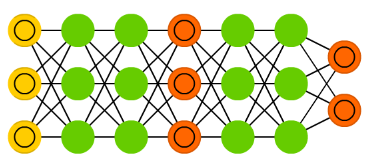} \newline\\\hline
    \multicolumn{3}{l}{} \\
    \multicolumn{6}{l}{\leavevmode\newline\includegraphics[scale=0.3]{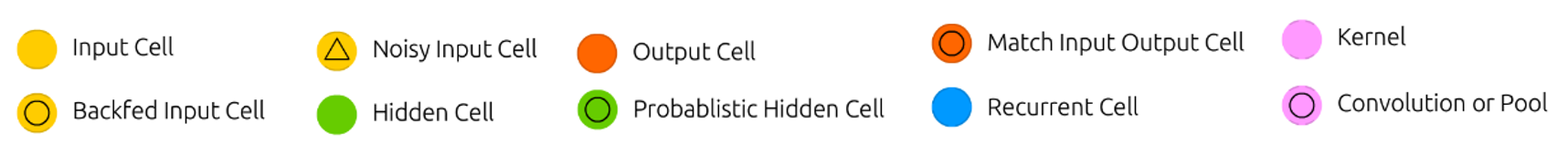}} \\
    \multicolumn{3}{l}{$^*$Legend for the notation} \\
    \multicolumn{3}{l}{$^\dagger$Topology figures are courtesy of \cite{veen_2019} } 
\end{tabular}
\end{table*}

\subsection{Why Deep Learning in RF Sensing}

Mapping RF signals to humans and their activities is a complex task as the signals can reflect from many objects and people in the surrounding. In most cases, the problem is mathematically not tractable, which motivated researchers to adopt machine learning as an effective tool for RF human sensing. Conventional machine learning algorithms, however, are limited in their capacity to fully capture the rich information contained in complex unstructured RF data. Deep learning provides the researchers exceptional flexibility to tune the `depth' of the learning networks until the necessary features are successfully learned for a given sensing application. With the emergence of more powerful radio hardware and protocols, such as multi-input-multi-output (MIMO) systems, multi-antenna radar sensors, and so on, researchers now have the ability to generate a vast amount of RF data for any given human scene, which help train deep neural networks. Deep learning therefore becomes a new tool to push the boundaries of RF sensing on multiple fronts such as enhancing existing sensing applications in terms of accuracy and scale, realizing completely new applications, and achieving more generalized models that work reliably across many different, and even unseen, environments.

\begin{figure}
\centering

\begin{tikzpicture}
\begin{axis}[
    ybar,
    enlargelimits=0.15,
    ylabel={Accuracy (\%)},
    symbolic x coords={
    Activity\cite{xue2020deepmv},
    Fall\cite{rf-fall},
    Gesture\cite{ma18signfi}
    },
    xtick=data,
    tick label style={font=\footnotesize},
    legend style={at={(0.5,-0.15)},
    anchor=north,legend columns=-1},
    nodes near coords,
    every node near coord/.append style={font=\normalfont},
   nodes near coords align={vertical},
    ]
\addplot coordinates {
(Activity\cite{xue2020deepmv}, 23) 
(Fall\cite{rf-fall}, 16)
(Gesture\cite{ma18signfi}, 68) 
};
\addplot coordinates {
(Activity\cite{xue2020deepmv},  84)
(Fall\cite{rf-fall}, 94) 
(Gesture\cite{ma18signfi}, 97) 
};
\legend{Shallow Learning, Deep Learning}
\end{axis}
\end{tikzpicture}
\caption{Evidence of deep learning's capability to significantly enhance the detection accuracy for popular RF sensing applications. Specialized versions of CNN were used for deep learning in all these experiments. Random Forest, SVM, and kNN were used as the baseline shallow learning for Activity, Fall, and Gesture detection, respectively.}
\label{fig:performance_gain}
\end{figure}
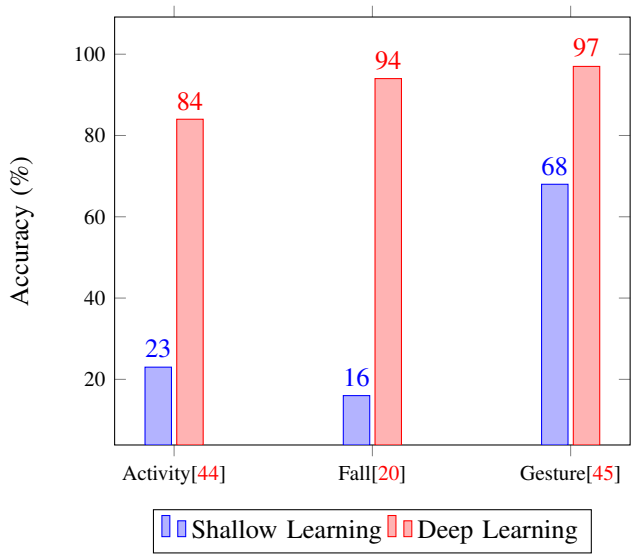

\begin{figure}[h]
    \centering
    \includegraphics[width=6cm]{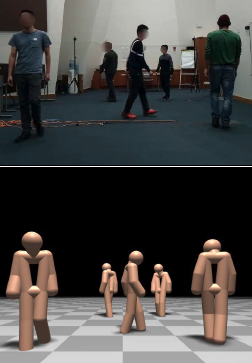}
    \centering
    \caption{Generating and tracking 3D human skeletons from RF signals by leveraging the power of deep learning. The top figure is a camera capture of five people while the bottom figure shows the 3D skeletons of all of these persons constructed from FMCW radio data with the help of a specialized CNN model designed by Zhao et al.~\cite{zhao3dpose} (Figure courtesy of ~\cite{zhao3dpose}). }
    \label{fig:rf_pose_timeline}
\end{figure}

In Figure \ref{fig:performance_gain}, we highlight evidence from the recent literature confirming the capability of deep learning in enhancing the detection accuracy significantly compared to the conventional shallow learning for three popular RF sensing applications.  Figure \ref{fig:rf_pose_timeline} shows a completely new RF sensing application, namely RF-Pose3D~\cite{zhao3dpose}, which uses a specialized CNN architecture to estimate simultaneously the 3D locations of 14 keypoints on the body to generate and track humans in 3D. Finally, researchers are now discovering deep learning solutions that can remove the environment
and subject specific information contained in the RF
data to generalize RF-based human sensing for ubiquitous deployments~\cite{jiang2018towards}. In the following section, we are going to survey many more recent advances in deep learning for RF sensing.

\section{Survey of Deep Learning based RF Sensing} \label{deeplearningworks}
In this section, we survey device-free RF human sensing research that used deep learning to analyse RF signals. In particular, we survey a total of 84 research publications and classify them according to the main deep learning architecture employed. Table \ref{table:DLApplications} provides a summary of this classification with rows indicating the deep learning architecture and the columns showing the application domain of the research. The table reveals that deep learning has expanded its footprint across all popular sensing domains, from localization through to gait recognition, using a good mix of neural network architectures. We cover each deep learning architecture (each row of Table \ref{table:DLApplications}) in a separate subsection, where we further compare and contrast the ways the architecture is implemented and investigated by different researchers.

\begin{table*}[h!]
    \caption{Categorization of reviewed publications based on their deep learning techniques.}
    \label{table:DLApplications}
    \centering
    \setlength\extrarowheight{2pt}
    \begin{tabular}{|p{1.5cm}|p{1.5cm}|p{2cm}|p{2cm}|p{2cm}|p{2cm}|p{1cm}|p{1.5cm}|} 
    \hline
    \textbf{Category} &  \textbf{Localization}&\textbf{Activity \newline Recognition}&\textbf{Gesture \newline Recognition} & \textbf{Human Detection} & \textbf{Vital Sign Monitoring } &\textbf{Pose\newline Estimation}&\textbf{Gait \newline Recognition}\\ \hline
    \hline
        MLP&
        \cite{Liu2019AnalysisAV}\cite{Zhou2019}&
        \cite{Wu2018}&
        \cite{Zhang2018}&
        \cite{Liu2017WiCountAD}\cite{chengcount17}\cite{fang19enhanced}&
        -&
        -&
        \cite{Zhang2018} \\ \hline
        RBM&
        \cite{Zhao2018}&
        -&
        \cite{Zhou2016}&
        -&
        -&
        -&
        -\\ \hline
        CNN&
        \cite{Cai2018PILCPI}\cite{Wang2019JointAR}&
        \cite{Lv2017QualitativeAR}\cite{xue2020deepmv}\cite{hsu2019enabling}\cite{li2019making}\newline \cite{fall18}\cite{Wang2019JointAR}\newline \cite{noprocess19}\cite{Zou2019}&
        \cite{Ma2020} \cite{finger19}\cite{skaria19}\newline  \cite{Zhou2018}\cite{Zou2018WiFienabledDG}\cite{santhalingam2020mmasl}\newline \cite{ma18signfi}&
        \cite{vander18} \cite{sobron18}\cite{Huang2018WiDetWB}\cite{fan2020learning}\cite{wang2019person} &
        \cite{khanpassive17}\cite{zhao2017learning}\cite{rf-fall}&
        \cite{unet18}\cite{zhaopose18}&
        -\\\hline
        RNN&
        -&
        \cite{shi18}\cite{Wang2019}\cite{Chen2019} \newline \cite{feng2019wi}\cite{Yousefi2017}\cite{Shi2019} &
        \cite{haseeb2017wisture}\cite{Kong2019} &
        \cite{Ibrahim2019}&
        -&
        -&
        \cite{wanglimb18}\cite{Ming2019}\\
        \hline
        AE&
        \cite{wanglocact17}\cite{Chen2017TamingTI}\cite{Khatab2018} \newline \cite{Chang2018DeviceFreeIL}\cite{gao17}\cite{zhang16}\newline \cite{Zhao2019}    \cite{Chen2020}&
        \cite{wanglocact17}\cite{gao17}\cite{zhang16}&
        -&
        \cite{shi17}&
        -&
        \cite{zhaopose18}\cite{wang2019person}&
        \cite{Xu2018AttentionbasedWG}\\
        \hline
        AN&
        -&
        \cite{CsiGAN2019}\cite{shu2018dirt}\cite{wang2019wicar,wang2020multi}\cite{jiang2018towards}&
        \cite{zou2018joint}\cite{yu2019rfid}\cite{Wang2020}&
        -&
        \cite{zhao2017learning}&
        -&
        -\\\hline
        Hybrid&
        -&
        \cite{Xue2019}\cite{Zhang2019}\cite{khan19}\newline \cite{yao2019stfnets}\cite{singh2019radhar}\cite{Zou2018}\newline \cite{Fan2019WhenRM} \cite{Wang2019DeepApproach}\cite{Fan2018TagFreeAI}\newline \cite{zou2018towards}&
        \cite{WIHF-infocom2020}\cite{soli16}\newline \cite{zheng2019zero}\cite{latern}& \cite{HuangAuID19}\cite{liu2019deepcount}&
        -&
        \cite{zhao3dpose}\cite{Jiang2020Towards3H}&
        -\\\hline

\end{tabular}
\end{table*}

\subsection{RF Sensing with MLP}
\label{sec:mlp}

\begin{table*}
    \centering
    \caption{MLP architectures in RF sensing}
    \label{table:MLPApp}
    \begin{tabular}{|p{0.6cm}|p{3cm}|p{2cm}|p{1.5cm}|p{4cm}|}
    \hline
    \textbf{Paper}&\textbf{Application}&\textbf{Radio Measurement}&\textbf{Layers} & \textbf{Performance}\\\hline
    \hline
    \cite{chengcount17}&Human Detection&CSI&3 &Accuracy \newline Fixed Location 0.96  \newline Arbitrary Location  0.88\\
    \hline
    \cite{fang19enhanced}&Human Detection&CSI& - & Accuracy  0.93 \\
    \hline
    \cite{Liu2019AnalysisAV}&Localization&CSI& 3 & Precision 0.86  \\
    \hline
    \cite{Zhou2019}&Localization&CSI& 5 & Mean Distance Error 0.54 m \\
    \hline
    \cite{Liu2017WiCountAD}&Human Detection&CSI& 2 & Accuracy 0.823 \\
    \hline
    \cite{Zhang2018}&Gait\addslash Gesture Recognition&CSI& 7 & Accuracy  Gait\addslash Gesture 0.94\addslash0.98 \\
    \hline   
    \cite{Wu2018}&Activity Recognition&CSI& 1 & Accuracy 0.94\\
    \hline
\end{tabular}
\end{table*}

Among major works in MLP based localization which consider the deep neural network training as a black box, Liu et al.~\cite{Liu2019AnalysisAV} conducted a visual analysis to understand the signature features of wireless localization  using visual analytics techniques, namely, dimensionality reduction visualization and visual analytics and information visualization  to better understand the learning process of MLP for localizing a human subject. The activations of deep models last layers (for a 3-hidden layer MLP) have shown well separated clusters of the learned weights (using $t$-SNE) after training process,For 16 predefined target locations, 86.06\%  average precision was achieved. 

Among a large number of object localization based on wireless sensing literature, FreeTrack\cite{Zhou2019} presented a MLP based localization approach for moving targets. Denoised CSI amplitude information is used taken as inputs to the MLP model (5 fully connected layers) which achieve 86 cm mean distance error and reduced to 54 cm with particle filter and map matching which are able to detect the obstacles in the environment.Extensive tests were introduced including multiple walking speeds, subjects, sampling rates  have proven the extendability and robustness of the model with state-of-the art. 

WiCount \cite{Liu2017WiCountAD} utilized a MLP to count the crowd using WiFi devices in environment. Its noteworthy to mention that the multi user sensing is rarely researched area due to its difficulty. WiCount used both amplitude and phase information of the WiFi signal. CSI data is preprocessed by using a Butterworth filter and moving average before being input to the DNN that consists of 2 hidden layers with 300 and 100 neurons respectively. The accuracy of 82.3\% for up to five people were observed for total of 6 activities in multi user environment. 

Cheng et al. \cite{chengcount17} achieved 88\% accuracy with up to 9 people in an indoor environment in arbitrary locations. The authors claimed that the
conventional de-noising and feature extraction methods were susceptible to information loss. They thus proposed a new feature called ``difference between the value and sample mean'' and appended it as an additional feature to the CSI feature vector. This scheme has significantly improved the performance of 3-layer MLP model. 

Fang et al.\cite{fang19enhanced} proposed a hybrid feature involving both amplitude and phase to learn three human modes, i.e., absence, working, and sleeping, in an indoor environment. The hybrid feature reduced the need of training data and the model achieved 93\% accuracy with 6\% training samples only. The first hybrid feature contained the magnitudes of the complex numbers in a given CSI vector and the second hybrid feature contained the calibrated amplitudes and phases. The authors tested model's robustness against the cross environment but the model failed to perform in unseen environment without retraining. 

Among the other notable works, TW-See \cite{Wu2018} proposed a through-wall activity recognition system which used MLP with one hidden layer for the activity recognition task. The model classified 7 activities in two environments where the senders and the receivers were separated by walls. The authors studied the model robustness with different wall materials, and TW-See achieved above 90\% classification accuracy for different wall materials. 

CrossSense \cite{Zhang2018} tried to address problem of domain generalization by incorporating MLP into a deep transnational model. Also the large scale sensing which includes numerous subjects and domains are not supported by many works. To this end, Crosssense used MLP for generating the virtual samples for a given target domain using a feed forward fully connected network with 7 hidden layers which uses data from two domains in order to learn the mapping relation between them. The trained network is then used to generate the virtual samples.

The summary of MLP related literature is shown in Table~\ref{table:MLPApp}. 
MLP has shown a simple yet powerful deep learning approach for feature learning from CSI data.It was applied to both classification and regression tasks. Large scale sensing applications like \cite{Zhang2018} also proved the MLP's ability in tranferable feature learning  between domains  from CSI data. 

Denoising and sanitizing of both amplitude and phase were given major attention but some works \cite{Zhou2019} only choose amplitude due to challenges in phase sanitation.  

Deep model optimization was a given a major part in model evaluation using hyper parameter tuning to maximize the models performance. However, the model training time is not reported by many works. 

\subsection{RF Sensing with RBN}
There are only two works so far that used RBM for RF sensing. For number (0 to 9) gesture recognition, DeNum~\cite{Zhou2016} stacked multiple RBMs, i.e., the output of one RBM was fed as input to the next, to extract the discriminating features from the complex WiFi CSI data. At the end, an SVM was used for the classification task using the features extracted by the stacked RBM. The average classification accuracy reported was 94\%. Although this was an interesting use of deep learning for gesture recognition, no benchmark results were available to gauge the utility of stacked RBM against conventional machine learning.

Zhao et al.\cite{Zhao2018} used RBM in a special way to address the challenging problem of localization using only the RSS of WiFi signal, which is easily accessible but known to be very unstable. Instead of using the basic RBM, which allows only binary input, the authors considered a variant called Gaussian Bernoulli RBM (GBRBM)\cite{cho2011improved} to input real values of RSS. They designed a series
of GBRBM blocks to extract features from the raw RSS data, which is then used as input to further train an autoencoder (AE) for location classification. The combined GBRBM-AE deep learning model achieved 97.1\% classification accuracy and outperformed conventional AEs, i.e., when the AE is not augmented with GBRBM in the pre-training stage, in both location accuracy and robustness against noise.

\begin{table*}[h!]
\caption{CNN Representative Architectures in RF Sensing}
\begin{tabular}{p{5cm} p{5.5cm} p{5.5cm}}
    \toprule
    Representative Architecture& Architecture Key Features & Example Usage in RF Sensing Context \\
    \midrule
    
    Encoder (E) & \textbf{invariance to translations} in space and time & extracting \textbf{spatio-temporal features} from CSI in sign language recognition system \cite{ma18signfi}
    \\
    
    & \textbf{aggregating} information over \textbf{temporal} dimension & tolerating \textbf{temporally missing reflections} from limbs when capturing body pose \cite{zhaopose18,zhao2017learning}\\
    
    Cascaded Encoder & multistage \textbf{robust} classification& dealing with \textbf{sample unbalance and scarcity} of fall data in fall detection system \cite{rf-fall}
    \\\hline

    Encoder with Attention (EA) & encoding \textbf{importance weights} of features relevant to sensing task  &  \textbf{focus on feature} representations from spectrogram relevant to ASL signs \cite{santhalingam2020mmasl}
    \\\hline
    
    Multistream Encoder (ME) & encoding features across \textbf{different channels} & channel-wise feature concatenation of RF heatmaps from \textbf{horizontal and vertical antennas} \cite{rf-fall}\\\hline

     Multistream Encoder with Attention (MEA)&\textbf{weighted aggregation} of features from \textbf{different channels} & combine features from different \textbf{receiving antennas} based on \textbf{quality weights} for activity detection \cite{xue2020deepmv}\\\hline
    
    Encoder with Sequence Model (ES) & tracking \textbf{state change} in classifier predictions & estimate \textbf{fall state duration} in a fall detection system \cite{rf-fall}
    \\
    \bottomrule
\end{tabular}

\label{tab:cnn_rf}
\end{table*}

\subsection{RF Sensing with CNN}

RF data, when organized properly, convey visual features with spatial pattern similar to those in real images. In RF heatmaps \cite{adib2014witrack}, reflections from a specific object tend to be clustered together while those from different objects appear as distanced blobs. Similarly in  spatial spectrograms \cite{santhalingam2020mmasl}, motions from different sources have their corresponding energy spatially distributed on beam sectors, and in CSI variations, neighbouring sub-carriers are correlated. Such behaviour aligns with the locality assumption of CNN and make CNN a favourable option for RF representation learning. Additionally, temporal features can be acquired as well by restructuring the input to be continuous sequence of RF samples rather than individual samples. This allows CNN convolutions to aggregate temporal information in the input sequence hence extending its role to spatio-temporal processing \cite{rf-fall}. These reasons indeed drive the popularity of CNN among RF sensing systems.

\begin{figure}[h!]
    \centering
    \includegraphics[trim ={0cm 0cm 0cm 0.05cm},clip,width= 0.42\textwidth]{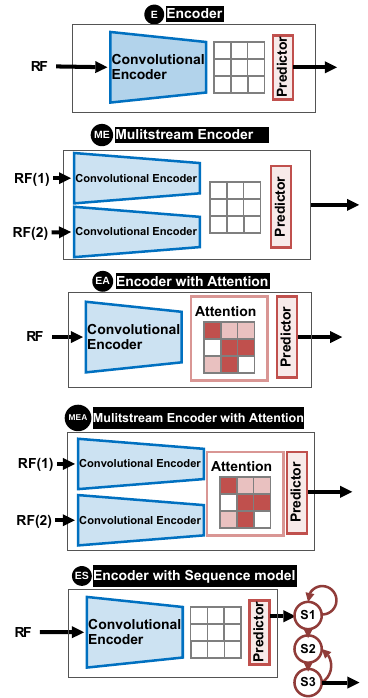}
    \caption{Representative Uni-Modal CNN architectures used in RF sensing}
    \label{fig:cnn_architectural_blocks}
\end{figure}

\begin{figure}[h!]
    \centering
    \includegraphics[width= 0.42\textwidth]{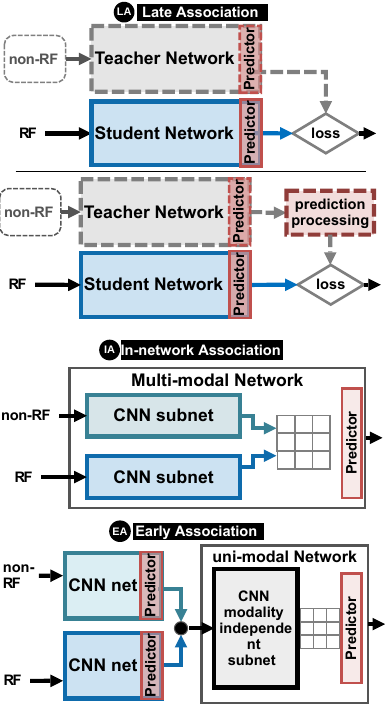}
    \caption{Representative Multi-Modal CNN architectures used in RF sensing. Dashed blocks denote training-time only processing.  }
    \label{fig:rf_cnn_multimodal}
\end{figure}

CNN architectural patterns can be broadly grouped into two categories; Uni-Modal CNN (see Figure \ref{fig:cnn_architectural_blocks}) that handles only RF input data and Multi-Modal CNN (see Figure \ref{fig:rf_cnn_multimodal}) which exploits support from another modality such as vision mostly during the learning process. We discuss the representative architectures in each category. In the literature, however, one can see that complex sensing systems tend to aggregate some of these architectures as building blocks into a larger complicated architecture. This is motivated by the need to combine the features offered by each architecture (see Table \ref{tab:cnn_rf}) to suit the sensing task. As an example, the CNN \textbf{Encoder} (E in Figure \ref{fig:cnn_architectural_blocks}) alone was sufficient for SignFi~\cite{ma18signfi} to achieve 86.6\% gesture recognition accuracy on a dataset of 150 sign gestures. In contrast, Aryokee \cite{rf-fall} combines the features of Multistream Encoder (ME) and Encoder with Sequence Model (ES) for robust fall detection in real world settings. 

\subsubsection{Uni-Modal CNN}

The vanilla CNN architecture \textbf{(Encoder (E))} consists of a few convolutional layers that encodes the extracted features into a latent space followed by a predictor. The predictor can produce either a single output as shown in most of the published papers, or multiple outputs. Despite its simplicity, the \textbf{Encoder} architecture can achieve great success in many practical applications. This was first demonstrated by SignFi \cite{ma18signfi} that successfully managed to significantly expand the classification capability of RF systems to accommodate 150 gestures. Also, Aryokee \cite{rf-fall} was able to reliably detect falls among 40 activities on a large scale dataset collected in 57 environments from 140 people. By cascading two \textbf{Encoders} sequentially, it built a two-stage fall detection classifier., which enhanced the performance of the classifier by allowing it to reject non-fall samples that resemble fall samples (hard negatives). As a result, a dramatic improvement in the precision by more than 29\% was achieved.

In some cases, a single RF sensor can export multiple independent measurements. Stacking them in a single input vector is not favourable as the measurement contains independent information. Alternatively, a \textbf{Multistream Encoder (ME)} could be used to extract the unique features of each measurement stream independently and subsequently combine them into latent feature vectors. For example, vertical and horizontal RF heatmaps \cite{rf-fall} \cite{hsu2019enabling} are processed by a two-stream CNN Encoder for fall detection and person identification, respectively. Similarly, DeepMV's \cite{xue2020deepmv}  multi-stream Encoder  processed CSI measurements from nine WiFi antennas distributed spatially across the room for activity recognition.

RF-ReID~\cite{li2019making} employs a deep learning framework that predicts human action from the 3D skeletons produced from RF heatmaps. The framework is based on a version of the two-stream Hierarchical Co-occurrence Network (HCN) \cite{zhu2016co} augmented with an attention module to tolerate inaccuracies in the input skeletons estimated from RF. This is an example of \textbf{Multistream Encoder with Attention (MAE)}, which is employed to allow the model to focus on keypoints with higher prediction confidence in RF heatmap snapshots when making predictions. In cases where RF samples contains various types of information that are relevant to the sensing task, Hierarchical Attention was employed. For example, in Person Re-identication systems~\cite{fan2020learning}, the body shape (short temporal window) and the walking style (long temporal window) are both relevant to the sensing task. Thus a hierarchical two-level attention blocks can be integrated to attend to each information type.

\subsubsection{Multi-Modal CNN}

Moving to Multi-modal CNN architectures, one can see three key approaches followed in order to fuse information from RF and supporting modalities (i.e. non-RF modalities). The key difference between them is the stage at which data from supporting modality is utilized in the learning.  In \textbf{Late Association (LA)}, the supporting modality is handled separately by different model called Teacher Network (usually pre-trained) and the output is used for providing labels for the RF model (Student Netwrok).This was adopted as a way to tackle the difficulty of labelling RF data.  For example, RF-Pose~\cite{zhaopose18} uses this techniques to train RF pose prediction network (student network) with human pose heatmap labels acquired from AlphaPose~\cite{cao2017realtime} on RGB frames of synchronized camera. Since the RF samples are synchronized with the camera samples, the teacher network predictions can be used as labels for RF samples. A similar approach was followed by CSI-UNet~\cite{unet18} and Person-in-WiFi \cite{wang2019person}. It should be noted that data from the supporting modality is utilized only during the learning process and not at the run time.

\textbf{In-network Association(IA)} fuses information directly from the RF and the supporting modality in a single architecture. For instance, in the behavioural tracking system, Marko ~\cite{hsu2019enabling}, tracklets from synchronized accelerometer was used for continuous masking of RF samples that carry extra information irrelevant to the user's actions. 

Finally,  the \textbf{Early Association(EA)} scheme depends on a unimodal network that can process intermediate representation produced from either RF or the supporting modality. RF-Action \cite{li2019making} systems for human action recognition is a representative example of this scheme. The intermediate representation is the 3D human skeleton and can be produced from either RF radar or RGB camera using deep CNN nets. The uni-modal network is agnostic to the original input type as it accepts the intermediate representation. A main advantage of this approach is that the uni-modal network can be trained and fine-tuned using data only from the supporting modality without the need for collecting additional RF data. In fact, RF-Action \cite{li2019making} leverages 20K additional samples from  PKU-MDD multimodal dataset \cite{liu2017pku} to improve the system performance.

\subsection{RF Sensing with Recurrent Neural Networks}

As explained in Section \ref{s:RFsensing}, RF sensing often use time series RF data, such as RSS and CSI obtained from successive frames, to detect changes during a human activity. Such time series data contain important temporal information about human behavior. Shallow learning techniques and conventional machine learning algorithms do not take this temporal factor into account when the data is provided as inputs, which leads to poor performance of the models. RNNs have proven their ability to produce promising results in speech recognition and human behaviour recognition in video as they are inherently designed to work with temporal data. RF sensing research has also recognized this benefit of RNN. Recently, RNN variants like LSTM and GRU have become popular in RF-based localization and human activity recognition applications. Table \ref{table:RNNSumary} summarizes the RNN-based RF sensing works we survey in this section.

LSTM, which has a gated structure for forgetting and remembering control, has dominated the state-of-the-art of recurrent networks. Yousefi et al.~\cite{Yousefi2017} were the first to explore the benefit of LSTM-based classification against the conventional machine learning algorithms using CSI for human activity recognition. In their experiments, LSTM significantly outperformed two popular machine learning methods, Random Forest (RF) and Hidden Markov Model (HMM). Later, Shi et al.~\cite{shi18,Shi2019} further improved this process with two feature extraction techniques, namely \textit{Local Mean} and \textit{Differential Method}, that removed unrelated static information from the CSI data. As a result, accuracy improvements were observed up to 18\% against the original method of~\cite{Yousefi2017}. 

LSTM quickly became a popular choice for detecting many other human contexts.  HumanFi~\cite{Ming2019} achieved 96\% accuracy in detecting human gaits using LSTM; Haseeb et al.~\cite{haseeb2017wisture} utilized an LSTM with 2 hidden layers to detect gestures with mobile phone's WiFi RSSI achieving recognition accuracy up to 94\%; WiMulti~\cite{feng2019wi} used LSTM for multi-person activity recognition with an overall accuracy of 96.1\%. Ibrahim et al.~\cite{Ibrahim2019} proposed a human counting system, called CrossCount, that leverages an LSTM to map a sequence of link-blockage temporal pattern to the human count using a single WiFi link. The
intuition behind this success is that, the higher the number of people in an area of interest, the shorter the time between blocking a single WiFi link and vice versa.

CSAR~\cite{Wang2019} proposed a channel hopping mechanism that continuously switches to less noisy channels for improved human activity recognition. They proposed an LSTM network as a classifier, which takes the Time-Frequency features generated from Discrete Wavelet Transform (DWT) spectrograms as the model inputs. LSTM is designed to work with inherent relationships in the frequency changes in the spectrogram data in long time intervals. As in most of deep learning architecture, LSTM can work with bigger data sets effectively. Along with a 200 hidden unit LSTM layer with 2 other fully connected layers,  CSAR  achieved 90\% accuracy for detecting 8 different activities.

Bidirectional LSTM (BLSTM) is a variant of conventional LSTM mode, which has two LSTM layers to  represent the sequence data in both forwards and backwards simultaneously. This enables the network to learn about both the forward and backward information of a given data point at a given time instance. BLSTM has been successfully applied to activity recognition model by Chen et al.\cite{Chen2019} along with an \textbf{attention-based deep learning module}. Rather than assigning the same level of importance, attention-based modules assign higher weights to the features that are more critical for the activity recognition task. 

GRU, a variant of LSTM, contains only 3 gates and connections, which makes it simpler and easier to train than LSTM. For effective sequential information learning, Wang et al.~\cite{wanglimb18} utilized two GRU layers stacked together to achieve 98.45\% average accuracy compared with a baseline shallow CNN network (with 2 layers), which achieved an accuracy of only 79.59\%.

\begin{table*}[h!]
    \caption{RNN architectures in RF sensing}
    \label{table:RNNSumary}
    \centering
    \begin{tabular}{|p{0.6cm}|p{4cm}|p{2.5cm}|p{3cm}|p{3.5cm}|}
    \hline
    \textbf{Paper} &\textbf{Application} &\textbf{Radio Measurement}  & \textbf{RNN Varient/Layer(s)} & \textbf{Performance}\\\hline
    \hline
    
    \cite{shi18}   & Activity Recognition&CSI & LSTM 1 & Best Accuracy  0.975 \\
    \hline
    \cite{wanglimb18} & Gait Recognition& CW Radar&GRU 2 & Avg. Accuracy   0.9177\\
    \hline
    \cite{Wang2019} & Activity Recognition& CSI& LSTM 1 & Accuracy $~$ 0.95  \\
    \hline
    
    \cite{Ibrahim2019} & Human Counting& RSSI & LSTM 1 & Accuracy \newline Upto 2 persons 1.0 \newline  Up to 10 persons 0.59 \\
    \hline
    
    \cite{feng2019wi} & Multi Activity Recognition&CSI& LSTM 3& Avg. Accuracy  0.962 \\
    \hline
    
    \cite{Yousefi2017} & Activity Recognition&CSI&LSTM 1& Best Accuracy  0.97 \\
    \hline
    
    \cite{Shi2019} & Activity Recognition&CSI&LSTM1& Best Accuracy  0.991 \\
    \hline
    
    \cite{haseeb2017wisture} & Gesture Recognition&RSSI&LSTM 2& Best Accuracy 0.91   \\
    \hline
    
    \cite{Chen2019}&Activity Recognition& CSI&BLSTM 2& Best Accuracy 0.973 \\ 
    \hline
    
    \cite{Ming2019}& Gait Recognition&CSI&LSTM 1& Accuracy 0.96\\ 
    \hline
\end{tabular}
\end{table*}

\begin{figure*}
\centering
\begin{forest}
  direction switch,
  for tree={fork sep=1em}
  [Autoencoder usage in RF Sensing,yshift=3em,alias=LP
    [Pretraining 
      [Layer-by-Layer~\cite{shi17,Chang2018DeviceFreeIL,gao17}]
      [Whole Autoencoder~\cite{Zhao2019}]
    ]
    [Data Augmentation
      [FiDo~\cite{Chen2020}]
    ]
    [Domain Adaptation
      [Auto-Fi~\cite{Chen2017TamingTI}]
    ]
    [Unconventional Encoding-Decoding
      [RNN Encoder-Decoder~\cite{Xu2018AttentionbasedWG}]
      [RFPose~\cite{zhaopose18}]
     [Person-in-WiFi~\cite{wang2019person}]
    ]
    ]
\end{forest}
\caption{Taxonomy of autoencoder usage in RF sensing.}
\label{f:AEuses}
\end{figure*}
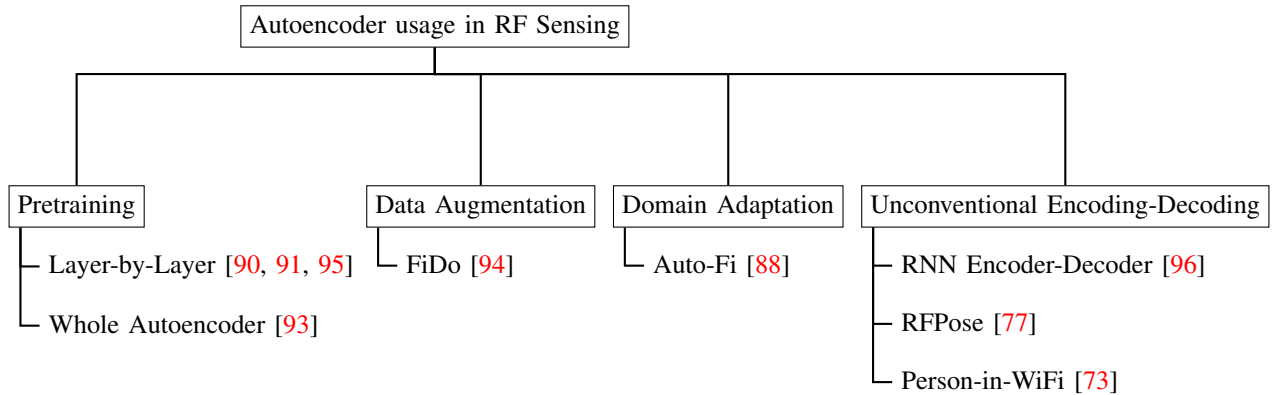

\subsection{RF sensing with Autoencoder}
\begin{figure}[h]
\includegraphics[width=0.7\linewidth]{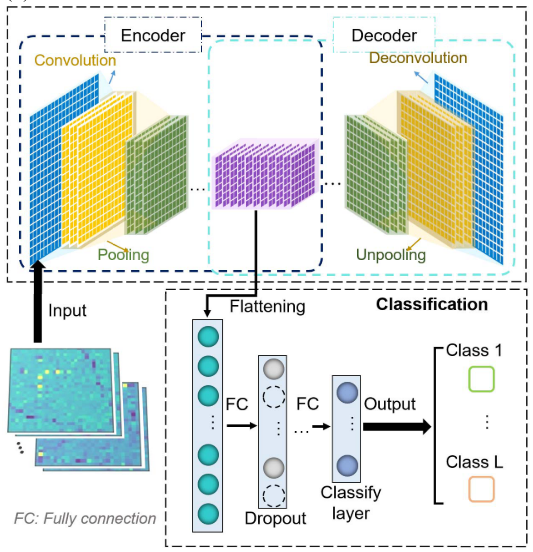}
    \centering
    \caption{Convolutional Autoencoder used in~\cite{Zhao2019}}
    \label{fig:convae}
\end{figure}
As explained in Section \ref{s:RFsensing}, autoencoder is fundamentally a deep learning technique to extract a \textbf{compressed knowledge representation} of the original input. In recent years, this property of autoencoders has been exploited by researchers in different ways to accomplish different RF sensing tasks. In this survey, we propose the taxonomy shown in Figure \ref{f:AEuses} to analyze state-of-the-art contributions under four different categories: \textbf{unsupervised pretraining, data augmentation, domain translation, and unconventional encoding-decoding}. In the following, we briefly review the works in each of these categories. 

\subsubsection{Unsupervised Pretraining}
Training deep neural networks from a completely random state requires large labelled datasets, which is a fundamental challenge for RF human sensing researchers. Also, with random initial weights, deep learning faces the well-known \textbf{vanishing gradient} problem, which basically means that the gradient descent used for backpropagation fails to update the layers closer to the input layer when the network is deep. This in turn increases the risk of not finding a good local minimum for the non-convex cost function. It turns out that autoencoders can help address these problems through a two-phase learning protocol called \textbf{unsupervised pretraining}, which basically builds up an unsupervised autoencoder model first using only unlabelled data, and later drops the decoder part of the autoencoder but adds a supervised output layer to the encoder part for classification.  The supervised learning phase may involve training a simple classifier on top of the compressed features learned by the autoencoder in the pretraining phase, or it may involve supervised fine-tuning of the entire network learned in the pretraining phase. Research has shown that such unsupervised pretraining~\cite{Dumitru2010,hinton2006reducing} can significantly improve the performance of deep learning models in some domains. 

In RF sensing domain, several researchers reported good results with autoencoder-based pretraining. Shi et al.~\cite{shi17} employed a deep neural network (DNN) with 3 hidden layers to detect user activities and authenticate the user at the same time based on the unique ways a user performs each activity. WiFi CSI was used as the input for the deep learning. The DNN was first pretrained with only unlabelled CSI layer-by-layer using a stacked autoencoder~\cite{Vincent2010} where a trained hidden layer became the input for the next autoencoder.  In the supervised learning phase, each of the layer is appended with a softmax classifier, where the first layer is used to detect whether the user is stationary or active, the second layer for detecting the activities of the user, and the final layer to identify the user based on the user behavior during her activities.  With the pretraining, the authors of~\cite{shi17} reported that over 90\% accuracy on user identification and activity recognition could be achieved for 11 subjects even with the training size of only 4 labelled examples per user. 

Similar to~\cite{shi17}, Chang et al.~\cite{Chang2018DeviceFreeIL} also used autoencoder for layer-by-layer unsupervised  pretraininhg of a 3-layer DNN for localization based on CSI, which achieved good performance for two different environments. In another localization work based on RSS inputs, Khatab et al.~\cite{Khatab2018} confirmed that layer-by-layer autoencoder-based pretraining of a 2-layer extreme learning machine (ELM) improves performance compared to that the case when the ELM is initialized with random weights. Finally, in a CSI-based deep learning for localization, Gao et al.~\cite{gao17,wanglocact17,zhang16} also demonstrated positive outcomes when using layer-by-layer pretraining with sparse autoencoders, which works even when the the successive hidden layers of the deep neural architecture do not reduce in size.

Zhao et al.~\cite{Zhao2019} combines the merits of convolutional spatial learning of CNNs with the unsupervised pretraining capability of autoencoders to design a so called convolutional autoencoder (CAE) to localize a user on a grid layout based on 2D RSS images. Unlike the layer-by-layer pretraining implemented with stacked autoencoders in~\cite{shi17,Chang2018DeviceFreeIL,Khatab2018}, Zhao et al.~\cite{Zhao2019} pretrained the entire CAE, after which the decoder part is dropped and the fully connected layers together with a Softmax layer are added for localization. The CAE architecture and its two-phase pretraining process are illustrated in Figure \ref{fig:convae}.

\subsubsection{Data Augmentation}
One of the challenges in WiFi-based localization is that WiFi location fingerprints experience significant inconsistency across different users. This means that deep networks trained on RF data collected from one user may not produce good accuracy when used for other users. Chen et al.,~\cite{Chen2020} trained a variational autoencoder (VAE) on a real user and then generated a large number (10 times the original data) of synthetic CSI data to further train a classifier. The proposed VAE-augmented classifier, called FiDo, resulted in 20\% accuracy improvement compared to the classifier that was trained without the VAE outputs.   

\subsubsection{Domain Adaptation}
WiFi CSI profiles are significantly affected by environment changes, which makes it challenging to generalize a trained model across many domains ('domain' refers to 'environment'). Chen et al.,~\cite{Chen2017TamingTI} used an autoencoder to `preserve' the critical features of the original environment where the initial training CSI data is collected from. Such feature preservation is achieved during the training phase by training the autoencoder with unlabelled CSI data. During the inference phase at another environment, the previously trained autoencoder is used to convert the CSI vector from the new environment to another vector that now inherits the features of the previous environment. By using the converted CSI vector, instead of the actual CSI, as an input to the pretrained classifier, the detection accuracy for WiFi localization is significantly improved.   

\subsubsection{Unconventional Encoding-Decoding}
\begin{figure}[h]
    \includegraphics[scale=0.4]{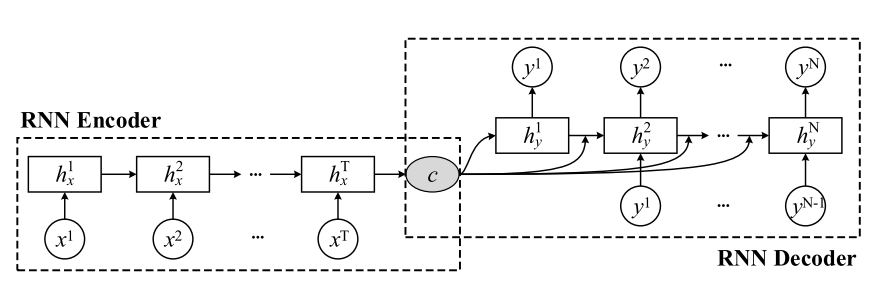}
    \centering
    \caption{The RNN encoder-decoder architecture used in~\cite{Xu2018AttentionbasedWG} }
    \label{fig:yangxu}
\end{figure}

Xu et al.\cite{Xu2018AttentionbasedWG} propose an attention based RNN encoder-decoder model for the classification of direction and gait recognition.  Attention based machine translations can further improve the accuracy as it mimics the human visual attention only to the vital parts when recognition occurs, which improves the performance of the models when the data collected are noisy. Attention based systems do not give equal importance to all features; instead, it focuses more on the important features which significantly leverage the training effort as well. As depicted in Figure \ref{fig:yangxu}, the encoder part consists of a bi-directional RNN with GRU cells to maintain the simplicity.

\begin{figure*}
    \includegraphics[scale=0.4]{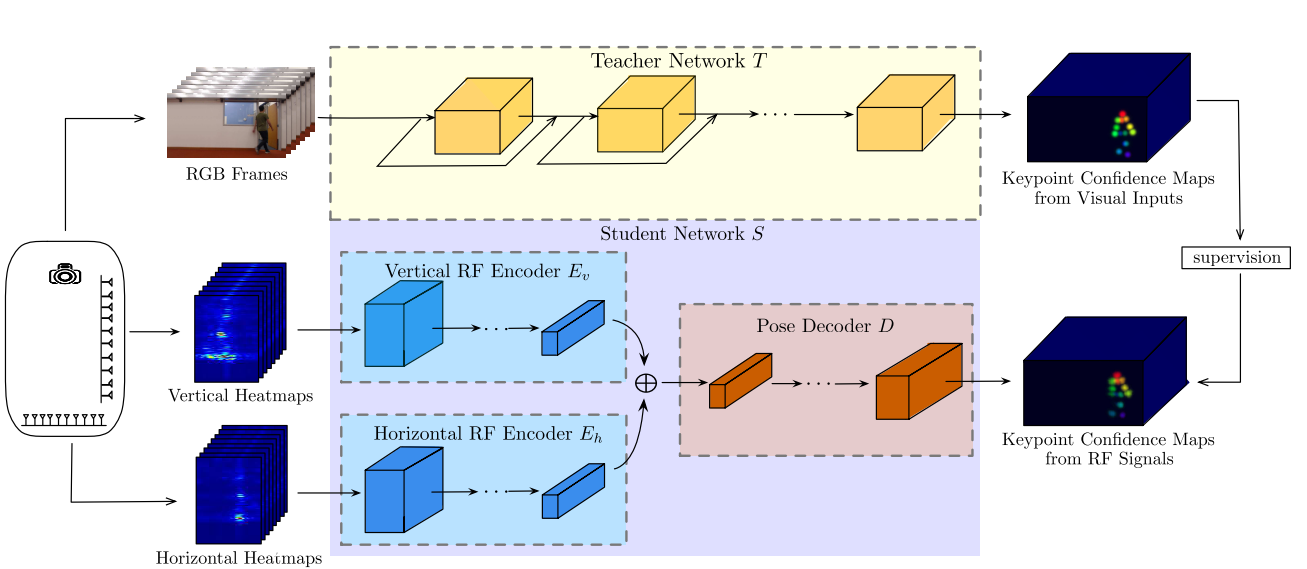}
    \centering
    \caption{The Teacher-Student Network used in~\cite{zhaopose18}}
    \label{fig:rfpose}
\end{figure*}

RF-Pose \cite{zhaopose18} propose encoder-decoder based deep learning architecture for human pose estimation. It made use of a cross model approach by first generating 2D-human skeletal images using RGB images from camera which works as a teacher network and radio heat maps images captured from the FMCW horizontal and vertical arrays as the student network.The teacher network facilitates annotation of the radio signal from RGB stream to the key point confidence maps. The proposed student network consists of two autoencoders to correspond with the vertical and horizontal RF images and concatenate the outputs at the end. The student network uses fractionaly strided convolutional\cite{deconvolutions2010} layers which are used for upscaling the low resolution inputs to a higher resolutions while preserving the abstract details of the output. This serves as the decoder part of the proposed architecture where the up sampling process is learned by the network itself rather than Hard coding the process.  The architecture of the proposed network is depicted in Figure \ref{fig:rfpose}. The Teacher-Student design of the deep learning architecture facilitate the cross model pose estimation which achieves  62.4\% average precision compared with the baseline 68.8\% but through the wall scenario achieves 58.1\% precision where the vision based baseline system completely fails. More importantly, RF-Pose tracks multiple persons simultaneously.  

Person-in-WiFi\cite{wang2019person} utilizes a U-net style autoencoder to map CSI data captures by 3 $\times$ 3 MIMO WiFi setup with corresponding 2D-pose of people in the sensing area. CSI is concurrently being mapped to 3 pose representations to the body Segmentation Mask (SM), Joint Heatmaps (JHMs) and Part Affinity Fields (PAFs) consecutively. SMs and JMMs share one U-net and PAFs share another thus the architecture contains two autoencoders. It is noteworthy to mention that the loss function, Mathew weight to optimize the learning process of JMMs and PAFs is chosen such a way that more attention is payed for improving the skeletal representation of the body than background of the image (which is black). The solution proves that the person's 2D pose can be perceived through 1D WiFi data.

\subsection{RF Sensing with Adversarial Networks} \label{sec:adv_training}

\begin{table*}[h!]

\caption{Summary of works involving adversarial networks in RF sensing}
    \centering
    \begin{tabular}{|p{2.2cm}|p{1.6cm}|p{2cm}|p{4.5cm}|p{1.5cm}|p{1.5cm}|}
    \hline
  \textbf{Paper} &\textbf{Monitoring \newline Application} &\textbf{RF \newline Measurement} &\textbf{Domain Adaptation Across} &  \multicolumn{2}{c|}{\textbf{Accuracy \%}}\\ \cline{5-6}
     & & & & \textbf{w/o Adapt.} & \textbf{w/ Adapt.}\\ 
    \hline

    \hline

     DIRT-T \cite{shu2018dirt} & Activity &  CSI (WiFi) & 2 rooms  & 35.7 & 53   \\
    \hline
    
     EIGUR \cite{yu2019rfid}  & Gesture & RSS \& Phase \newline (RFID) &3 rooms and 15 subjects &  87.2 (Prec.) \newline 86.2 (Recall) & 96.6 (Prec.) \newline 96 (Recall)\\
    \hline
    
    RF-Sleep \cite{zhao2017learning}& Sleep & FMCW Radar  & 25 subjects & - & 79.8 \\
    \hline
    
     DeepMV \cite{xue2020deepmv} & Activity & CSI (WiFi)  & 3 rooms and 8 subjects & - &  83.7 \\
    \hline

     EI \cite{jiang2018towards} &Activity  &
    CSI(WiFi) \newline CIR(mmWave)&
    3 rooms \& 11 subjects (WiFi) \newline 4 rooms \& 10 subjects (mmWave)
    &  - & 78 \newline 65    \\
    \hline

     WiCAR \cite{wang2019wicar,wang2020multi}   & Activity & CSI (WiFi)  &
    4 cars, 4 subs \& 4 driving conditions 
     &  53
    &83\\
    \hline
     CsiGAN~\cite{CsiGAN2019} & Gesture \addslash Fall & CSI (WiFi) & 5 subs (Gest.) \addslash 3 subs (Fall) &  \Centerstack{-\\-} &  84.17 (Gest.) \newline 86.27 (Fall)  \\
    \hline  

\end{tabular}
\label{tab:advereserial_net_rf}
\end{table*}

RF measurements of human activities usually contain significant information that is specific to the user, i.e., the body shape, position, and orientation to the radio receiver, as well as the physical environment, i.e., walls, furniture etc. Consequently, an activity classifier trained with one user in a specific environment do not perform reliably when tested with another person in another environment. In the literature, the user-environment combination is often referred to as a \textbf{domain}.  

To achieve ubiquitous RF sensing models that can be deployed across different domains, it is imperative to extract features from the `noisy' RF measurements that only represent the activities of the user without being influenced by domain specific properties as much as possible. One way to achieve this is to design hand-crafted features to model the motion or velocity components of the activity, which clearly do not depend on the domain yet can identify activities based on their unique motion profiles. Examples of this approach include CARM~\cite{wang2015understanding}, Widar 3.0~\cite{zheng2019zero}, and WiPose~\cite{Jiang2020Towards3H}. While these modeling-based solutions can achieve generalization across domains (a.k.a. \textbf{domain adaptation}) to some extent, they require rather precise knowledge of the physical layout in terms of the user location/orientation and the radio transmitters and receivers. In some cases~\cite{zheng2019zero}, they work well only when multiple RF receivers are installed in the sensing area. 

In recent years, researchers have demonstrated that adversarial networks can be an effective deep learning tool to realize RF domain adaptation without having to worry about the specific positions and orientations of the users and the RF receivers. For RF domain adaptation, adversarial networks were used in two  different ways, \textbf{unsupervised adversarial training} and \textbf{semi-supervised generative adversarial network (SGAN)}. 

\begin{figure}[h]
    \includegraphics[width=8cm]{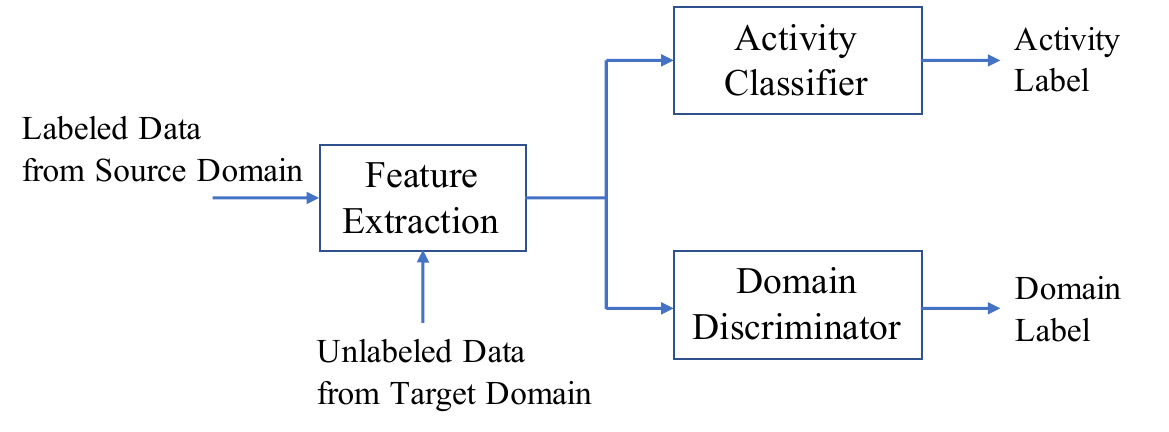}
    \centering
    \caption{Principles of unsupervised adversarial training.}  
    \label{f:adv-training}
\end{figure}

\begin{figure}[h]
    \includegraphics[width=8cm]{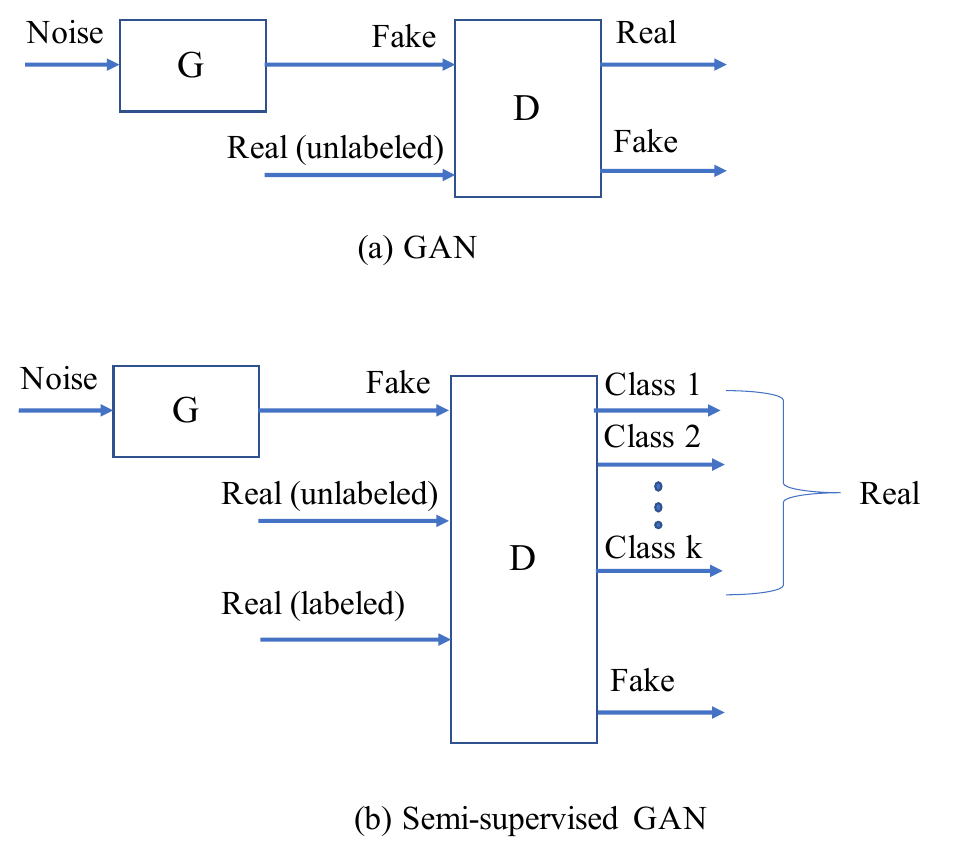}
    \centering
    \caption{GAN vs. semi-supervised GAN.}  
    \label{f:SGAN}
    \end{figure}
\subsubsection{Unsupervised Adversarial Training}

Unsupervised adversarial training is a well-known domain adaptation technique used in many fields~\cite{unsupervisedsurvey,ganin2014unsupervised}. Its basic principle is illustrated in Figure \ref{f:adv-training}. There are three main interconnected components: \textbf{feature extractor}, \textbf{activity classifier}, and \textbf{domain discriminator}. The feature extractor takes labeled input from the source domain, but only unlabeled data from the target domain. The goal of the classifier is to predict the activity, while the discriminator tries to predict the domain label. The feature extractor tries its best to cheat the domain discriminator, i.e., minimize the accuracy for domain prediction, and at the same time maximize the predictive performance of the activity classifier. By playing this \textbf{minimax game}, the network eventually learns the features for all the activities that are domain invariant. Table~\ref{tab:advereserial_net_rf} compares several works that employed the basic philosophy of Figure~\ref{f:adv-training} to generalize RF sensing classifiers across multiple domains.

\subsubsection{Semi-supervised GAN}

In Section \ref{s:DL}, we have learned that GAN is a special kind of adversarial network that trains a generator to produce realistic fake samples. Although the generator of a GAN is trained with the help of a discriminator, it is the generator that is used eventually to fake samples while the discriminator is of no further use in the post-training phase.

Semi-supervised GAN (SGAN)~\cite{salimans2016} is a recent proposal that extends GAN to achieve classification as an added functionality in addition to the generation of fake samples. As illustrated in Figure~\ref{f:SGAN}, only the discriminator is extended while the generator remains intact. In terms of its input, the discriminator now takes some \textit{labeled} real samples in addition to the unlabeled real samples. The discriminator network is extended to classify the samples detected as real into $k$ classes by learning these classes from the labeled samples. A key benefit of SGAN as a classifier is to learn to classify reliably with only a small amount of labeled samples as it can still learn significantly from the vast amount of unlabeled samples while playing the minimax game with the generator. 

In their proposed RF sensing system called CSI-GAN, Xiao et al.~\cite{CsiGAN2019} successfully applied the concept of SGAN to realize domain adaptation across unseen (target) users. The main challenge of this application was that the amount of unlabeled CSI samples that could be collected from the target user is severely limited due to the need for avoiding lengthy training for new users. It was observed though that the performance of SGAN deteriorated in the case of limited unlabeled data, because the generator could produce fake samples of only limited diversity due to the limited unlabeled data available from the target user. 

CSI-GAN addressed the limited unlabeled data issue in SGAN by adding a second complement generator that used the concept of CycleGAN~\cite{zhu2017unpaired} to transfer the CSI from the source user to the target user style, thus creating additional fake samples. It was shown that such fake sample boosting method could effectively overcome the issue of limited unlabeled data in SGAN.  

\subsection{Hybrid Deep Learning Models}
For complex tasks, the basic deep learning models are often combined in a hybrid model. In this section, we summarise the existing hybrid models that proved to be effective in RF sensing. 

\textbf{Convolutional Recurrent Models}. This category of models stacks convolutions and recurrent blocks sequentially in the same architecture as a way to combine the best of the two worlds, i.e., the spatial pattern extraction property of CNNs and the temporal modelling capability of RNNs. Empirical studies~\cite{yin2017comparative} have confirmed the effectiveness of such hybrid models across tasks as diverse as car tracking from motion sensors, human activity recognition, and user identification. Moreover, by dividing the input layer into multiple subnets for each input sensor tensor, the model can be used for sensor fusion as well. These attractive features were leveraged by several researchers for various RF sensing applications. DeepSoli~\cite{soli16} uses a CNN followed by an LSTM to map a sequence of radar frames into a prediction of the gesture performed by the user. The model can recognize 11 complex micro-gestures collected from 10 subjects with 87\% accuracy. RadHar~\cite{singh2019radhar} uses a similar architecture composed of a CNN followed by a Bi-directional LSTM to predict human activities from point clouds collected by a mmWave Radar. 

While the basic Convolutional Recurrent model worked well across various tasks, it was further enhanced in some works to enable additional input or output processing to enhance the accuracy. WiPose~\cite{Jiang2020Towards3H} uses the Convolutional Recurrent model enhanced with post-processing component to map 3D Body-coordinate Velocity Profile (BVP) to human poses. In addition to CNN and RNN components, the model in WiPose was supported by a ``Forward Kinematics Layer'' that recursively estimates the rotation of the body segments, which provide a smooth skeleton reconstruction of the human body. Zhou et. al.~\cite{Zou2018} prefix the architecture with an auto-encoder as a pre-processing component for reconstructing a de-noised version of the input CSI measurements before forwarding it to the core Convolutional Recurrent model.

\textbf{Domain Specialized Neural Models}. STFNets~\cite{yao2019stfnets} introduced a novel Short-Time Fourier Neural Network that integrates neural networks with time-frequency analysis, which allows the network to learn frequency domain representations. It was shown that it improves the learning performance for applications that deal with measurements that are fundamentally a function of signal frequencies such as the signals from motion sensors, WiFi, ultrasound and visible light. The architecture was used for several recognition tasks including CSI-based human activity recognition and the evaluation showed that STFNets significantly outperformed the state-of-the-art deep learning
models. 

\section{Review of Public Datasets}
\label{sec:datasets}

Deep learning research requires access to large amount of data for training and evaluating proposed neural networks. Unfortunately, collecting and labeling radio data for various human activities is a labor-intensive task.  Although most researchers are currently collecting their own datasets to evaluate their deep learning algorithms, access to public datasets would help accelerate the future research in this area. Besides, due to the sensitiveness of radio signals to the actual experimental settings and equipment, comparison of different related works based on different datasets becomes problematic. Fortunately, some researchers have released their datasets in recent years, creating an opportunity for future researchers to reuse them in their deep learning work.    

We perform a survey of the publicly available datasets
that have already been used in radio-based human sensing publications. Our survey only analyzes those datasets that we were able to download and look into. Table \ref{table:BenchmarkDataSets} reviews the source of the surveyed datasets, year of creation, size of the data, radio signal feature collected, hardware used for data collection, and the scope of the data in terms of types and numbers of human activities, data collection environment, number of human participants and so on. We also indicate any additional materials, such as codes implementing deep learning models that use the datasets, that may have been released along with the datasets.  Important observations from this survey are summarized as follows:  

\begin{itemize}
    \item There are already 20 different datasets from 18 separate research groups that are publicly available for any researchers. Some datasets are released without any licenses, while others are under different licensing acts mostly for restricting non-academic use.  All these datasets were created only in recent years.
    \item Activity and gesture are the dominant applications targeted by these datasets. Other applications include location/tracking, fall detection, respiratory monitoring, and people counting.  
    \item The size of these datasets vary widely from mere 18MB to 325GB.
    \item Number of human participants vary from a single subject to 20 subjects. 
    \item Although half of the datasets collected data from a single environment, there are several offering data from five or more different environments with the maximum being seven.
    \item WiFi CSI collected by Intel 5300 NIC is the most common data type. 
    \item Codes implementing the authors' proposed deep learning models are also released for most datasets.  
\end{itemize}

While the availability of these datasets is certainly encouraging for deep learning research in RF human sensing, we identify several limitations and learn some lessons as follows:

\begin{itemize}
     \item Number of participants in the datasets were rather low. Although the associated publication for the CrossSense~\cite{Zhang2018} dataset reports deep learning training with data collected from 100 subjects, the publicly released dataset actually contains data from only 20 subjects. 
    \item Many datasets do not mention the gender and age distribution of the participants. Even when they are mentioned, the actual data is not labeled with gender and age, making it difficult to study gender and age specific characteristics of RF sensing.
    \item Although our survey in Table \ref{table:DLApplications} shows that RF device-free localization is a popular application for deep learning, there appears to be only 2 localization datasets available for public use and both are from the same research group. 
    \item All the 20 public datasets were mainly used by the creators themselves. Cross-use of the datasets is still rare with the exception of CSIGAN~\cite{CsiGAN2019}, \cite{noprocess19}, \cite{Chen2019} and \cite{WIHF-infocom2020} which used the public datasets SignFi~\cite{ma18signfi}, FallDeFi~\cite{Palipana2018}, \cite{Yousefi2017} and \cite{zheng2019zero}, respectively.

\end{itemize}

\begin{table*}[h!]
\caption{Public datasets of labeled radio signal measurements for human activities}
\label{table:BenchmarkDataSets}
\begin{tabular}{|p{2.5cm}|p{0.5cm}|p{2.5cm}|p{2.5cm}|p{0.8cm}|p{1cm}|p{2cm}|p{3cm}|}
\hline
\rowcolor{Gray}

\textbf{Source\addslash Year\newline License\addslash Repository } &  \textbf{\#Envs}&\textbf{Applications}&\textbf{Subjects} & \textbf{\#Activities} & \textbf{Size} &\textbf{Data-type \addslash Hardware}&\textbf{Additional Items}\\\hline
\hline
    CrossSense \cite{Zhang2018} 2018 \newline  Apache license v2 \newline GitHub&
    3 & 
    Gait \&\newline Gesture \newline Recognition &
    20&
    40&
    4.21 GB&
    Intel5300(CSI)/
    \newline XiaoMI Note2 Smartphone (RSSI)&
    \checkmark Implementation code  \\ \hline

    Widar 1.0 \cite{Qian2017} \newline 2017 \newline Private repo&
    1&
    Localization \& \newline Tracking&
    5[M4,F1]\newline Age[20-25]&
    5&
    2.76GB&
    Intel5300(CSI)&
    \checkmark Implementation code\\ \hline
    
    Widar 2.0 \cite{Qian2018} \newline 2018\newline Private repo&
    3&
    Localization \& \newline Tracking&
    6[M4,F2]&
    3&
    303MB &
    Intel5300(CSI)&
    \checkmark Implementation code\\ \hline
    
    Widar 3.0 \cite{zheng2019zero} \newline 2019 \newline Private repo&
    3&
    Gesture \newline Recognition&
    [M12,F4]\newline Age[23-28]&
    22&
    325GB&
    Intel5300(CSI)&
    \checkmark Implementation code \newline \checkmark Performing videos\\ \hline
    
    WiAG\cite{Virmani2017} \newline2017\newline Private repo &
    3&
    Gesture \newline Recognition&
    1&
    6&
    3GB &
    Intel5300(CSI)&
    -\\ \hline
    SignFi\cite{ma18signfi} \newline 2018 \newline  University license \newline GitHub&
    2&
    Sign Language\newline Gesture Recognition & 
    M5&
    276  &
    6.07GB &
    Intel5300(CSI)&
    \checkmark Implementation code\newline  \checkmark Performing videos\\ \hline
    
    \cite{Wang2019JointAR} \newline 2019\newline GitHub&
    1&
    Activity \newline Recognition&
    1&
    6&
    300MB &
    USRP N210 (CSI)&\checkmark Implementation code\\ \hline
    
    Wisture\cite{haseeb2017wisture} \newline 2017\newline GitHub&
    2&
    Gesture \newline Recognition& 
    1&
    3&
    58MB &
    Android Smartphone (RSS)&
    \checkmark Implementation code\\ \hline

    \cite{Hillyard2018}\newline 2018\newline IEEE DataPort&
    1&
    Respiratory \newline Monitoring&
    20[M11,F9] \newline Average Age[M:55,F:60]  &
    1 &
    60GB& 
      CC1200 Radio(sub-1 dB RSS) 
     \newline CC2530 Radio(RSS)
     \newline Atheros AR9462(WiFi CSI) 
     \newline Decawave EVB1000(CIR)
    & \\\hline

    FallDeFi \cite{Palipana2018} \newline 2018 \newline  MIT license \newline Harvard Dataverse&
    7&
    Fall \newline Detection&
    3\newline Age[27-30]&
    11&
    2.1GB &
    Intel5300(CSI)&
    \checkmark Implementation code  \\ \hline
    
    \cite{Yousefi2017}2017 \newline  GNU General Public License&
    1&
    Activity \newline Recognition&
    6&
    6&
    3.59GB&
    Intel5300(CSI) &
    \checkmark Implementation code \\ \hline

    \cite{Brinke2019} 2019 \newline CC BY-NC-SA license\newline GitHub&
    1&
    Activity \newline Recognition&
    9&
    6&
    2.04GB&
    Intel5300(CSI)&
    \checkmark Visualization code \\ \hline

    RadHAR\cite{singh2019radhar} \newline 2019\newline BSD 3-Clause license\newline GitHub&
    1&
    Activity \newline Recognition&
    M2&
    5&
    881MB&
    FMCW Radar  IWR1443BOOST\newline(Point cloud) &
    \checkmark Implementation code \\ \hline

    WiAR\cite{Wiar2019} \newline 2019\newline DATA4U&
    1&
    Activity \newline Recognition&
    10[M5,F5]&
    16&
    667MB&
    Intel5300(CSI) &
    \checkmark Implementation code \newline\\ \hline

    CSI-net\cite{unet18} \newline 2018 \newline  MIT license\newline GitHub&
    1&
    Sign Recognition \newline Falling Detection &
    1 &
    10 &
    18.9 MB&
    Intel5300(CSI) &
    \checkmark Implementation code \newline \\ \hline

\end{tabular}
\label{tab:datasets}
\end{table*}

\begin{table*}[h!]
\caption*{Public datasets of labeled radio signal measurements for human activities continued from Table \ref{tab:datasets}}

\begin{tabular}{|p{2.5cm}|p{0.5cm}|p{2.5cm}|p{2.5cm}|p{0.8cm}|p{1cm}|p{2cm}|p{3cm}|}
\hline
\rowcolor{Gray}

\textbf{Source\addslash Year\newline License\addslash Repository } &  \textbf{\#Envs}&\textbf{Applications}&\textbf{Subjects} & \textbf{\#Activities} & \textbf{Size} &\textbf{Data-type \addslash Hardware}&\textbf{Additional Items}\\\hline
\hline

EHUCOUNT\cite{sobron2018} \newline 2018\newline Private repo&6&People \newline Counting &5&1&183MB&Anritsu MS2690A(CSI)&-\\ \hline

mmGaitNet\cite{menggait} \newline 2020\newline GitHub&2&Gait \newline Recognition &95[M45,F50]\newline Age[19-27] \newline Height[150-185]cm\newline weight[45-115]kg&1&913MB&IWR 1443(Point cloud)&-\\ \hline

\cite{gu2020wife} \newline 2020\newline Google Drive.&1&Facial emotions recognition & 10 [M7,F3]\newline Age[23-25]&7&43GB&Intel5300(CSI) \newline Laptop webcam(video) &\checkmark Annotated video\\ \hline

\cite{Alazrai2020} 2020 \newline CC BY 4.0 licence\newline Mendeley Data &1& Human-to-Human Interaction recognition& 66 [M6,F3] \newline Age(avg$\pm$std) [22.1 $\pm$ 3.7]&12 &4.3GB&Intel5300(CSI)& \checkmark Well documented with interaction steps\\ \hline

\cite{Shi2020ADO} 2020 \newline GitHub &1& Respiratory \& Vital signs recognition&11[M7,F4]\newline Age \newline [34.73$\pm$ 15.94]\newline BMI\newline[23.19$\pm$3.61] kg/$m^2$&1&558MB&Six-Port-based radar system (24 GHz)&\checkmark Monitoring is performed under multiple scenarios\newline \checkmark Implementation code \\ \hline

\end{tabular}
\end{table*}

\section{Lessons Learned and Future Directions}
\label{futuredirections}

Although deep learning is proving to be an effective tool for enhancing RF-based human sensing beyond the state-of-the-art, there still exist several roadblocks to fully benefit from it. In this section we discuss some lessons learned and potential future research directions to combat them.

\subsection{The Scale of Human Sensing Experiments}

A clear lesson learned from the recent works is that the shallow machine learning algorithms cannot cope with human sensing tasks at larger scale, where deep learning exhibits great potentials (see Figure~\ref{fig:performance_gain}). Human sensing can scale in many dimensions, i.e., the practical RF sensing systems will be expected to work reliably over a large user population, activities, physical environments, and RF devices. Deep learning research therefore must explore all of these dimensions. However, recent deep learning research considered the scaling only along one of these dimensions. For example, SignFi~\cite{ma18signfi} experiments with 276 sign language gestures, but recruits only 5 subjects working in 2 different physical environments. Similarly, FallDeFi~\cite{Palipana2018} increases the number of physical environments to 7, but recruits only 3 subjects for the experiments. An important future direction, therefore, would be to conduct truly large-scale experiments with scaling achieved simultaneously in multiple dimensions of the sensing problem.  

\subsection{Automatic Labeling}

Manual labeling of RF sensing data is extremely inefficient because, unlike vision data labeling which can be done off-line by watching camera recordings, RF data usually is not intuitive and humans cannot directly interpret it through visual inspection. This forces RF labeling to be done on-line either by external persons observing the experiments, or the subjects carefully following explicit instructions to perform the activities, which increases labeling effort and reduces the quality\footnote{Data  collected naturally in the wild is preferred over data collected in controlled laboratory settings.} of the data considerably. To facilitate large-scale deep learning research for human sensing, a future direction should focus on developing novel tools and techniques that can automatically label RF data collected passively in the wild from many environments capturing data from a vast population performing a myriad of activities as part of their daily routines. 

One option for automatic labeling could be the use of a non-RF modality to record the same scene at the same time as observed by the RF. Then, if the events and activities could be labeled automatically from the non-RF sensor data, then the same labels could be used for the RF data as well. Zhao et al.~\cite{zhaopose18,zhao3dpose} has recently pursued this philosophy successfully using camera as the non-RF modality, where multiple cameras were installed in the RF environment, synchronized with the RF recording device, and human pose was later detected from camera output automatically using image processing to generate the labels for the RF source. This is clearly a promising direction and worthy of further exploration. 

\subsection{Learning from Unlabeled CSI Data}
A fundamental pitfall of deep learning is that it requires massive amount of training data to adequately learn the latent features. As acquisition of vast amount of \textit{labeled} RF data incurs significant difficulty and overhead, in addition to automatic labeling, future research should also investigate efficient exploitation of \textit{unlabeled} data, which is much easier to collect or may be already available elsewhere. Indeed, over the years, the machine learning community has discovered efficient methods for exploiting freely available unlabeled data to reduce the burden of labeled data collection. As these methods have proven very successful in image, audio and text classifications, it would be worth exploring them for WiFi sensing. 

Semi-supervised learning is a machine learning approach that combines a small amount of labeled data with a large amount of unlabeled data during training. In this approach, the knowledge gained from the vast amount of rather easily obtainable unlabeled data can significantly help the supervised classification task at hand, which consequently requires only a small amount of labeled data to achieve good performance. Although typical semi-supervised learning methods would help reduce the burden of collecting massive amount of labeled data to some extend, they usually require~\cite{Nigam2000} the unlabeled data to contain the same label classes that the classifier is trained to classify. For CSI-based activity classification, this means that the unlabeled data must also collect CSI when the humans in the area are performing some specific set of activities of interest, such as falling to the ground if fall detection is the sensing task. Conventional semi-supervised learning therefore is not applicable to WiFi sensing tasks responsible for detecting rare events, such as falls, or have a very large number of activity classes, such as detection of sign language. 

There is a particular type of semi-supervised learning, known as \textit{self-taught learning} (STL)~\cite{ICML_STL_2007}, that relaxes the requirement of the unlabeled data to contain the same classes as used in the classification task. This has vastly enhanced the applicability of unlabeled data for challenging classification tasks in domains such as image, audio, and text. Using STL, the authors of \cite{ICML_STL_2007} have demonstrated that rhino and elephants can be accurately classified with the knowledge gathered from the vast amount of random images, not of elephants or rhinos, freely available in the Internet. They have also successfully applied STL to audio classification, by downloading random speech data to classify speakers, and text classification. STL for WiFi sensing would mean that any available CSI data, irrespective of the actual human activities involved in the data, could be potentially used by any other activity classification applications. For example, unlabeled CSI collected passively when arbitrary people are simply carrying out their usual activities, such as walking, sitting, etc., may provide valuable knowledge when training a deep neural classifier to detect rare and specialised activities, such as fall or sign language. 

\subsection{Deep Learning on Multi-modal RF Sensing}

The vast majority of recent works explored learning from a single RF mode, such as WiFi CSI, mmWave FMCW radar, or even the sub-GHz LoRa signals~\cite{zhang-IMWUT2020}. Since these RF modes work on different spectrum and operate on different principles, opportunities exist to improve human sensing by training deep learning networks on the combination of such multiple RF data streams. It is also worthwhile to investigate deep learning networks that can learn from the combination of RF and other signals, e.g., acoustic and infrared. To achieve power automation, many Internet of Things products in future smart homes are expected to be fitted with solar cells~\cite{youssef2019}. Researchers~\cite{solargest2019} have recently demonstrated that photovoltaic (PV) signals generated by such solar cells contain discriminating features to detect hand gestures. Thus, deep learning that can be simultaneously trained from both RF and PV may lead to more robust human sensing neural networks for ubiquitous deployments.

\subsection{Privacy and Security for WiFi Sensing}
Deep learning is enhancing WiFi sensing capability on multiple fronts. First, it helps to recognise human actions with greater accuracy. Second, more detailed and fine-grained  information about humans, such as smooth tracking of 3D pose~\cite{Jiang2020Towards3H}, can be detected with deep learning. Finally, researchers are now exploring deep learning solutions that make cross-domain sensing less strenuous. While the combined effect of these deep learning advancements no doubt will make WiFi a powerful human sensing tool, they will unfortunately also pose a serious privacy threat. For example, armed with a cross-domain deep learning classifier trained elsewhere, a burglar can easily detect whether any target house is currently empty (no one in the house), and if not empty then where in the house the occupants are located etc. without raising an alarm. Similarly, given the WiFi signals can penetrate walls, windows, and fabrics, neighbours can pry on us even with curtains shut. 

Privacy protection against WiFi sensing, therefore, could be an important future research direction. This is a challenging problem though, because any solution to foil the sensing attempt of an attacker should neither affect any legitimate sensing nor any ongoing data communication over the very WiFi signals used for sensing. Work on this topic is rare with the exception of \cite{phycloak,zhou2020}. For a single antenna system, authors of \cite{phycloak} showed that it is possible for a legitimate sensing device to regenerate a carefully distorted version of the signal to obfuscate the physical information in the WiFi signal without affecting the logical information, i.e., the actual bits carried in the signal. This is a promising direction, but more work is required to make such techniques work for multi-antenna systems, which are becoming increasingly available in commodity hardware. It would be also an interesting research to explore deep learning architectures that can fool such signal obfuscating  and still detect human activities to some extend. This would further push researchers to design more advanced obfuscation techniques resilient to even highly sophisticated attackers. To this end, specialised adversarial networks, as explored in~\cite{zhou2020}, could be designed to effectively prevent such adversarial sensing. Zhou et al.~\cite{zhou2020} have shown that with proper design of the loss function, an adversarial network can only reveal some target human behaviour from the CSI data, such as falling of a person, while not allowing the detection of other private behaviours, such as bathing. These are encouraging developments confirming the privacy protection capabilities of deep learning.

\subsection{Deep Learning for Wide Area RF Sensing}

Existing literature on RF sensing is heavily centred around WiFi mainly because of its ubiquity. However, WiFi is mostly used indoors and severely limited in range, hindering its use for many wide area and outdoor human sensing applications, such as gesture control for outdoor utilities (e.g., a vending machine), search and rescue of human survivors in disaster zones, terrorist spotting and activity tracking, and so on. 

Wide area RF sensing traditionally had not been considered practical due to very weak reflections off the human targets for the signals that were generated from a distant radio tower. Some recent technological developments, however, are creating new opportunities for wide area RF sensing. Dense deployments of shorter-range cellular towers means that outdoor locations can receive cellular signals from a close-by radio tower, increasing the opportunity for a stronger reflection off the human body. To support wide area connectivity for various low-power Internet of Things (IoT) sensors, novel wide area wireless communications technologies, e.g., LoRa~\cite{Lora} and SigFox~\cite{Sigfox}, are being developed. A key distinguishing feature of these wide area IoT communications technologies is their capability to process very weak signals. For example, LoRa can decode signals as weak as $-148$dBm. Finally, it is now becoming possible to carry wireless base stations in low cost flying drones~\cite{UAV_survey_2019} providing further opportunity to extend the sensing coverage over a wide area. 

Indeed, researchers are beginning to explore wide area RF sensing by taking advantage of these new developments. Chen et.al.\cite{LTE_Gesture_IPSN2020} showed that gestures can be accurately detected in outdoor areas using LTE signals, and using a drone-mounted LoRa transmitter-receiver pair, Chen et al.\cite{WideSee_SenSys2019} demonstrated feasibility of outdoor human localization using LoRa signals. While these experiments clearly indicate the feasibility of wide area RF sensing, they also highlight the severe challenges it is facing. LTE-based gesture was only possible if the user was located at some specific spots between the tower and the terminal\cite{LTE_Gesture_IPSN2020}, which severely reduces the quality of user experience. Similarly, the LoRa-based outdoor localisation accuracy was limited to 4.6m, which may not be adequate for some applications. Finally, for drone-mounted LoRa transceivers, the authors\cite{LTE_Gesture_IPSN2020} found that drone vibrations cause significant interference to the LoRa signals, which had to be addressed using algorithms specifically designed for the drone in use. These challenges highlight the potential benefit of deep learning in improving the performance and generalization of wide area RF sensing for a wider range of use case and hardware scenarios. 

\subsection{RF Sensing in Programmable Wireless Environment}

Programmable wireless environment (PWE) \cite{Akyildiz_TON2019} is a novel concept rapidly gaining attention in the wireless communications research community. According to the PWE, the walls or any object surface can be coated with a special artificial metamaterial, that can arbitrarily control the reflection, i.e., the amplitude, phase, and angle, of impinging electromagnetic waves under software control. These surfaces are often dubbed as intelligent reflective surfaces (IRSs). Thus, with IRS, the multipath of any environment can be precisely controlled to realise the desired effects at the intended receivers promising unprecedented performance improvements for wireless communications. Indeed, many research works have recently confirmed that IRS-assisted solutions can significantly improve the capacity, coverage and energy efficiency of existing mobile networks~\cite{Jornet_INFOCOM2018,Rui_Zhang_ICASSP2019,Dong_WCNC2020,RFOCUS_NSDI2020}. 

While current research in PWE is mainly focusing on enhancing the communication performance, the dynamic control of the multipath will also affect any sensing task that relies on wireless multipath for sensing. We envisage the following challenges and future research opportunities for WiFi-based human sensing in PWE. 

\subsubsection{Deep learning for IRS-affected CSI}
Current WiFi sensing research largely assumes that the multipath reflections from the environment is rather stationary because they bounce from fixed surfaces, such as walls, tables, and chairs. It makes it easier to detect human activities from the CSI by focusing on the dynamic elements of the multipath created by the moving human body parts. However, in PWE, the reflections from walls and other environmental surfaces can be highly non-stationary due to the dynamic control of their reflection properties. As a result, the 
amplitude and phase of CSI measurements will be affected not only due to the movement of the human, but also due to the specific control patterns of the IRSs in the environment. This will make it more challenging to classify human activities, which will require more advanced learning and classification techniques to separate the IRS-related effect on CSI from the ones caused by human activity. New deep learning algorithms may be designed that can be trained to separate such IRS effects from the CSI measurements.

\subsubsection{IRS as a sensor for detecting human activities}
The PWE vision indicates that an entire wall may be an IRS with massive number of passive elements that can record the angle, amplitude, and phase of the impinging electromagnetic waves. Thus as the reflections from the human body impinge on the IRS-coated wall, the wall will have a high-resolution view of the human activity and hence can assist in detecting fine-grained human movements with much greater accuracy and ease compared to a single WiFi receiver often considered in conventional research. How to design the human activity detection intelligence for the IRS would be an interesting new research direction, which is likely to benefit from the power of deep learning. 

\subsection{Deep learning for multi-person and complex activity recognition}
To date, RF has been successfully used to detect only single person and simple (atomic) activities, such as sitting, walking, falling, etc. To take RF sensing to the next level where it can used to analyse high level human behaviour, such as whether a person is having dinner in a restaurant or having a conversation with another person, more sophisticated deep learning would be required. Such deep learning would be capable of detecting activities of multiple person simultaneously. Deep multi-task learning, a technique that can learn multiple tasks jointly, has been used by Peng et al.~\cite{Peng2018} successfully to detect complex human behaviour from wearable sensors. It would be an interesting future direction to extend such models to work with RF signal data, such WiFi CSI.   

\section{Conclusion}
\label{conclusion}
We have presented a comprehensive survey of deep learning techniques, architectures, and algorithms recently applied to radio-based device-free human sensing. Our survey has revealed that although the utilization of deep learning in RF sensing is a relatively new trend, significant exploration has already been achieved. It has become clear that deep learning can be an effective tool for improving both the accuracy and scope of device-free RF sensing. Researchers have demonstrated deep learning capabilities for sensing new phenomena that were not possible with conventional methods. Despite these important achievements, progress on domain or environment independent deep learning models has been slow, limiting their ubiquitous use. Dependency on large amounts of labeled data for training is another major drawback of current deep learning models that must be overcome. Through this survey, we have also unveiled the existence of many publicly available datasets for labeled radio signal measurements corresponding to various human activities. With many new deep learning algorithms being discovered each year, these datasets can be readily used in future studies to evaluate and compare new algorithms for RF sensing. We also believe that to further catalyse the deep learning research for RF sensing, researchers should come forward and release more comprehensive datasets for public use.

\section{Acknowledgment}

This work is partially supported by a research grant from Cisco Systems, Inc.

\printbibliography

@INPROCEEDINGS{khanpassive17, 
author={U. M. {Khan} and Z. {Kabir} and S. A. {Hassan} and S. H. {Ahmed}}, 
booktitle={GLOBECOM 2017 - 2017 IEEE Global Communications Conference}, 
title={A Deep Learning Framework Using Passive WiFi Sensing for Respiration Monitoring}, 
year={2017}, 
volume={}, 
number={}, 
pages={1-6}, 
 
month={Dec}


}

@ARTICLE{wanglocact17, 
author={J. {Wang} and X. {Zhang} and Q. {Gao} and H. {Yue} and H. {Wang}}, 
journal={IEEE Transactions on Vehicular Technology}, 
title={Device-Free Wireless Localization and Activity Recognition: A Deep Learning Approach}, 
year={2017}, 
volume={66}, 
number={7}, 
pages={6258-6267}, 

month={July}

}

@INPROCEEDINGS{zhang16, 
author={X. {Zhang} and J. {Wang} and Q. {Gao} and X. {Ma} and H. {Wang}}, 
booktitle={2016 IEEE International Conference on Pervasive Computing and Communication Workshops (PerCom Workshops)}, 
title={Device-free wireless localization and activity recognition with deep learning}, 
year={2016}, 
volume={}, 
number={}, 
pages={1-5}, 

month={March}


}

@INPROCEEDINGS{shi18, 
author={Z. {Shi} and J. A. {Zhang} and R. {Xu} and G. {Fang}}, 
booktitle={2018 IEEE Globecom Workshops (GC Wkshps)}, 
title={Human Activity Recognition Using Deep Learning Networks with Enhanced Channel State Information}, 
year={2018}, 
volume={}, 
number={}, 
pages={1-6}, 
month={Dec}

}

@ARTICLE{gao17, 
author={Q. {Gao} and J. {Wang} and X. {Ma} and X. {Feng} and H. {Wang}}, 
journal={IEEE Transactions on Vehicular Technology}, 
title={CSI-Based Device-Free Wireless Localization and Activity Recognition Using Radio Image Features}, 
year={2017}, 
volume={66}, 
number={11}, 
pages={10346-10356}, 
month={Nov}

}

@article{ma18signfi,
 author = {Ma, Yongsen and Zhou, Gang and Wang, Shuangquan and Zhao, Hongyang and Jung, Woosub},
 title = {SignFi: Sign Language Recognition Using WiFi},
 journal = {Proc. ACM Interact. Mob. Wearable Ubiquitous Technol.},
 issue_date = {March 2018},
 volume = {2},
 number = {1},
 month = mar,
 year = {2018},
 url = {https://github.com/yongsen/SignFi},
 note = {(Accessed on 13/03/2020)},
 pages = {23:1--23:21},
 articleno = {23},
 numpages = {21},
 acmid = {3191755},
 publisher = {ACM},
 address = {New York, NY, USA}

}

@INPROCEEDINGS{khan19, 
author={D. A. {Khan} and S. {Razak} and B. {Raj} and R. {Singh}}, 
booktitle={ICASSP 2019 - 2019 IEEE International Conference on Acoustics, Speech and Signal Processing (ICASSP)}, 
title={Human Behaviour Recognition Using Wifi Channel State Information}, 
year={2019}, 
volume={}, 
number={}, 
pages={7625-7629}, 
month={May}



}

@inproceedings{shi17,
 author = {Shi, Cong and Liu, Jian and Liu, Hongbo and Chen, Yingying},
 title = {Smart User Authentication Through Actuation of Daily Activities Leveraging WiFi-enabled IoT},
 booktitle = {Proceedings of the 18th ACM International Symposium on Mobile Ad Hoc Networking and Computing},
 series = {Mobihoc '17},
 year = {2017},
 isbn = {978-1-4503-4912-3},
 location = {Chennai, India},
 pages = {5:1--5:10},
 articleno = {5},
 numpages = {10},
 publisher = {ACM},
 address = {New York, NY, USA}

 
}

@INPROCEEDINGS{zhaopose18, 
author={M. {Zhao} and T. {Li} and M. A. {Alsheikh} and Y. {Tian} and H. {Zhao} and A. {Torralba} and D. {Katabi}}, 
booktitle={2018 IEEE/CVF Conference on Computer Vision and Pattern Recognition}, 
title={Through-Wall Human Pose Estimation Using Radio Signals}, 
year={2018}, 
volume={}, 
number={}, 
pages={7356-7365}, 
month={June}


}

@INPROCEEDINGS{chengcount17, 
author={Y. {Cheng} and R. Y. {Chang}}, 
booktitle={GLOBECOM 2017 - 2017 IEEE Global Communications Conference}, 
title={Device-Free Indoor People Counting Using Wi-Fi Channel State Information for Internet of Things}, 
year={2017}, 
volume={}, 
number={}, 
pages={1-6}, 
month={Dec}

}

@inproceedings{zhao3dpose,
 author = {Zhao, Mingmin and Tian, Yonglong and Zhao, Hang and Alsheikh, Mohammad Abu and Li, Tianhong and Hristov, Rumen and Kabelac, Zachary and Katabi, Dina and Torralba, Antonio},
 title = {RF-based 3D Skeletons},
 booktitle = {Proceedings of the 2018 Conference of the ACM Special Interest Group on Data Communication},
 series = {SIGCOMM '18},
 year = {2018},
 isbn = {978-1-4503-5567-4},
 location = {Budapest, Hungary},
 pages = {267--281},
 numpages = {15},
 publisher = {ACM},
 address = {New York, NY, USA}

 
}

@inproceedings{soli16,
 author = {Wang, Saiwen and Song, Jie and Lien, Jaime and Poupyrev, Ivan and Hilliges, Otmar},
 title = {Interacting with Soli: Exploring Fine-Grained Dynamic Gesture Recognition in the Radio-Frequency Spectrum},
 booktitle = {Proceedings of the 29th Annual Symposium on User Interface Software and Technology},
 series = {UIST '16},
 year = {2016},
 isbn = {978-1-4503-4189-9},
 location = {Tokyo, Japan},
 pages = {851--860},
 numpages = {10},
 publisher = {ACM},
 address = {New York, NY, USA}

}

@ARTICLE{latern, 
author={Z. {Zhang} and Z. {Tian} and M. {Zhou}}, 
journal={IEEE Sensors Journal}, 
title={Latern: Dynamic Continuous Hand Gesture Recognition Using FMCW Radar Sensor}, 
year={2018}, 
volume={18}, 
number={8}, 
pages={3278-3289}, 
month={April}


}

@ARTICLE{wanglimb18, 
author={M. {Wang} and G. {Cui} and X. {Yang} and L. {Kong}}, 
journal={IET Radar, Sonar Navigation}, 
title={Human body and limb motion recognition via stacked gated recurrent units network}, 
year={2018}, 
volume={12}, 
number={9}, 
pages={1046-1051}, 
month={}


}

@ARTICLE{vander18, 
author={B. {Vandersmissen} and N. {Knudde} and A. {Jalalvand} and I. {Couckuyt} and A. {Bourdoux} and W. {De Neve} and T. {Dhaene}}, 
journal={IEEE Transactions on Geoscience and Remote Sensing}, 
title={Indoor Person Identification Using a Low-Power FMCW Radar}, 
year={2018}, 
volume={56}, 
number={7}, 
pages={3941-3952}, 
month={July}


}

@inproceedings{adib2014witrack,
  title={WiTrack: motion tracking via radio reflections off the body},
  author={Adib, Fadel and Kabelac, Zachary and Katabi, Dina and Miller, Rob},
  booktitle={Proc. of NSDI},
  year={2014}

}

@Article{Zhou2018,
   Author="Zhou, Q.  and Xing, J.  and Chen, W.  and Zhang, X.  and Yang, Q. ",
   Title="{{F}rom {S}ignal to {I}mage: {E}nabling {F}ine-{G}rained {G}esture {R}ecognition with {C}ommercial {W}i-{F}i {D}evices}",
   Journal="Sensors (Basel)",
   Year={2018},
   Volume="18",
   Number="9",
   Month="Sep"
}

@Article{finger19,
AUTHOR = {Ahmed, Shahzad and Khan, Faheem and Ghaffar, Asim and Hussain, Farhan and Cho, Sung Ho},
TITLE = {Finger-Counting-Based Gesture Recognition within Cars Using Impulse Radar with Convolutional Neural Network},
JOURNAL = {Sensors},
VOLUME = {19},
YEAR = {2019},
NUMBER = {6},
ARTICLE-NUMBER = {1429}

}

@ARTICLE{skaria19, 
author={S. {Skaria} and A. {Al-Hourani} and M. {Lech} and R. J. {Evans}}, 
journal={IEEE Sensors Journal}, 
title={Hand-Gesture Recognition Using Two-Antenna Doppler Radar With Deep Convolutional Neural Networks}, 
year={2019}, 
volume={19}, 
number={8}, 
pages={3041-3048}, 
month={April}

}

@InProceedings{sobron18,
author="Sobron, Iker
and Del Ser, Javier
and Eizmendi, I{\~{n}}aki
and Velez, Manuel",
editor="Del Ser, Javier
and Osaba, Eneko
and Bilbao, Miren Nekane
and Sanchez-Medina, Javier J.
and Vecchio, Massimo
and Yang, Xin-She",
title="A Deep Learning Approach to Device-Free People Counting from WiFi Signals",
booktitle="Intelligent Distributed Computing XII",
year={2018},
publisher="Springer International Publishing",
address="Cham",
pages="275--286"

}

@ARTICLE{Wang2019DeepApproach, 
author={F. {Wang} and W. {Gong} and J. {Liu}}, 
journal={IEEE Internet of Things Journal}, 
title={On Spatial Diversity in WiFi-Based Human Activity Recognition: A Deep Learning-Based Approach}, 
year={2019}, 
volume={6}, 
number={2}, 
pages={2035-2047}, 
month={April}

}

@article{fall18,
author = {Xu Yang and Fangyuan Xiong and Yuan Shao and Qiang Niu},
title ={WmFall: WiFi-based multistage fall detection with channel state information},
journal = {International Journal of Distributed Sensor Networks},
volume = {14},
number = {10},
pages = {1550147718805718},
year = {2018}

}

@misc{unet18,
Author = {Fei Wang and Jinsong Han and Shiyuan Zhang and Xu He and Dong Huang},
Title = {CSI-Net: Unified Human Body Characterization and Pose Recognition},
Year = {2018},
Eprint = {arXiv:1810.03064},
url = {https://github.com/geekfeiw/CSI-Net},
note = {(Accessed on 13/03/2020)}

}

@ARTICLE{fang19enhanced, 
author={S. {Fang} and C. {Li} and W. {Lu} and Z. {Xu} and Y. {Chien}}, 
journal={IEEE Transactions on Vehicular Technology}, 
title={Enhanced Device-Free Human Detection: Efficient Learning From Phase and Amplitude of Channel State Information}, 
year={2019}, 
volume={68}, 
number={3}, 
pages={3048-3051}, 
month={March}

}

@inproceedings{zhao2017learning,
  title={Learning sleep stages from radio signals: A conditional adversarial architecture},
  author={Zhao, Mingmin and Yue, Shichao and Katabi, Dina and Jaakkola, Tommi S and Bianchi, Matt T},
  booktitle={Proceedings of the 34th International Conference on Machine Learning-Volume 70},
  pages={4100--4109},
  year={2017},
  organization={JMLR. org}

}

@inproceedings{wang2019wicar,
author = {Wang, Fangxin and Liu, Jiangchuan and Gong, Wei},
title = {WiCAR: Wifi-Based in-Car Activity Recognition with Multi-Adversarial Domain Adaptation},
year = {2019},
isbn = {9781450367783},
publisher = {Association for Computing Machinery},
address = {New York, NY, USA},
url = {https://doi.org/10.1145/3326285.3329054},
doi = {10.1145/3326285.3329054},
booktitle = {Proceedings of the International Symposium on Quality of Service},
articleno = {19},
numpages = {10},
location = {Phoenix, Arizona},
series = {IWQoS '19}
}

@inproceedings{zheng2019zero,
  title={Zero-Effort Cross-Domain Gesture Recognition with Wi-Fi},
  author={Zheng, Yue and Zhang, Yi and Qian, Kun and Zhang, Guidong and Liu, Yunhao and Wu, Chenshu and Yang, Zheng},
  booktitle={Proceedings of the 17th Annual International Conference on Mobile Systems, Applications, and Services},
  pages={313--325},
  year={2019},
  organization={ACM},
  url = {http://tns.thss.tsinghua.edu.cn/widar3.0/index.html},
  note = {(Accessed on 13/03/2020)},

}

@inproceedings{jiang2018towards,
  title={Towards environment independent device free human activity recognition},
  author={Jiang, Wenjun and Miao, Chenglin and Ma, Fenglong and Yao, Shuochao and Wang, Yaqing and Yuan, Ye and Xue, Hongfei and Song, Chen and Ma, Xin and Koutsonikolas, Dimitrios and others},
  booktitle={Proceedings of the 24th Annual International Conference on Mobile Computing and Networking},
  pages={289--304},
  year={2018},
  organization={ACM}

}

@inproceedings{shu2018dirt,
  author    = {Rui Shu and
               Hung H. Bui and
               Hirokazu Narui and
               Stefano Ermon},
  title     = {A {DIRT-T} Approach to Unsupervised Domain Adaptation},
  booktitle = {6th International Conference on Learning Representations, {ICLR} 2018,
               Vancouver, BC, Canada, April 30 - May 3, 2018, Conference Track Proceedings},
  publisher = {OpenReview.net},
  year      = {2018},
  url       = {https://openreview.net/forum?id=H1q-TM-AW}

}

@article{HuangAuID19,
 author = {Huang, Anna and Wang, Dong and Zhao, Run and Zhang, Qian},
 title = {Au-Id: Automatic User Identification and Authentication Through the Motions Captured from Sequential Human Activities Using RFID},
 journal = {Proc. ACM Interact. Mob. Wearable Ubiquitous Technol.},
 issue_date = {June 2019},
 volume = {3},
 number = {2},
 month = jun,
 year = {2019},
 pages = {48:1--48:26},
 articleno = {48},
 numpages = {26},
 acmid = {3328919},
 publisher = {ACM},
 address = {New York, NY, USA},

}

@ARTICLE{Wang2019, 
author={F. {Wang} and W. {Gong} and J. {Liu} and K. {Wu}}, 
journal={IEEE Transactions on Network Science and Engineering}, 
title={Channel Selective Activity Recognition with WiFi: A Deep Learning Approach Exploring Wideband Information}, 
year={2019}, 
volume={}, 
number={}, 
pages={1-1}, 
month={},


}

@ARTICLE{Yousefi2017, 
author={Yousefi,Siamak  and Narui, Hirokazu  and Dayal,Sankalp  and Ermon,Stefano  and Valaee, Shahrokh }, 
journal={IEEE Communications Magazine}, 
title={A Survey on Behavior Recognition Using WiFi Channel State Information}, 
year={2017}, 
volume={55}, 
number={10}, 
pages={98-104}, 
month={Oct},
url = {https://github.com/ermongroup/Wifi_Activity_Recognition},
note = {(Accessed on 13/03/2020)},

}

@inproceedings{xie2019mdtrack,
  title={mD-Track: Leveraging multi-dimensionality for passive indoor Wi-Fi tracking},
  author={Xie, Yaxiong and Xiong, Jie and Li, Mo and Jamieson, Kyle},
  booktitle={The 25th Annual International Conference on Mobile Computing and Networking},
  pages={1--16},
  year={2019},
  organization={ACM},

}

@INPROCEEDINGS{Zou2018, 
author={H. {Zou} and Y. {Zhou} and J. {Yang} and H. {Jiang} and L. {Xie} and C. J. {Spanos}}, 
booktitle={2018 IEEE International Conference on Communications (ICC)}, 
title={DeepSense: Device-Free Human Activity Recognition via Autoencoder Long-Term Recurrent Convolutional Network}, 
year={2018}, 
volume={}, 
number={}, 
pages={1-6}, 
month={May},

}

@article{haseeb2017wisture,
  title={Wisture: Rnn-based learning of wireless signals for gesture recognition in unmodified smartphones},
  author={Haseeb, Mohamed Abudulaziz Ali and Parasuraman, Ramviyas},
  journal={arXiv preprint arXiv:1707.08569},
  year={2017},
  url = {https://ieee-dataport.org/documents/wi-fi-signal-strength-measurements-smartphone-various-hand-gestures},
  note = {(Accessed on 13/03/2020)},

}

@article{Chen2017TamingTI,
  title={Taming the inconsistency of Wi-Fi fingerprints for device-free passive indoor localization},
  author={Xi Chen and Chen Ma and Michel Allegue and Xue Liu},
  journal={IEEE INFOCOM 2017 - IEEE Conference on Computer Communications},
  year={2017},
  pages={1-9},

}

@INPROCEEDINGS{Shi2019, 
author={Z. {Shi} and J. A. {Zhang} and R. {Xu} and Q. {Cheng}}, 
booktitle={ICC 2019 - 2019 IEEE International Conference on Communications (ICC)}, 
title={Deep Learning Networks for Human Activity Recognition with CSI Correlation Feature Extraction}, 
year={2019}, 
volume={}, 
number={}, 
pages={1-6}, 
month={May},

}

@ARTICLE{Zhao2019, 
author={L. {Zhao} and H. {Huang} and X. {Li} and S. {Ding} and H. {Zhao} and Z. {Han}}, 
journal={IEEE Internet of Things Journal}, 
title={An Accurate and Robust Approach of Device-Free Localization With Convolutional Autoencoder}, 
year={2019}, 
volume={6}, 
number={3}, 
pages={5825-5840}, 
month={June},

}

@article{Fan2018TagFreeAI,
  title={TagFree Activity Identification with RFIDs},
  author={Xiaoyi Fan and Wei Gong and Jiangchuan Liu},
  journal={IMWUT},
  year={2018},
  volume={2},
  pages={7:1-7:23},

}

@article{Xu2018AttentionbasedWG,
  title={Attention-based Walking Gait and Direction Recognition in Wi-Fi Networks},
  author={Yang Xu and Min Chen and Wei Yang and Siguang Chen and Liusheng Huang},
  journal={ArXiv},
  year={2018},
  volume={abs/1811.07162},

}

@article{Fan2019WhenRM,
  title={When RFID Meets Deep Learning: Exploring Cognitive Intelligence for Activity Identification},
  author={Xiaoyi Fan and Fangxin Wang and Fei Wang and Wei Gong and Jiangchuan Liu},
  journal={IEEE Wireless Communications},
  year={2019},
  volume={26},
  pages={19-25},

}

@INPROCEEDINGS{Zhao2018, 
author={L. {Zhao} and H. {Huang} and S. {Ding} and X. {Li}}, 
booktitle={2018 IEEE International Conference on Systems, Man, and Cybernetics (SMC)}, 
title={An Accurate and Efficient Device-Free Localization Approach Based on Gaussian Bernoulli Restricted Boltzmann Machine}, 
year={2018}, 
volume={}, 
number={}, 
pages={2323-2328}, 
month={Oct},

}

@inproceedings{Huang2018WiDetWB,
author = {Huang, Hua and Lin, Shan},
title = {WiDet: Wi-Fi Based Device-Free Passive Person Detection with Deep Convolutional Neural Networks},
year = {2018},
isbn = {9781450359603},
publisher = {Association for Computing Machinery},
address = {New York, NY, USA},
url = {https://doi.org/10.1145/3242102.3242119},
doi = {10.1145/3242102.3242119},
booktitle = {Proceedings of the 21st ACM International Conference on Modeling, Analysis and Simulation of Wireless and Mobile Systems},
pages = {53–60},
numpages = {8},
location = {Montreal, QC, Canada},
series = {MSWIM ’18}
}

@article{Cai2018PILCPI,
  title={PILC: Passive Indoor Localization Based on Convolutional Neural Networks},
  author={Chenwei Cai and Li Juan Deng and Mingyang Zheng and Shufang Li},
  journal={2018 Ubiquitous Positioning, Indoor Navigation and Location-Based Services (UPINLBS)},
  year={2018},
  pages={1-6}
}

@article{Zou2018WiFienabledDG,
  title={WiFi-enabled Device-free Gesture Recognition for Smart Home Automation},
  author={Han Zou and Yuxun Zhou and Jianfei Yang and Hao Lin Jiang and Lihua Xie and Costas J. Spanos},
  journal={2018 IEEE 14th International Conference on Control and Automation (ICCA)},
  year={2018},
  pages={476-481}
}

@article{Lv2017QualitativeAR,
  title={Qualitative Action Recognition by Wireless Radio Signals in Human–Machine Systems},
  author={Shaohe Lv and Yong Lu and Mianxiong Dong and Xiaodong Wang and Yong Dou and Weihua Zhuang},
  journal={IEEE Transactions on Human-Machine Systems},
  year={2017},
  volume={47},
  pages={789-800}
  
}

@article{Wang2019JointAR,
  title={Joint Activity Recognition and Indoor Localization With WiFi Fingerprints},
  author={Feng Wang and Jianwei Feng and Yinliang Zhao and Xiaobin Zhang and Shiyuan Zhang and Jinsong Han},
  journal={IEEE Access},
  year={2019},
  volume={7},
  pages={80058-80068},
  url = {https://github.com/geekfeiw/ARIL},
  note = {(Accessed on 13/03/2020)}
}

@ARTICLE{Khatab2018, 
author={Z. E. {Khatab} and A. {Hajihoseini} and S. A. {Ghorashi}}, 
journal={IEEE Sensors Letters}, 
title={A Fingerprint Method for Indoor Localization Using Autoencoder Based Deep Extreme Learning Machine}, 
year={2018}, 
volume={2}, 
number={1}, 
pages={1-4}, 
month={March}

}

@article{Liu2017WiCountAD,
  title={WiCount: A Deep Learning Approach for Crowd Counting Using WiFi Signals},
  author={Shangqing Liu and Yanchao Zhao and Bingzhang Chen},
  journal={2017 IEEE International Symposium on Parallel and Distributed Processing with Applications and 2017 IEEE International Conference on Ubiquitous Computing and Communications (ISPA/IUCC)},
  year={2017},
  pages={967-974}

}

@article{zou2018towards,
  title={Towards occupant activity driven smart buildings via WiFi-enabled IoT devices and deep learning},
  author={Zou, Han and Zhou, Yuxun and Yang, Jianfei and Spanos, Costas J},
  journal={Energy and Buildings},
  volume={177},
  year={2018},
  publisher={Elsevier}
}

@ARTICLE{Zhou2019, 
    author={R. {Zhou} and M. {Tang} and Z. {Gong} and M. {Hao}}, 
    journal={IEEE Systems Journal}, 
    title={FreeTrack: Device-Free Human Tracking With Deep Neural Networks and Particle Filtering}, 
    year={2019}, 
    volume={}, 
    number={}, 
    pages={1-11}, 
    month={}
}

@INPROCEEDINGS{Zhou2016, 
author={Q. {Zhou} and J. {Xing} and J. {Li} and Q. {Yang}}, 
booktitle={2016 12th International Conference on Computational Intelligence and Security (CIS)}, 
title={A Device-Free Number Gesture Recognition Approach Based on Deep Learning}, 
year={2016}, 
volume={}, 
number={}, 
pages={57-63}, 
month={Dec}



}

@article{Liu2019AnalysisAV,
  title={Analysis and Visualization of Deep Neural Networks in Device-Free Wi-Fi Indoor Localization},
  author={Shing-Jiuan Liu and Ronald Y. Chang and Feng-Tsun Chien},
  journal={IEEE Access},
  year={2019},
  volume={7},
  pages={69379-69392}

}

@inproceedings{noprocess19,
  title={No Need of Data Pre-processing: A General Framework for Radio-Based Device-Free Context Awareness},
  author={Bo Wei and Kai Li and Chengwen Luo and Weitao Xu and Jin Zhang},
  year={2019},
  eprint={1908.03398},
  archivePrefix={arXiv}
}

@article{Chang2018DeviceFreeIL,
  title={Device-Free Indoor Localization Using Wi-Fi Channel State Information for Internet of Things},
  author={Ronald Y. Chang and Shing-Jiuan Liu and Yen-Kai Cheng},
  journal={2018 IEEE Global Communications Conference (GLOBECOM)},
  year={2018},
  pages={1-7}

}

@ARTICLE{Wang2019Survey, 
author={Z. {Wang} and K. {Jiang} and Y. {Hou} and Z. {Huang} and W. {Dou} and C. {Zhang} and Y. {Guo}}, 
journal={IEEE Access}, 
title={A Survey on CSI-Based Human Behavior Recognition in Through-the-Wall Scenario}, 
year={2019}, 
volume={7}, 
number={}, 
pages={78772-78793}, 
month={}


}

@inproceedings{Zhang2018,
 author = {Zhang, Jie and Tang, Zhanyong and Li, Meng and Fang, Dingyi and Nurmi, Petteri and Wang, Zheng},
 title = {CrossSense: Towards Cross-Site and Large-Scale WiFi Sensing},
 booktitle = {Proceedings of the 24th Annual International Conference on Mobile Computing and Networking},
 series = {MobiCom '18},
 year = {2018},
 isbn = {978-1-4503-5903-0},
 location = {New Delhi, India},
 pages = {305--320},
 numpages = {16},
 acmid = {3241570},
 publisher = {ACM},
 address = {New York, NY, USA},
 url = {https://github.com/nwuzj/CrossSense},
 note = {(Accessed on 13/03/2020)}


}

@inproceedings{zou2018joint,
  title={Joint Adversarial Domain Adaptation for Resilient WiFi-Enabled Device-Free Gesture Recognition},
  author={Zou, Han and Yang, Jianfei and Zhou, Yuxun and Spanos, Costas J},
  booktitle={2018 17th IEEE International Conference on Machine Learning and Applications (ICMLA)},
  pages={202--207},
  year={2018},
  organization={IEEE}

}

@inproceedings{Virmani2017,
 author = {Virmani, Aditya and Shahzad, Muhammad},
 title = {Position and Orientation Agnostic Gesture Recognition Using WiFi},
 booktitle = {Proceedings of the 15th Annual International Conference on Mobile Systems, Applications, and Services},
 series = {MobiSys '17},
 year = {2017},
 isbn = {978-1-4503-4928-4},
 location = {Niagara Falls, New York, USA},
 pages = {252--264},
 numpages = {13},
 acmid = {3081340},
 publisher = {ACM},
 address = {New York, NY, USA},
 url = {https://people.engr.ncsu.edu/mshahza/publications/Datasets/WiAGData.zip},
 note = {(Accessed on 13/03/2020)}

}

@inproceedings{Hillyard2018,
 author = {Hillyard, Peter and Luong, Anh and Abrar, Alemayehu Solomon and Patwari, Neal and Sundar, Krishna and Farney, Robert and Burch, Jason and Porucznik, Christina and Pollard, Sarah Hatch},
 title = {Experience: Cross-Technology Radio Respiratory Monitoring Performance Study},
 booktitle = {Proceedings of the 24th Annual International Conference on Mobile Computing and Networking},
 series = {MobiCom '18},
 year = {2018},
 isbn = {978-1-4503-5903-0},
 location = {New Delhi, India},
 pages = {487--496},
 numpages = {10},
 acmid = {3241560},
 publisher = {ACM},
 url = {https://dataverse.harvard.edu/dataverse/rf_respiration_monitoring},
 note = {(Accessed on 13/03/2020)}

}

@inproceedings{cho2011improved,
  title={Improved learning of Gaussian-Bernoulli restricted Boltzmann machines},
  author={Cho, KyungHyun and Ilin, Alexander and Raiko, Tapani},
  booktitle={International conference on artificial neural networks},
  pages={10--17},
  year={2011},
  organization={Springer}

}

@article{wang2017tensorbeat,
  title={TensorBeat: Tensor decomposition for monitoring multiperson breathing beats with commodity WiFi},
  author={Wang, Xuyu and Yang, Chao and Mao, Shiwen},
  journal={ACM Transactions on Intelligent Systems and Technology (TIST)},
  volume={9},
  number={1},
  pages={8},
  year={2017},
  publisher={ACM}

}

@article{rf-fall,
  title={RF-based fall monitoring using convolutional neural networks},
  author={Tian, Yonglong and Lee, Guang-He and He, Hao and Hsu, Chen-Yu and Katabi, Dina},
  journal={Proceedings of the ACM on Interactive, Mobile, Wearable and Ubiquitous Technologies},
  volume={2},
  number={3},
  pages={137},
  year={2018},
  publisher={ACM}

}

@article{ma2019wifisurvey,
 author = {Ma, Yongsen and Zhou, Gang and Wang, Shuangquan},
 title = {WiFi Sensing with Channel State Information: A Survey},
 journal = {ACM Comput. Surv.},
 issue_date = {July 2019},
 volume = {52},
 number = {3},
 month = jun,
 year = {2019},
 pages = {46:1--46:36},
 articleno = {46},
 numpages = {36},
 acmid = {3310194},
 publisher = {ACM},
 address = {New York, NY, USA}

}

@article{unsupervisedsurvey,
   title={A Survey of Unsupervised Deep Domain Adaptation},
   volume={11},
   ISSN={2157-6912},
   url={http://dx.doi.org/10.1145/3400066},
   DOI={10.1145/3400066},
   number={5},
   journal={ACM Transactions on Intelligent Systems and Technology},
   publisher={Association for Computing Machinery (ACM)},
   author={Wilson, Garrett and Cook, Diane J.},
   year={2020},
   month={Aug},
   pages={1–46}
}

@ARTICLE{Ibrahim2019,
author={O. T. {Ibrahim} and W. {Gomaa} and M. {Youssef}},
journal={IEEE Sensors Journal},
title={CrossCount: A Deep Learning System for Device-Free Human Counting Using WiFi},
year={2019},
volume={19},
number={21},
pages={9921-9928},
ISSN={},
month={Nov}


}

@article{feng2019wi,
  title={Wi-multi: A Three-phase System for Multiple Human Activity Recognition with Commercial WiFi Devices},
  author={Feng, Chunhai and Arshad, Sheheryar and Zhou, Siwang and Cao, Dun and Liu, Yonghe},
  journal={IEEE Internet of Things Journal},
  year={2019},
  publisher={IEEE}
}

@article{Palipana2018,
 author = {Palipana, Sameera and Rojas, David and Agrawal, Piyush and Pesch, Dirk},
 title = {FallDeFi: Ubiquitous Fall Detection Using Commodity Wi-Fi Devices},
 journal = {Proc. ACM Interact. Mob. Wearable Ubiquitous Technol.},
 issue_date = {December 2017},
 volume = {1},
 number = {4},
 month = jan,
 year = {2018},
 issn = {2474-9567},
 pages = {155:1--155:25},
 articleno = {155},
 numpages = {25},
 url = {https://github.com/dmsp123/FallDeFi},
 note = {(Accessed on 13/03/2020)},
 doi = {10.1145/3161183},
 acmid = {3161183},
 publisher = {ACM},
 address = {New York, NY, USA}

}

@inproceedings{Qian2017,
 author = {Qian, Kun and Wu, Chenshu and Yang, Zheng and Liu, Yunhao and Jamieson, Kyle},
 title = {Widar: Decimeter-Level Passive Tracking via Velocity Monitoring with Commodity Wi-Fi},
 booktitle = {Proceedings of the 18th ACM International Symposium on Mobile Ad Hoc Networking and Computing},
 series = {Mobihoc '17},
 year = {2017},
 isbn = {978-1-4503-4912-3},
 location = {Chennai, India},
 pages = {6:1--6:10},
 articleno = {6},
 numpages = {10},
 url = {http://tns.thss.tsinghua.edu.cn/wifiradar/WidarProject.zip},
 note = {(Accessed on 13/03/2020)},
 doi = {10.1145/3084041.3084067},
 acmid = {3084067},
 publisher = {ACM},
 address = {New York, NY, USA}

}

@inproceedings{Qian2018,
 author = {Qian, Kun and Wu, Chenshu and Zhang, Yi and Zhang, Guidong and Yang, Zheng and Liu, Yunhao},
 title = {Widar2.0: Passive Human Tracking with a Single Wi-Fi Link},
 booktitle = {Proceedings of the 16th Annual International Conference on Mobile Systems, Applications, and Services},
 series = {MobiSys '18},
 year = {2018},
 isbn = {978-1-4503-5720-3},
 location = {Munich, Germany},
 pages = {350--361},
 numpages = {12},
 url = {http://tns.thss.tsinghua.edu.cn/wifiradar/Widar2.0Project.zip},
 note = {(Accessed on 13/03/2020)},
 doi = {10.1145/3210240.3210314},
 acmid = {3210314},
 publisher = {ACM},
 address = {New York, NY, USA}

}

@ARTICLE{Wu2018,
author={X. {Wu} and Z. {Chu} and P. {Yang} and C. {Xiang} and X. {Zheng} and W. {Huang}},
journal={IEEE Transactions on Vehicular Technology},
title={TW-See: Human Activity Recognition Through the Wall With Commodity Wi-Fi Devices},
year={2019},
volume={68},
number={1},
pages={306-319},
doi={10.1109/TVT.2018.2878754},
ISSN={},
month={Jan}
}

@inproceedings{singh2019radhar,
  title={RadHAR: Human Activity Recognition from Point Clouds Generated through a Millimeter-wave Radar},
  author={Singh, Akash Deep and Sandha, Sandeep Singh and Garcia, Luis and Srivastava, Mani},
  booktitle={Proceedings of the 3rd ACM Workshop on Millimeter-wave Networks and Sensing Systems},
  pages={51--56},
  year={2019},
  organization={ACM},
  url = {https://github.com/nesl/RadHAR},
  note = {(Accessed on 13/03/2020)}


  
}

@ARTICLE{Wiar2019,
author={L. {Guo} and L. {Wang} and C. {Lin} and J. {Liu} and B. {Lu} and J. {Fang} and Z. {Liu} and Z. {Shan} and J. {Yang} and S. {Guo}},
journal={IEEE Access},
title={Wiar: A Public Dataset for Wifi-Based Activity Recognition},
year={2019},
volume={7},
number={},
pages={154935-154945},
ISSN={2169-3536},
month={},
url = {https://github.com/linteresa/WiAR},
note = {(Accessed on 13/03/2020)}

}

@article{Vincent2010,
 author = {Vincent, Pascal and Larochelle, Hugo and Lajoie, Isabelle and Bengio, Yoshua and Manzagol, Pierre-Antoine},
 title = {Stacked Denoising Autoencoders: Learning Useful Representations in a Deep Network with a Local Denoising Criterion},
 journal = {J. Mach. Learn. Res.},
 issue_date = {3/1/2010},
 volume = {11},
 month = dec,
 year = {2010},
 issn = {1532-4435},
 pages = {3371--3408},
 numpages = {38},
 url = {http://dl.acm.org/citation.cfm?id=1756006.1953039},
 acmid = {1953039},
 publisher = {JMLR.org}
}

@inproceedings{cao2017realtime,
  author = {Zhe Cao and Tomas Simon and Shih-En Wei and Yaser Sheikh},
  booktitle = {CVPR},
  title = {Realtime Multi-Person 2D Pose Estimation using Part Affinity Fields},
  year = {2017}
  }

@inproceedings{deconvolutions2010,
title = "Deconvolutional networks",
author = "Zeiler, {Matthew D.} and Dilip Krishnan and Taylor, {Graham W.} and Rob Fergus",
year = "2010",
doi = "10.1109/CVPR.2010.5539957",
language = "English (US)",
isbn = "9781424469840",
pages = "2528--2535",
booktitle = "2010 IEEE Computer Society Conference on Computer Vision and Pattern Recognition, CVPR 2010"
}

@ARTICLE{CsiGAN2019,
author={C. {Xiao} and D. {Han} and Y. {Ma} and Z. {Qin}},
journal={IEEE Internet of Things Journal},
title={CsiGAN: Robust Channel State Information-Based Activity Recognition With GANs},
year={2019},
volume={6},
number={6},
pages={10191-10204},
doi={10.1109/JIOT.2019.2936580},
ISSN={2372-2541},
month={Dec}
}

@inproceedings{yang2020muid,
  title={MU-ID: Multi-user Identification Through Gaits Using Millimeter Wave Radios},
  author={Yang, Xin and Liu, Jian and Chen, Yingying and Guo, Xiaonan and Xie, Yucheng  },
  booktitle={IEEE INFOCOM 2020-IEEE Conference on Computer Communications},
  year={2020},
  organization={IEEE}
}

@inproceedings{wang2017phasebeat,
  title={PhaseBeat: Exploiting CSI phase data for vital sign monitoring with commodity WiFi devices},
  author={Wang, Xuyu and Yang, Chao and Mao, Shiwen},
  booktitle={2017 IEEE 37th International Conference on Distributed Computing Systems (ICDCS)},
  pages={1230--1239},
  year={2017},
  organization={IEEE}
}

@inproceedings{khamis2018cardiofi,
  title={Cardiofi: Enabling heart rate monitoring on unmodified cots wifi devices},
  author={Khamis, Abdelwahed and Chou, Chun Tung and Kusy, Branislav and Hu, Wen},
  booktitle={Proceedings of the 15th EAI International Conference on Mobile and Ubiquitous Systems: Computing, Networking and Services},
  pages={97--106},
  year={2018}
}

@article{zeng2018fullbreathe,
  title={FullBreathe: Full human respiration detection exploiting complementarity of CSI phase and amplitude of WiFi signals},
  author={Zeng, Youwei and Wu, Dan and Gao, Ruiyang and Gu, Tao and Zhang, Daqing},
  journal={Proceedings of the ACM on Interactive, Mobile, Wearable and Ubiquitous Technologies},
  volume={2},
  number={3},
  pages={1--19},
  year={2018},
  publisher={ACM New York, NY, USA}

}

@ARTICLE{sobron2018,
author={I. {Sobron} and J. {Del Ser} and I. {Eizmendi} and M. {Vélez}},
journal={IEEE Internet of Things Journal},
title={Device-Free People Counting in IoT Environments: New Insights, Results, and Open Challenges},
year={2018},
volume={5},
number={6},
pages={4396-4408},
doi={10.1109/JIOT.2018.2806990},
ISSN={2372-2541},
month={Dec},
url = {http://www.ehu.eus/tsr_radio/index.php/research-areas/data-analytics-in-wireless-networks},
note = {(Accessed on 13/03/2020)}

}

@ARTICLE{Liu2019,
author={J. {Liu} and H. {Liu} and Y. {Chen} and Y. {Wang} and C. {Wang}}, 
journal={IEEE Communications Surveys and Tutorials}, 
title={Wireless Sensing for Human Activity: A Survey}, 
year={2020}, 
volume={22}, 
number={3}, 
pages={1629-1645},
month={},

}

@ARTICLE{Wang2019survey2,
author={Z. {Wang} and K. {Jiang} and Y. {Hou} and W. {Dou} and C. {Zhang} and Z. {Huang} and Y. {Guo}},
journal={IEEE Access},
title={A Survey on Human Behavior Recognition Using Channel State Information},
year={2019},
volume={7},
number={},
pages={155986-156024},
ISSN={2169-3536},
month={}
}

@inproceedings{wang2015understanding,
  title={Understanding and modeling of wifi signal based human activity recognition},
  author={Wang, Wei and Liu, Alex X and Shahzad, Muhammad and Ling, Kang and Lu, Sanglu},
  booktitle={Proceedings of the 21st annual international conference on mobile computing and networking},
  pages={65--76},
  year={2015}

}

@article{ganin2014unsupervised,
  title={Unsupervised domain adaptation by backpropagation},
  author={Ganin, Yaroslav and Lempitsky, Victor},
  journal={arXiv preprint arXiv:1409.7495},
  year={2014}

}

@inproceedings{Brinke2019,
author = {Brinke, Jeroen Klein and Meratnia, Nirvana},
title = {Dataset: Channel State Information for Different Activities, Participants and Days},
year = {2019},
isbn = {9781450369930},
publisher = {Association for Computing Machinery},
address = {New York, NY, USA},
url = {https://data.4tu.nl/repository/uuid:42bffa4c-113c-46eb-84a1-c87b6a31a99f},
note = {(Accessed on 13/03/2020)},
booktitle = {Proceedings of the 2nd Workshop on Data Acquisition To Analysis},
pages = {61–64},
numpages = {4},
location = {New York, NY, USA},
series = {DATA’19}

}

@inproceedings{Jiang2020Towards3H,
author = {Jiang, Wenjun and Xue, Hongfei and Miao, Chenglin and Wang, Shiyang and Lin, Sen and Tian, Chong and Murali, Srinivasan and Hu, Haochen and Sun, Zhi and Su, Lu},
title = {Towards 3D Human Pose Construction Using Wifi},
year = {2020},
isbn = {9781450370851},
publisher = {Association for Computing Machinery},
address = {New York, NY, USA},
url = {https://doi.org/10.1145/3372224.3380900},
doi = {10.1145/3372224.3380900},
booktitle = {Proceedings of the 26th Annual International Conference on Mobile Computing and Networking},
articleno = {Article 23},
numpages = {14},
location = {London, United Kingdom},
series = {MobiCom ’20}

}

@inproceedings {phycloak,
author = {Yue Qiao and Ouyang Zhang and Wenjie Zhou and Kannan Srinivasan and Anish Arora},
title = {PhyCloak: Obfuscating Sensing from Communication Signals},
booktitle = {13th {USENIX} Symposium on Networked Systems Design and Implementation ({NSDI} 16)},
year = {2016},
isbn = {978-1-931971-29-4},
address = {Santa Clara, CA},
pages = {685--699},
url = {https://www.usenix.org/conference/nsdi16/technical-sessions/presentation/qiao},
publisher = {{USENIX} Association},
month = mar

}

@INPROCEEDINGS{Jornet_INFOCOM2018, author={X. {Tan} and Z. {Sun} and D. {Koutsonikolas} and J. M. {Jornet}}, booktitle={IEEE INFOCOM 2018 - IEEE Conference on Computer Communications},
 title={Enabling Indoor Mobile Millimeter-wave Networks Based on Smart Reflect-arrays}, year={2018}, volume={}, number={}, pages={270-278}

 }

@INPROCEEDINGS{Rui_Zhang_ICASSP2019, author={Q. {Wu} and R. {Zhang}}, booktitle={ICASSP 2019 - 2019 IEEE International Conference on Acoustics, Speech and Signal Processing (ICASSP)}, 
 title={Beamforming Optimization for Intelligent Reflecting Surface with Discrete Phase Shifts}, year={2019}, volume={}, number={}, pages={7830-7833}

 }

@INPROCEEDINGS{Dong_WCNC2020,
  title={Enhancing Cellular Communications for UAVs via Intelligent Reflective Surface},
  author={Ma, Dong and Ding, Ming and Hassan, Mahbub},
  booktitle={IEEE WCNC 2020 - IEEE Wireless Communications and Networking Conference},
  year={2020}

}

@ARTICLE{Akyildiz_TON2019, author={C. {Liaskos} and A. {Tsioliaridou} and S. {Nie} and A. {Pitsillides} and S. {Ioannidis} and I. F. {Akyildiz}}, journal={IEEE/ACM Transactions on Networking}, title={On the Network-Layer Modeling and Configuration of Programmable Wireless Environments}, year={2019}, volume={27}, number={4}, pages={1696-1713}

}

@inproceedings{ICML_STL_2007,
author = {Raina, Rajat and Battle, Alexis and Lee, Honglak and Packer, Benjamin and Ng, Andrew Y.},
title = {Self-Taught Learning: Transfer Learning from Unlabeled Data},
year = {2007},
isbn = {9781595937933},
publisher = {Association for Computing Machinery},
address = {New York, NY, USA},
url = {https://doi.org/10.1145/1273496.1273592},
doi = {10.1145/1273496.1273592},
booktitle = {Proceedings of the 24th International Conference on Machine Learning},
pages = {759–766},
numpages = {8},
location = {Corvalis, Oregon, USA},
series = {ICML ’07}

}

@article{Nigam2000,
  title={Text classification from labeled and unlabeled documents using EM},
  author={Nigam, Kamal and McCallum, Andrew Kachites and Thrun, Sebastian and Mitchell, Tom},
  journal={Machine learning},
  volume={39},
  number={2-3},
  pages={103--134},
  year={2000},
  publisher={Springer}

}

@article{LTE_Gesture_IPSN2020,
author = {Chen, Weiyan and Niu, Kai and Zhao, Deng and Zheng, Rong and Wu , Dan and Wang , Wei and Wang , Leye and Zhang, Leye },
doi = {10.1109/IPSN48710.2020.00017},
journal = {IPSN 2020 - Proceedings of the 2020 Information Processing in Sensor Networks},
title = {Robust Dynamic Hand Gesture Interaction using LTE Terminals},
year = {2020}

}

@inproceedings{WideSee_SenSys2019,
author = {Chen, Lili and Xiong, Jie and Chen, Xiaojiang and Lee, Sunghoon Ivan and Chen, Kai and Han, Dianhe and Fang, Dingyi and Tang, Zhanyong and Wang, Zheng},
title = {WideSee: Towards Wide-Area Contactless Wireless Sensing},
year = {2019},
isbn = {9781450369503},
publisher = {Association for Computing Machinery},
address = {New York, NY, USA},
url = {https://doi.org/10.1145/3356250.3360031},
doi = {10.1145/3356250.3360031},
booktitle = {Proceedings of the 17th Conference on Embedded Networked Sensor Systems},
pages = {258–270},
numpages = {13},
location = {New York, New York},
series = {SenSys ’19}

}

@ARTICLE{UAV_survey_2019,  author={A. {Fotouhi} and H. {Qiang} and M. {Ding} and M. {Hassan} and L. G. {Giordano} and A. {Garcia-Rodriguez} and J. {Yuan}},  journal={IEEE Communications Surveys   Tutorials},   
title={Survey on UAV Cellular Communications: Practical Aspects, Standardization Advancements, Regulation, and Security Challenges},   
year={2019},  volume={21},  number={4},  pages={3417-3442}


}

@manual{Lora,
  title        = {LoRa™	Modulation	Basics},
  organization = {Semtech Corporation},
  year         = 2015,
  url         = {https://web.archive.org/web/20190718200516/https://www.semtech.com/uploads/documents/an1200.22.pdf}
}

@manual{Sigfox,
  title        = {Sigfox Device Radio Specifications},
  organization = {Sigfox },
  year         = 2020,
  url         = {https://build.sigfox.com/sigfox-device-radio-specifications}
}

@inproceedings{RFOCUS_NSDI2020,
  title={Smart Radio Environments Empowered by Reconfigurable Intelligent Surfaces: How it Works, State of Research, and Road Ahead},
  author={Di Renzo, Marco and Zappone, Alessio and Debbah, Merouane and Alouini, Mohamed-Slim and Yuen, Chau and de Rosny, Julien and Tretyakov, Sergei},
  booktitle={17th USENIX Symposium on Networked Systems Design and Implementation (NSDI ’20)},
  year={2020}

}

@inproceedings{menggait,
  title={Gait Recognition for Co-Existing Multiple People Using Millimeter Wave Sensing},
  author={Meng, Zhen and Fu, Song and Yan, Jie and Liang, Hongyuan and Zhou, Anfu and Zhu, Shilin and Ma, Huadong and Liu, Jianhua and Yang, Ning},
  booktitle={Proceedings of the AAAI Conference on Artificial Intelligence},
  volume={34},
  number={01},
  pages={849--856},
  year={2020},
 url = {https://github.com/mmGait/people-gait}

}

@article{gu2020wife,
  title={WiFE: WiFi and Vision based Intelligent Facial-Gesture Emotion Recognition},
  author={Gu, Yu and Zhang, Xiang and Liu, Zhi and Ren, Fuji},
  journal={arXiv preprint arXiv:2004.09889},
  url = {https://drive.google.com/drive/folders/1OdNhCWDS28qT21V8YHdCNRjHLbe042eG},
  year={2020}

}

@article{Ganin2016,
author = {Ganin, Yaroslav and Ustinova, Evgeniya and Ajakan, Hana and Germain, Pascal and Larochelle, Hugo and Laviolette, Fran\c{c}ois and Marchand, Mario and Lempitsky, Victor},
title = {Domain-Adversarial Training of Neural Networks},
year = {2016},
issue_date = {January 2016},
publisher = {JMLR.org},
volume = {17},
number = {1},
issn = {1532-4435},
journal = {J. Mach. Learn. Res.},
month = jan,
pages = {2096–2030},
numpages = {35}

}

@inproceedings{zhu2017unpaired,
  title={Unpaired image-to-image translation using cycle-consistent adversarial networks},
  author={Zhu, Jun-Yan and Park, Taesung and Isola, Phillip and Efros, Alexei A},
  booktitle={Proceedings of the IEEE international conference on computer vision},
  pages={2223--2232},
  year={2017}

}

@inproceedings{WIHF-infocom2020,
  title={WiHF: Enable User Identified Gesture Recognition
with WiFi},
  author={Li, Chenning   and Liu,  Manni and  Cao, Zhichao },
  booktitle={IEEE INFOCOM 2020-IEEE Conference on Computer Communications},
  year={2020},
  organization={IEEE}
}

@inproceedings{qian2017inferring,
  title={Inferring motion direction using commodity wi-fi for interactive exergames},
  author={Qian, Kun and Wu, Chenshu and Zhou, Zimu and Zheng, Yue and Yang, Zheng and Liu, Yunhao},
  booktitle={Proceedings of the 2017 CHI Conference on Human Factors in Computing Systems},
  pages={1961--1972},
  year={2017}
}

@inproceedings{wang2019person,
  title={Person-in-WiFi: Fine-grained person perception using WiFi},
  author={Wang, Fei and Zhou, Sanping and Panev, Stanislav and Han, Jinsong and Huang, Dong},
  booktitle={Proceedings of the IEEE International Conference on Computer Vision},
  pages={5452--5461},
  year={2019}
}

@inproceedings{li2019making,
  title={Making the invisible visible: Action recognition through walls and occlusions},
  author={Li, Tianhong and Fan, Lijie and Zhao, Mingmin and Liu, Yingcheng and Katabi, Dina},
  booktitle={Proceedings of the IEEE International Conference on Computer Vision},
  pages={872--881},
  year={2019}
}

@book{Goodfellow-et-al-2016,
    title={Deep Learning},
    author={Ian Goodfellow and Yoshua Bengio and Aaron Courville},
    publisher={MIT Press},
    note={\url{http://www.deeplearningbook.org}},
    year={2016}
}

@misc{veen_2019, title={The Neural Network Zoo}, url={https://www.asimovinstitute.org/neural-network-zoo/}, journal={The Asimov Institute}, author={Veen, Fjodor van}, year={2019}, month={Apr},urldate = {2020-08-100}}

@inproceedings{Chen2020,
author = {Chen, Xi and Li, Hang and Zhou, Chenyi and Liu, Xue and Wu, Di and Dudek, Gregory},
title = {FiDo: Ubiquitous Fine-Grained WiFi-Based Localization for Unlabelled Users via Domain Adaptation},
year = {2020},
isbn = {9781450370233},
publisher = {Association for Computing Machinery},
address = {New York, NY, USA},
url = {https://doi.org/10.1145/3366423.3380091},
doi = {10.1145/3366423.3380091},
booktitle = {Proceedings of The Web Conference 2020},
pages = {23–33},
numpages = {11},
location = {Taipei, Taiwan},
series = {WWW ’20}
}

@article{Alazrai2020,
title = "A dataset for Wi-Fi-based human-to-human interaction recognition",
journal = "Data in Brief",
volume = "31",
pages = "105668",
year = "2020",
issn = "2352-3409",
doi = "https://doi.org/10.1016/j.dib.2020.105668",
url = "http://www.sciencedirect.com/science/article/pii/S235234092030562X",
author = "Rami Alazrai and Ali Awad and Baha’A. Alsaify and Mohammad Hababeh and Mohammad I. Daoud"
}

@article{Dumitru2010,
author = {Erhan, Dumitru and Bengio, Yoshua and Courville, Aaron and Manzagol, Pierre-Antoine and Vincent, Pascal and Bengio, Samy},
title = {Why Does Unsupervised Pre-Training Help Deep Learning?},
year = {2010},
issue_date = {3/1/2010},
publisher = {JMLR.org},
volume = {11},
issn = {1532-4435},
journal = {J. Mach. Learn. Res.},
month = mar,
pages = {625–660},
numpages = {36}
}

@article{hinton2006reducing,
  title={Reducing the dimensionality of data with neural networks},
  author={Hinton, Geoffrey E and Salakhutdinov, Ruslan R},
  journal={science},
  volume={313},
  number={5786},
  pages={504--507},
  year={2006},
  publisher={American Association for the Advancement of Science}
}

@article{liu2017pku,
  title={PKU-MMD: A large scale benchmark for continuous multi-modal human action understanding},
  author={Liu, Chunhui and Hu, Yueyu and Li, Yanghao and Song, Sijie and Liu, Jiaying},
  journal={arXiv preprint arXiv:1703.07475},
  year={2017}
}

@inproceedings{hsu2019enabling,
  title={Enabling identification and behavioral sensing in homes using radio reflections},
  author={Hsu, Chen-Yu and Hristov, Rumen and Lee, Guang-He and Zhao, Mingmin and Katabi, Dina},
  booktitle={Proceedings of the 2019 CHI Conference on Human Factors in Computing Systems},
  pages={1--13},
  year={2019}

}

@article{xue2020deepmv,
  title={DeepMV: Multi-View Deep Learning for Device-Free Human Activity Recognition},
  author={Xue, Hongfei and Jiang, Wenjun and Miao, Chenglin and Ma, Fenglong and Wang, Shiyang and Yuan, Ye and Yao, Shuochao and Zhang, Aidong and Su, Lu},
  journal={Proceedings of the ACM on Interactive, Mobile, Wearable and Ubiquitous Technologies},
  volume={4},
  number={1},
  pages={1--26},
  year={2020},
  publisher={ACM New York, NY, USA}

}

@article{santhalingam2020mmasl,
  title={mmASL: Environment-Independent ASL Gesture Recognition Using 60 GHz Millimeter-wave Signals},
  author={Santhalingam, Panneer Selvam and Hosain, Al Amin and Zhang, Ding and Pathak, Parth and Rangwala, Huzefa and Kushalnagar, Raja},
  journal={Proceedings of the ACM on Interactive, Mobile, Wearable and Ubiquitous Technologies},
  volume={4},
  number={1},
  pages={1--30},
  year={2020},
  publisher={ACM New York, NY, USA}

}

@article{fan2020learning,
  title={Learning Longterm Representations for Person Re-Identification Using Radio Signals},
  author={Fan, Lijie and Li, Tianhong and Fang, Rongyao and Hristov, Rumen and Yuan, Yuan and Katabi, Dina},
  journal={arXiv preprint arXiv:2004.01091},
  year={2020}

}

@inproceedings{zhu2016co,
  title={Co-occurrence feature learning for skeleton based action recognition using regularized deep LSTM networks},
  author={Zhu, Wentao and Lan, Cuiling and Xing, Junliang and Zeng, Wenjun and Li, Yanghao and Shen, Li and Xie, Xiaohui},
  booktitle={Thirtieth AAAI Conference on Artificial Intelligence},
  year={2016}
}

@article{yin2017comparative,
  title={Comparative study of cnn and rnn for natural language processing},
  author={Yin, Wenpeng and Kann, Katharina and Yu, Mo and Sch{\"u}tze, Hinrich},
  journal={arXiv preprint arXiv:1702.01923},
  year={2017}

}

@inproceedings{yao2019stfnets,
  title={Stfnets: Learning sensing signals from the time-frequency perspective with short-time fourier neural networks},
  author={Yao, Shuochao and Piao, Ailing and Jiang, Wenjun and Zhao, Yiran and Shao, Huajie and Liu, Shengzhong and Liu, Dongxin and Li, Jinyang and Wang, Tianshi and Hu, Shaohan and others},
  booktitle={The World Wide Web Conference},
  pages={2192--2202},
  year={2019}

}

@inproceedings{yu2019rfid,
  title={RFID based real-time recognition of ongoing gesture with adversarial learning},
  author={Yu, Yinggang and Wang, Dong and Zhao, Run and Zhang, Qian},
  booktitle={Proceedings of the 17th Conference on Embedded Networked Sensor Systems},
  pages={298--310},
  year={2019}

}

@ARTICLE {wang2020multi,
author = {F. Wang and J. Liu and W. Gong},
journal = {IEEE Transactions on Mobile Computing},
title = {Multi-Adversarial In-Car Activity Recognition using RFIDs},
year = {2020},
volume = {Early Access},
issn = {1558-0660},
pages = {1-1},
doi = {10.1109/TMC.2020.2977902},
publisher = {IEEE Computer Society},
address = {Los Alamitos, CA, USA},
month = {mar}
}

@article{Shi2020ADO,
  title={A dataset of radar-recorded heart sounds and vital signs including synchronised reference sensor signals},
  author={Kilin Shi and Sven Schellenberger and Christoph Will and Tobias Steigleder and Fabian Michler and Jonas Fuchs and Robert Weigel and Christoph Ostgathe and Alexander Koelpin},
  journal={Scientific Data},
  year={2020},
  url={https://gitlab.com/kilinshi/scidata_vsmdb},
  volume={7}

}

@ARTICLE{Wang2020,
  author={J. {Wang} and Q. {Gao} and X. {Ma} and Y. {Zhao} and Y. {Fang}},
  journal={IEEE Wireless Communications}, 
  title={Learning to Sense: Deep Learning for Wireless Sensing with Less Training Efforts}, 
  year={2020},
  volume={},
  number={},
  pages={1-7}

  
  }

@ARTICLE{Chen2019,
  author={Z. {Chen} and L. {Zhang} and C. {Jiang} and Z. {Cao} and W. {Cui}},
  journal={IEEE Transactions on Mobile Computing}, 
  title={WiFi CSI Based Passive Human Activity Recognition Using Attention Based BLSTM}, 
  year={2019},
  volume={18},
  number={11},
  pages={2714-2724}

  }

@ARTICLE{Gu2010,
  author={T. {Gu} and L. {Wang} and Z. {Wu} and X. {Tao} and J. {Lu}},
  journal={IEEE Transactions on Knowledge and Data Engineering}, 
  title={A Pattern Mining Approach to Sensor-Based Human Activity Recognition}, 
  year={2011},
  volume={23},
  number={9},
  pages={1359-1372}

  
  }

@article{Autokey2020,
author = {Wu, Yuezhong and Lin, Qi and Jia, Hong and Hassan, Mahbub and Hu, Wen},
title = {Auto-Key: Using Autoencoder to Speed Up Gait-Based Key Generation in Body Area Networks},
year = {2020},
issue_date = {March 2020},
publisher = {Association for Computing Machinery},
address = {New York, NY, USA},
volume = {4},
number = {1},
url = {https://doi.org/10.1145/3381004},
doi = {10.1145/3381004},
journal = {Proc. ACM Interact. Mob. Wearable Ubiquitous Technol.},
month = mar,
articleno = {32},
numpages = {23}

}

@INPROCEEDINGS{Ejacket2020,
  author={Q. {Lin} and S. {Peng} and Y. {Wu} and J. {Liu} and W. {Hu} and M. {Hassan} and A. {Seneviratne} and C. H. {Wang}},
  booktitle={2020 19th ACM/IEEE International Conference on Information Processing in Sensor Networks (IPSN)}, 
  title={E-Jacket: Posture Detection with Loose-Fitting Garment using a Novel Strain Sensor}, 
  year={2020},
  volume={},
  number={},
  pages={49-60}

  }

@INPROCEEDINGS{Rana2014,
  author={R. L. {Shinmoto Torres} and D. C. {Ranasinghe} and  {Qinfeng Shi} and A. P. {Sample}},
  booktitle={2013 IEEE International Conference on RFID (RFID)}, 
  title={Sensor enabled wearable RFID technology for mitigating the risk of falls near beds}, 
  year={2013},
  volume={},
  number={},
  pages={191-198}

  
  }

@ARTICLE{iot2020,
  author={Y. {He} and Y. {Chen} and Y. {Hu} and B. {Zeng}},
  journal={IEEE Internet of Things Journal}, 
  title={WiFi Vision: Sensing, Recognition, and Detection with Commodity MIMO-OFDM WiFi}, 
  year={2020},
  volume={},
  number={},
  pages={1-1}

  }

@article{Farhana2020,
title = "Device free human gesture recognition using Wi-Fi CSI: A survey",
journal = "Engineering Applications of Artificial Intelligence",
volume = "87",
pages = "103281",
year = "2020",
issn = "0952-1976",
doi = "https://doi.org/10.1016/j.engappai.2019.103281",
url = "http://www.sciencedirect.com/science/article/pii/S0952197619302441",
author = "Hasmath Farhana Thariq Ahmed and Hafisoh Ahmad and Aravind C.V."


}

@ARTICLE{zhou2020,
  author={S. {Zhou} and W. {Zhang} and D. {Peng} and Y. {Liu} and X. {Liao} and H. {Jiang}},
  journal={IEEE Communications Letters}, 
  title={Adversarial WiFi Sensing for Privacy Preservation of Human Behaviors}, 
  year={2020},
  volume={24},
  number={2},
  pages={259-263}

  
  }

@article{Peng2018,
author = {Peng, Liangying and Chen, Ling and Ye, Zhenan and Zhang, Yi},
title = {AROMA: A Deep Multi-Task Learning Based Simple and Complex Human Activity Recognition Method Using Wearable Sensors},
year = {2018},
issue_date = {June 2018},
publisher = {Association for Computing Machinery},
address = {New York, NY, USA},
volume = {2},
number = {2},
url = {https://doi.org/10.1145/3214277},
doi = {10.1145/3214277},
journal = {Proc. ACM Interact. Mob. Wearable Ubiquitous Technol.},
month = jul,
articleno = {74},
numpages = {16}



}

@article{special_issue2019,
author = {Guo, Bin and Zhang, Yanyong and Zhang, Daqing and Wang, Zhu},
title = {Special Issue on Device-Free Sensing for Human Behavior Recognition},
year = {2019},
issue_date = {February  2019},
publisher = {Springer-Verlag},
address = {Berlin, Heidelberg},
volume = {23},
number = {1},
issn = {1617-4909},
url = {https://doi.org/10.1007/s00779-019-01201-8},
doi = {10.1007/s00779-019-01201-8},
journal = {Personal Ubiquitous Comput.},
month = feb,
pages = {1–2},
numpages = {2}

}

@inproceedings{Kong2019,
author = {Kong, Hao and Lu, Li and Yu, Jiadi and Chen, Yingying and Kong, Linghe and Li, Minglu},
title = {FingerPass: Finger Gesture-Based Continuous User Authentication for Smart Homes Using Commodity WiFi},
year = {2019},
isbn = {9781450367646},
publisher = {Association for Computing Machinery},
address = {New York, NY, USA},
url = {https://doi.org/10.1145/3323679.3326518},
doi = {10.1145/3323679.3326518},
booktitle = {Proceedings of the Twentieth ACM International Symposium on Mobile Ad Hoc Networking and Computing},
pages = {201–210},
numpages = {10},
location = {Catania, Italy},
series = {Mobihoc ’19}


}

@ARTICLE{Ma2020,
  author={X. {Ma} and Y. {Zhao} and L. {Zhang} and Q. {Gao} and M. {Pan} and J. {Wang}},
  journal={IEEE Transactions on Industrial Informatics}, 
  title={Practical Device-Free Gesture Recognition Using WiFi Signals Based on Metalearning}, 
  year={2020},
  volume={16},
  number={1},
  pages={228-237}
}

@INPROCEEDINGS{Zou2019,
  author={H. {Zou} and J. {Yang} and H. P. {Das} and H. {Liu} and Y. {Zhou} and C. J. {Spanos}},
  booktitle={2019 IEEE/CVF Conference on Computer Vision and Pattern Recognition Workshops (CVPRW)}, 
  title={WiFi and Vision Multimodal Learning for Accurate and Robust Device-Free Human Activity Recognition}, 
  year={2019},
  volume={},
  number={},
  pages={426-433}

  }

@misc{liu2019deepcount,
    title={DeepCount: Crowd Counting with WiFi via Deep Learning},
    author={Shangqing Liu and Yanchao Zhao and Fanggang Xue and Bing Chen and Xiang Chen},
    year={2019},
    eprint={1903.05316},
    archivePrefix={arXiv},
    primaryClass={cs.LG}

    
    
}

@inproceedings{Xue2019,
author = {Xue, Hongfei and Jiang, Wenjun and Miao, Chenglin and Yuan, Ye and Ma, Fenglong and Ma, Xin and Wang, Yijiang and Yao, Shuochao and Xu, Wenyao and Zhang, Aidong and Su, Lu},
title = {DeepFusion: A Deep Learning Framework for the Fusion of Heterogeneous Sensory Data},
year = {2019},
isbn = {9781450367646},
publisher = {Association for Computing Machinery},
address = {New York, NY, USA},
url = {https://doi.org/10.1145/3323679.3326513},
doi = {10.1145/3323679.3326513},
booktitle = {Proceedings of the Twentieth ACM International Symposium on Mobile Ad Hoc Networking and Computing},
pages = {151–160},
numpages = {10},
location = {Catania, Italy},
series = {Mobihoc ’19}

}

@INPROCEEDINGS{Zhang2019,
  author={J. {Zhang} and F. {Wu} and W. {Hu} and Q. {Zhang} and W. {Xu} and J. {Cheng}},
  booktitle={2019 15th International Conference on Mobile Ad-Hoc and Sensor Networks (MSN)}, 
  title={WiEnhance: Towards Data Augmentation in Human Activity Recognition Using WiFi Signal}, 
  year={2019},
  volume={},
  number={},
  pages={309-314}

  
  }

@INPROCEEDINGS{Ming2019,
  author={X. {Ming} and H. {Feng} and Q. {Bu} and J. {Zhang} and G. {Yang} and T. {Zhang}},
  booktitle={2019 IEEE SmartWorld, Ubiquitous Intelligence   Computing, Advanced   Trusted Computing, Scalable Computing   Communications, Cloud   Big Data Computing, Internet of People and Smart City Innovation (SmartWorld/SCALCOM/UIC/ATC/CBDCom/IOP/SCI)}, 
  title={HumanFi: WiFi-Based Human Identification Using Recurrent Neural Network}, 
  year={2019},
  volume={},
  number={},
  pages={640-647}
  }

@url{warp,
	Title = {WARP Project},
	Url = {http://warpproject.org}}

@inbook{usrp,
author = {Ettus, Matt and Braun, Martin},
publisher = {John Wiley and Sons, Ltd},
isbn = {9781119057246},
title = {The Universal Software Radio Peripheral (USRP) Family of Low-Cost SDRs},
booktitle = {Opportunistic Spectrum Sharing and White Space Access},
chapter = {1},
pages = {3-23},
doi = {10.1002/9781119057246.ch1},
url = {https://onlinelibrary.wiley.com/doi/abs/10.1002/9781119057246.ch1},
eprint = {https://onlinelibrary.wiley.com/doi/pdf/10.1002/9781119057246.ch1},
year = {2015},
}

@inproceedings{nexmon,
author = {Gringoli, Francesco and Schulz, Matthias and Link, Jakob and Hollick, Matthias},
title = {Free Your CSI: A Channel State Information Extraction Platform For Modern Wi-Fi Chipsets},
year = {2019},
isbn = {9781450369312},
publisher = {Association for Computing Machinery},
address = {New York, NY, USA},
url = {https://doi.org/10.1145/3349623.3355477},
doi = {10.1145/3349623.3355477},
booktitle = {Proceedings of the 13th International Workshop on Wireless Network Testbeds, Experimental Evaluation and Characterization},
pages = {21–28},
numpages = {8},
location = {Los Cabos, Mexico},
series = {WiNTECH ’19}
}

@INPROCEEDINGS{WiGest,
  author={H. {Abdelnasser} and M. {Youssef} and K. A. {Harras}},
  booktitle={2015 IEEE Conference on Computer Communications (INFOCOM)}, 
  title={WiGest: A ubiquitous WiFi-based gesture recognition system}, 
  year={2015},
  volume={},
  number={},
  pages={1472-1480},}

@inproceedings{WiSee,
author = {Pu, Qifan and Gupta, Sidhant and Gollakota, Shyamnath and Patel, Shwetak},
title = {Whole-Home Gesture Recognition Using Wireless Signals},
year = {2013},
isbn = {9781450319997},
publisher = {Association for Computing Machinery},
address = {New York, NY, USA},
url = {https://doi.org/10.1145/2500423.2500436},
doi = {10.1145/2500423.2500436},
booktitle = {Proceedings of the 19th Annual International Conference on Mobile Computing and Networking},
pages = {27–38},
numpages = {12},
location = {Miami, Florida, USA},
series = {MobiCom ’13}
}

@inproceedings{WiFiU,
author = {Wang, Wei and Liu, Alex X. and Shahzad, Muhammad},
title = {Gait Recognition Using Wifi Signals},
year = {2016},
isbn = {9781450344616},
publisher = {Association for Computing Machinery},
address = {New York, NY, USA},
url = {https://doi.org/10.1145/2971648.2971670},
doi = {10.1145/2971648.2971670},
booktitle = {Proceedings of the 2016 ACM International Joint Conference on Pervasive and Ubiquitous Computing},
pages = {363–373},
numpages = {11},
location = {Heidelberg, Germany},
series = {UbiComp ’16}
}

@inbook{goldsmith_2005, place={Cambridge}, title={Path Loss and Shadowing}, DOI={10.1017/CBO9780511841224.003}, booktitle={Wireless Communications}, publisher={Cambridge University Press}, author={Goldsmith, Andrea}, year={2005}, pages={27–63}}

@article{TI-FMCW,
  title={The fundamentals of millimeter wave sensors},
  author={Iovescu, Cesar and Rao, Sandeep},
  journal={Texas Instruments, SPYY005},
  year={2017}
  }

@misc{celeno, title={Wi-Fi Doppler Imaging: Celeno - Wi-Fi Beyond Connectivity}, url={https://www.celeno.com/wifi-doppler-imaging}, journal={celeno}, author={Celeno},year = {2020},urldate = {2020-08-10}}

@misc{emerald, title={Emerald}, url={https://www.emeraldinno.com/}, journal={Emerald},year = {2020}}

@misc{walabot, title={Walabot Fall Alert System: Detect Falls with No Wearables}, url={https://walabot.com/walabot-home}, journal={Walabot Fall Alert System | Detect Falls with No Wearables | Award Winning Wall Scanner and Stud Finder},year = {2020},urldate = {2020-08-10}}

@misc{origin, title={Wireless Artificial Intelligence}, url={https://www.originwirelessai.com/}, journal={Origin Wireless AI},year = {2020},urldate = {2020-08-100}}

@misc{xkcorp, url={https://xkcorp.com/}, journal={XandarKardian Inc}, author={Thompson, Kris and Writer, Tina Danelsen Sr. and Danelsen, Tina and Writer, Sr.}, year={2020}, month={Jul},urldate = {2020-08-10}}

@misc{linksys, title={Linksys Aware}, url={https://www.linksys.com/us/linksys-aware/}, journal={Linksys},year={2020}, month={Jul},urldate = {2020-08-10}}

@article{samira2018,
author = {Pouyanfar, Samira and Sadiq, Saad and Yan, Yilin and Tian, Haiman and Tao, Yudong and Reyes, Maria Presa and Shyu, Mei-Ling and Chen, Shu-Ching and Iyengar, S. S.},
title = {A Survey on Deep Learning: Algorithms, Techniques, and Applications},
year = {2018},
issue_date = {January 2019},
publisher = {Association for Computing Machinery},
address = {New York, NY, USA},
volume = {51},
number = {5},
issn = {0360-0300},
url = {https://doi.org/10.1145/3234150},
doi = {10.1145/3234150},
journal = {ACM Comput. Surv.},
month = sep,
articleno = {92},
numpages = {36}
}

@article{lecun2015,
  title={Deep learning},
  author={LeCun, Yann and Bengio, Yoshua and Hinton, Geoffrey},
  journal={nature},
  volume={521},
  number={7553},
  pages={436--444},
  year={2015},
  publisher={Nature Publishing Group}
}

@ARTICLE{youssef2019,
  author={M. {Youssef} and M. {Hassan}},
  journal={IEEE Pervasive Computing}, 
  title={Next Generation IoT: Toward Ubiquitous Autonomous Cost-Efficient IoT Devices}, 
  year={2019},
  volume={18},
  number={4},
  pages={8-11},}

@inproceedings{solargest2019,
author = {Ma, Dong and Lan, Guohao and Hassan, Mahbub and Hu, Wen and Upama, Mushfika B. and Uddin, Ashraf and Youssef, Moustafa},
title = {SolarGest: Ubiquitous and Battery-Free Gesture Recognition Using Solar Cells},
year = {2019},
isbn = {9781450361699},
publisher = {Association for Computing Machinery},
address = {New York, NY, USA},
url = {https://doi.org/10.1145/3300061.3300129},
doi = {10.1145/3300061.3300129},
booktitle = {The 25th Annual International Conference on Mobile Computing and Networking},
articleno = {12},
numpages = {15},
location = {Los Cabos, Mexico},
series = {MobiCom '19}
}

@article{zhang-IMWUT2020,
author = {Zhang, Fusang and Chang, Zhaoxin and Niu, Kai and Xiong, Jie and Jin, Beihong and Lv, Qin and Zhang, Daqing},
title = {Exploring LoRa for Long-Range Through-Wall Sensing},
year = {2020},
issue_date = {June 2020},
publisher = {Association for Computing Machinery},
address = {New York, NY, USA},
volume = {4},
number = {2},
url = {https://doi.org/10.1145/3397326},
doi = {10.1145/3397326},
journal = {Proc. ACM Interact. Mob. Wearable Ubiquitous Technol.},
month = jun,
articleno = {68},
numpages = {27}
}

@inproceedings{salimans2016,
  title={Improved techniques for training gans},
  author={Salimans, Tim and Goodfellow, Ian and Zaremba, Wojciech and Cheung, Vicki and Radford, Alec and Chen, Xi},
  booktitle={Advances in neural information processing systems},
  pages={2234--2242},
  year={2016}
}

\begin{IEEEbiography}[{\includegraphics[width=1in,height=1.25in,clip,keepaspectratio]{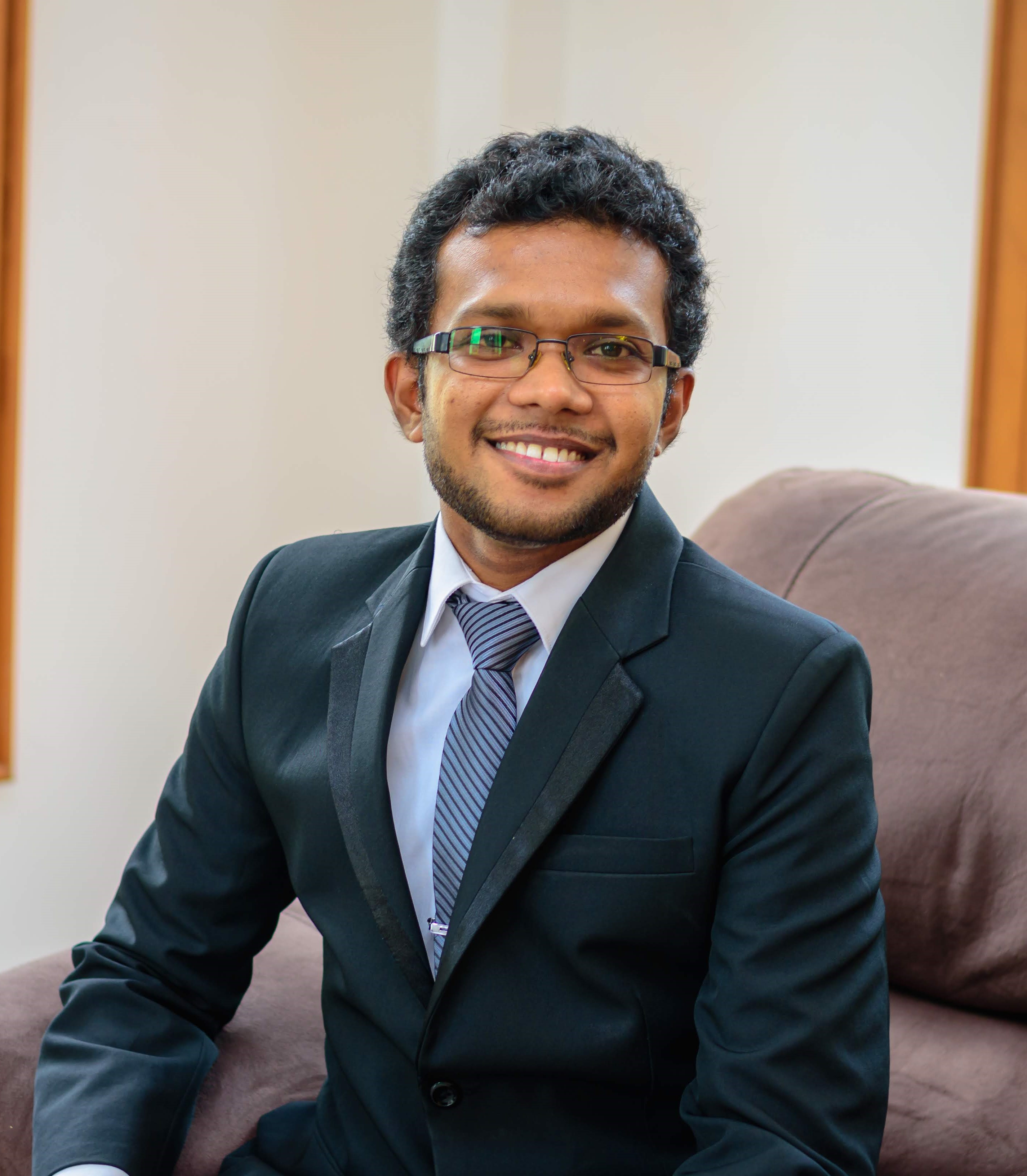}}]{Isura Nirmal}
is currently a PhD researcher in School of Computer Science and Engineering in University of New South Wales(UNSW), Sydney, Australia. He received his BSc in Information and Communication Technology from University of Colombo, Sri Lanka. His research interests are wireless sensing, IoT and deep learning which is the focus for his forthcoming PhD.
\end{IEEEbiography}
\begin{IEEEbiography}[{\includegraphics[width=1in,height=1.25in,clip,keepaspectratio]{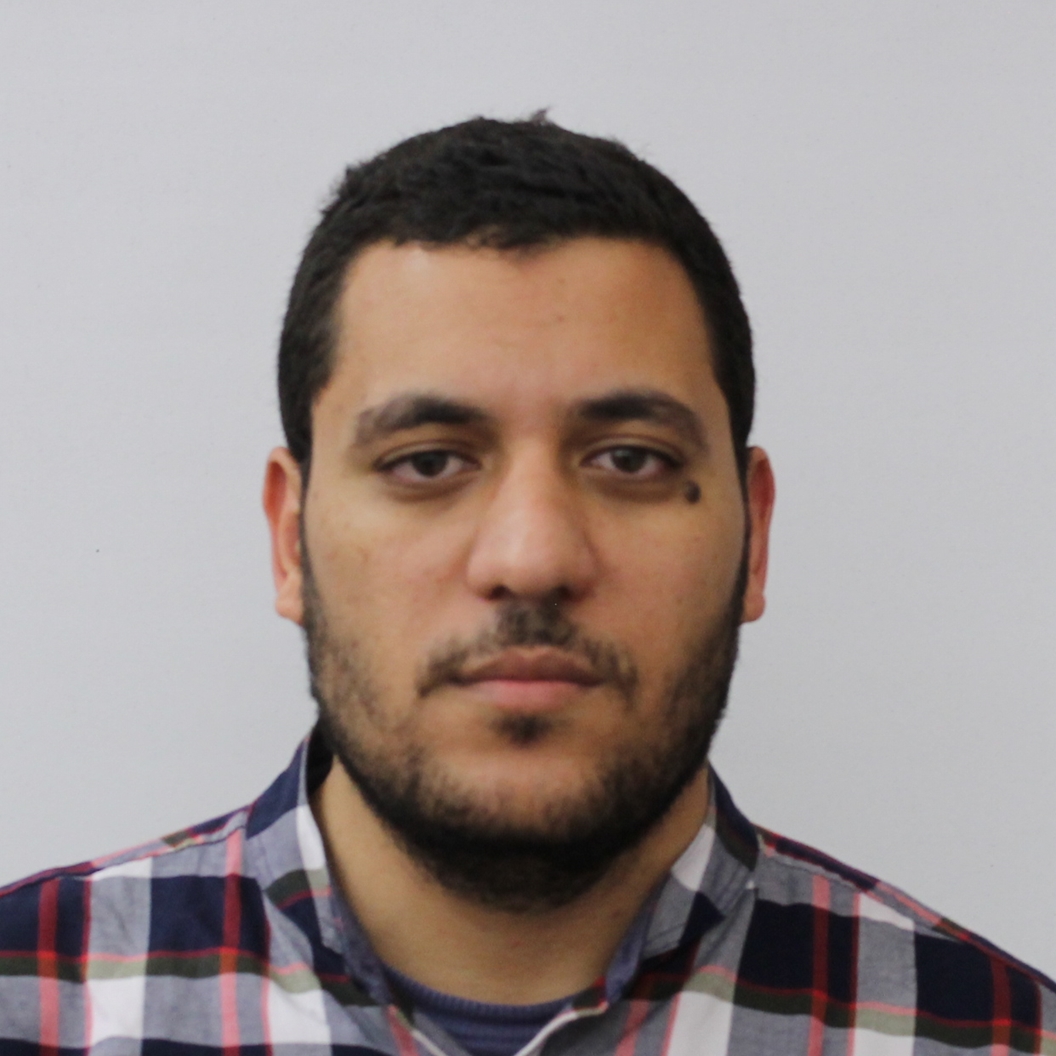}}]{Abdelwahed Khamis}
 is a Research Associate at the School of Computer Science and Engineering, UNSW, Sydney and a Visiting Scientist in Data61, CSIRO. His current research interests include ubiquitous and device-free sensing, mobile computing and wireless security. Abdelwahed completed his Ph.D in Computer Science and Engineering from UNSW, Australia in 2020. Prior to that he got Bsc and Msc in Computer Science from Zagazig University, Egypt. His PhD research focused on the use of RF technologies for medical sensing applications and resulted in a number of innovative contact-free sensing system for Hand Hygiene tracking and vital sign monitoring. His postdoctoral and doctoral research was sponsored by industry leading companies such as CISCO and Huawei.
\end{IEEEbiography}

\vskip -2\baselineskip plus -1fil
\begin{IEEEbiography}[{\includegraphics[width=1in,height=1.25in,clip,keepaspectratio]{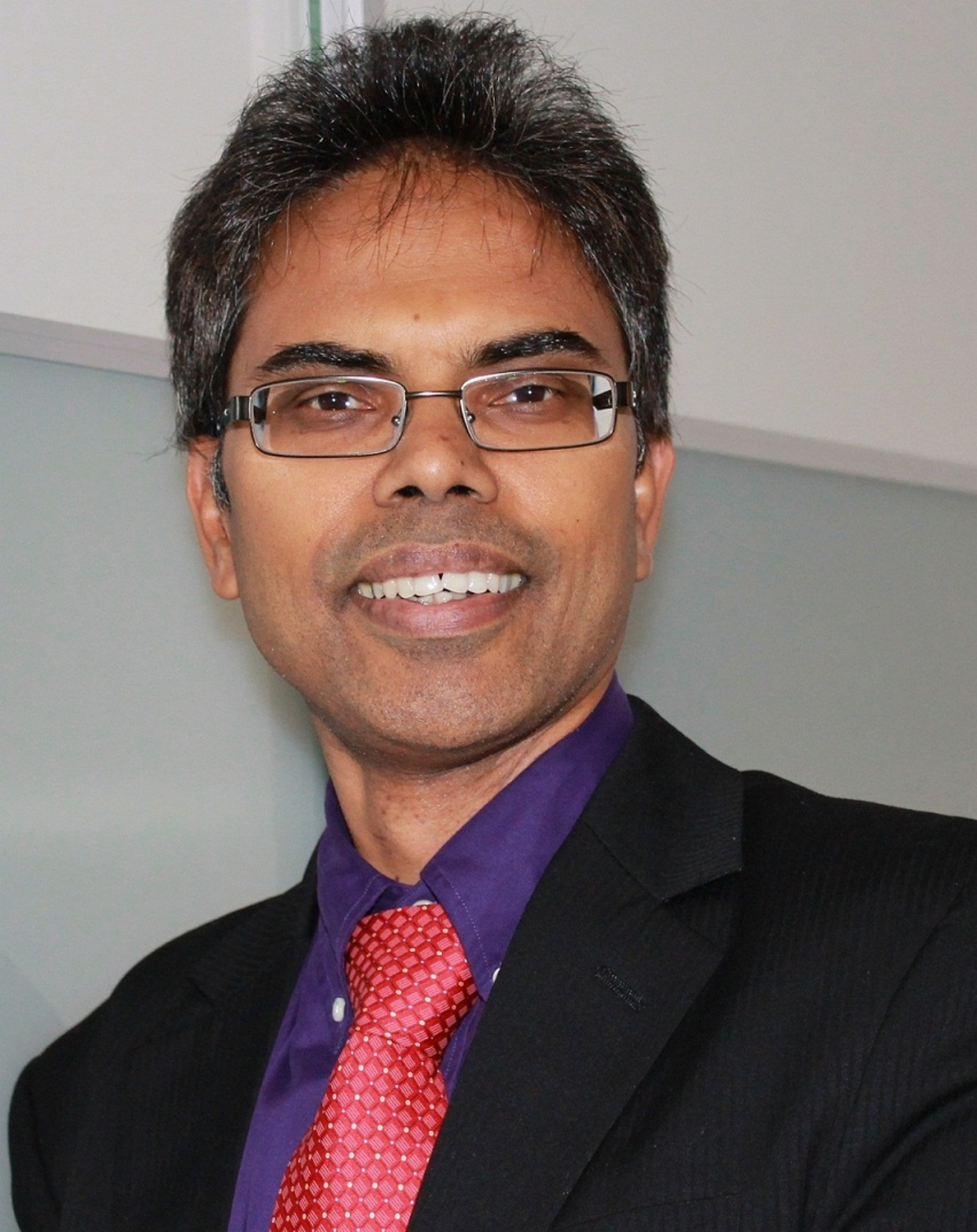}}]{Mahbub Hassan}
is a Full Professor in the School of Computer Science and
Engineering, University of New South Wales, Sydney, Australia. He has
PhD from Monash University, Australia, and MSc from University of
Victoria, Canada, both in Computer Science. He served as IEEE
Distinguished Lecturer and held visiting appointments at universities in USA, France, Japan, and Taiwan. He has co-authored three books, over 200 scientific articles, and a US patent. He served as editor or guest editor for many journals including IEEE Communications Magazine, IEEE Network, and IEEE Transactions on Multimedia. His current research interests
include Mobile Computing and Sensing, Nanoscale Communication, and Wireless Communication Networks.  More information is
available from \url{http://www.cse.unsw.edu.au/~mahbub}. 
\end{IEEEbiography}
\vskip -2\baselineskip plus -1fil

\begin{IEEEbiography}[{\includegraphics[width=1in,height=1.25in,clip,keepaspectratio]{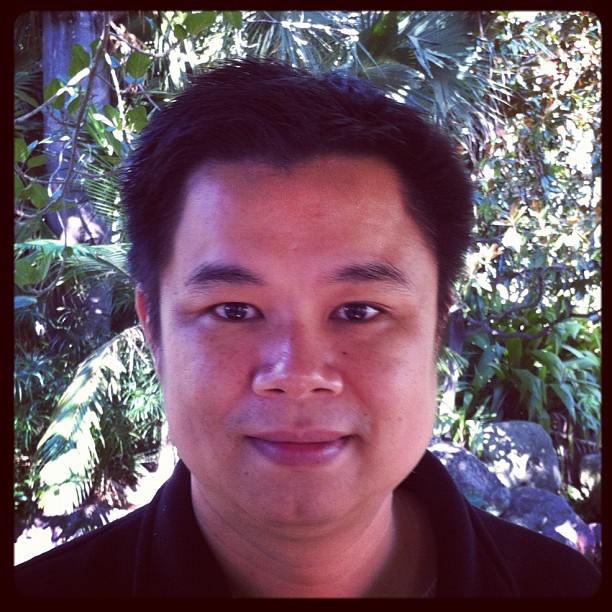}}]{Wen Hu}
is currently an Associate Professor with the
School of Computer Science and Engineering, University of New South Wales (UNSW). He has published regularly in the top-rated sensor network and mobile computing venues, such as  IPSN, SenSys, MobiCom, UbiComp, TOSN, the TMC, TIFS and the PROCEEDINGS OF THE
IEEE. His research interests focus on the novel applications, low-power communications, security and compressive sensing in sensor
network systems and the Internet of Things (IoT). He is a Senior Member of ACM and IEEE. He is an Associate Editor of ACM TOSN. He is the General Chair of CPS-IoT Week 2020, and serves on the organizing and program committees of networking conferences, including  IPSN,
SenSys, MobiSys,  MobiCom, and  IoTDI.
\end{IEEEbiography}
\vskip -2\baselineskip plus -1fil
\begin{IEEEbiography}[{\includegraphics[width=1in,height=1.25in,clip,keepaspectratio]{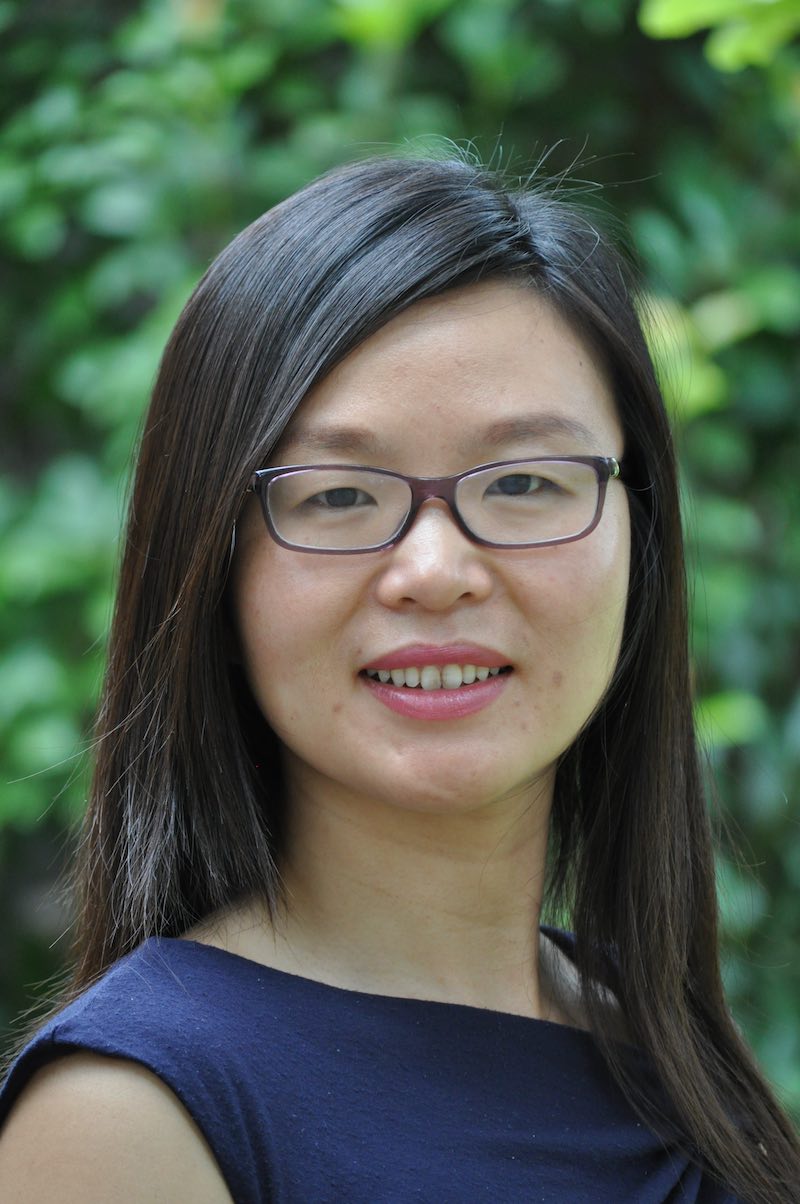}}]{Xiaoqing Zhu}
is a Sr. Technical Leader at the Innovation Labs of Cisco Systems, Inc. Her research interests include Internet video delivery, real-time interactive multimedia communications, distributed resource optimization, and machine learning for wireless. She holds a B.Eng. in Electronics Engineering from Tsinghua University, Beijing, China. She received both M.S. and Ph.D. degrees in Electrical Engineering from Stanford University, CA, USA. She has published over 80 peer-reviewed journal and conference papers, receiving the Best Student Paper Award at ACM Multimedia in 2007, the Best Presentation Award at IEEE Packet Video Workshop in 2013, and the Best Research Paper Award for VEHCOM 2017. She has over 30 granted  U.S. patents. Xiaoqing has served extensively within the multimedia research community, as TPC member and area chair for conferences, guest editor for special issues of leading journals. and more recently as chair of the MCDIG (Multimedia Content Distribution: Infrastructure and Algorithms) Interest Group in Multimedia Communication Technical Committee (MMTC) and Associate Editor for IEEE Transactions on Multimedia.
\end{IEEEbiography}

\end{document}